\def\bbibitem#1{\bibitem{#1}}

\documentstyle[prb,aps,amsfonts]{revtex}

\newcommand{\NN}{{\cal N}}

\newcommand{\skipp}[1]{}

\newcommand{\cmin}{\lesssim} 
\newcommand{\cmag}{\gtrsim}

\newcommand{\bfmath}[1]{\bbox{#1}} 

\def\d2{\mbox{$d_{x^2-y^2}$} }

\newcommand{\af}{an\-ti\-fer\-ro\-ma\-gne\-ti}

\def\cuo2{CuO$_2$ }

\newcommand{\rhs}{right-hand side }

\renewcommand{\v}[1]{\bfmath{#1}}

\newcommand{\E}{\varepsilon }

\newcommand{\mez}{{1  \over 2}}

\newcommand{\n}[2]{#1_{#2}^{\dag} #1_{#2}} 
 
\newcommand{\nn}[4]{#1_{#3}^{\dag} #2_{#4}}

\newcommand{\al}{\alpha}

\newcommand{\bet}{\beta}
\newcommand{\g}{\sigma}
\newcommand{\de}{\partial}
\newcommand{\lam}{\lambda}
\def\l0{\lambda_0}

\newcommand{\up}{\uparrow}
\newcommand{\down}{\downarrow}

\newcommand{\F}{{\cal F}}
\newcommand{\ZZ}{{\cal Z}}   

\def\b|{\Bigr |}
\newcommand{\wti}{\widetilde}

\newcommand{\R}{{\cal R}}
\newcommand{\unsuN}{{1 \over N}}

\renewcommand{\det}{\hbox{\rm det} \,}

\newcommand{\ord}{{\cal O}}

\newcommand{\MF}{mean-field~}
\newcommand{\MFL}{mean-field level~}
\newcommand{\FI}{functional integral~}
\newcommand{\CL}{continuum limit~}
\newcommand{\CFI}{contribution from infinity~}

\newcommand{\evi}{\bf}

\newcommand{\beq}{\begin{equation}}
\newcommand{\eeq}{\end{equation}}
\newcommand{\beqn}{\begin{eqnarraynn}}
\newcommand{\eeqn}{\end{eqnarraynn}}

\newcommand{\ffiglandvvvv}[6]{\def\figuradata{#5}
           \ifx\figurascelta\figuradata\figlandvvvv{#1}{#2}{#3}{#4}{#5}{#6}\fi}
\newcommand{\ffiglandvvv}[5]{\def\figuradata{#4}
           \ifx\figurascelta\figuradata\figlandvvv{#1}{#2}{#3}{#4}{#5}\fi}
\newcommand{\ffiglandvv}[4]{\def\figuradata{#3}
          \ifx\figurascelta\figuradata\figlandvv{#1}{#2}{#3}{#4}\fi}
\newcommand{\ffigportt}[4]{\def\figuradata{#3}
          \ifx\figurascelta\figuradata\figportt{#1}{#2}{#3}{#4}\fi}
\newcommand{\ffigland}[3]{\def\figuradata{#2}
          \ifx\figurascelta\figuradata\figland{#1}{#2}{#3}\fi}
\newcommand{\ffigport}[3]{\def\figuradata{#2}
          \ifx\figurascelta\figuradata\figport{#1}{#2}{#3}\fi}
\newcommand{\ffigportb}[3]{\def\figuradata{#2}
          \ifx\figurascelta\figuradata\figportb{#1}{#2}{#3}\fi}

\newcommand{\figref}[1]{\protect{\ref{#1}}}

\newcommand{\psland}[1]{\psfig{height=9truecm
            ,width=14truecm,bbllx=0pt
            ,bblly=0pt,bburx=612pt,bbury=792pt,angle=-90,file={#1}}}
\newcommand{\psport}[1]{\psfig{height=9truecm
            ,width=9truecm,bbllx=160pt
            ,bblly=100pt,bburx=660pt,bbury=600pt,angle=0,file={#1}}}
\newcommand{\psportb}[1]{\psfig{height=14truecm
            ,width=9truecm,bbllx=0pt
            ,bblly=0pt,bburx=620pt,bbury=800pt,angle=0,file={#1}}}

\newcommand{\pslands}[1]{\psfig{height=6truecm
            ,width=7.5truecm,bbllx=0pt
            ,bblly=0pt,bburx=612pt,bbury=792pt,angle=-90,file={#1}}}
\newcommand{\psports}[1]{\psfig{height=7.6truecm
            ,width=7.truecm,bbllx=60pt
            ,bblly=100pt,bburx=560pt,bbury=600pt,angle=0,file={#1}}}

\newcommand{\figland}[3]{
    \begin{figure}[h t b  ]
        \centerline{\psland{#1}}
        \caption{#3}
     \label{#2}
    \end{figure}
}
\newcommand{\figport}[3]{
    \begin{figure}[h t b  ]
        \centerline{\psport{#1}}
        \caption{#3}
     \label{#2}
    \end{figure}
}
\newcommand{\figportb}[3]{
    \begin{figure}[h t b  ]
        \centerline{\psportb{#1}}
        \caption{#3}
     \label{#2}
    \end{figure}
}

\newcommand{\figlandvvvv}[6]{
    \begin{figure}[h t b  ]
       \begin{tabular}{l l}
        \pslands{#1} & \pslands{#2} \\
        \pslands{#3} & \pslands{#4} 
        \end{tabular}
        \caption{#6}
     \label{#5}
    \end{figure}
       }
\newcommand{\figlandvvv}[5]{
    \begin{figure}[h t b  ]
       \begin{tabular}{l l}
        \pslands{#1} & \pslands{#2} \\
        \pslands{#3} & 
        \end{tabular}
        \caption{#5}
     \label{#4}
    \end{figure}
       }
\newcommand{\figlandvv}[4]{
    \begin{figure}[h t b  ]
       \begin{tabular}{l l}
        \pslands{#1} & \pslands{#2} 
        \end{tabular}
        \caption{#4}
     \label{#3}
    \end{figure}
       }

\newcommand{\figportt}[4]{
    \begin{figure}[h t b  ]
       \begin{tabular}{c c}
        \psports{#1}    &\psports{#2}
        \end{tabular}
        \caption{#4}
     \label{#3}
    \end{figure}
}

\newcommand{\eqref}[1]{(\ref{#1})}

\catcode`\@=11       
\newcommand{\uti}[1]{\mathop{\vtop{\ialign{##\crcr
$\hfil\displaystyle{#1}\hfil$\crcr\noalign{\kern3\p@\nointerlineskip}
      $\scriptstyle\sim$\crcr\noalign{\kern3\p@}}}}\limits\!}

\newcommand{\barw}[1]{\bar{#1}\!\,}

\newcommand{\baruti}[1]{\uti{\bar #1}}

\newcommand {\tr}{\hbox{\rm tr}\,}

\newcommand{\undersymbolsub}[2]{\mathop{\vtop{\ialign{##\crcr
$\hfil\displaystyle{#2}\hfil$\crcr\noalign
{\kern1pt\nointerlineskip}\hbox{$\hfil\scriptstyle{#1}\hfil$}\crcr
\noalign{\kern1pt}}}}}

\newcommand{\men}{- }

\newcommand{\wwbar}{\overline}

\newcommand{\fiem}{ \varphi_{i,m}^{(e)} }

\newcommand{\fidm}{ \varphi_{i,m}^{(d)} }
\newcommand{\fism}{ \varphi_{i,m}^{(s_{\g})} }

\newcommand{\fiemu}{ \varphi_{i,m-1}^{(e)} }
\newcommand{\fismu}{ \varphi_{i,m-1}^{(s_{\g})} }

\newcommand{\undel}{{{1\over\delta}\,} }

\newcommand{\is}{_{i,\g} }

\newcommand{\nene}{\v \Delta }

\newcommand{\js}{_{i+\nene,\g} }

\newcommand{\ism}{_{i,\g,m} }
\newcommand{\jsm}{_{i+\nene,\g,m} }
\newcommand{\im}{_{i,m} }
\newcommand{\jm}{_{i+\nene,m} }
\newcommand{\ismu}{_{{i,\g,m-1}\,} }
\newcommand{\imsm}{_{i,\men \g,m} }
\newcommand{\imsmu}{_{{i,\men \g,m-1}\,} }

\newcommand{\imu}{_{{i,m-1}\,} }
\newcommand{\jmu}{_{{i+\nene,m-1}\,} }

\newcommand{\cc}{^{\ast} }

\newcommand{\dg}{^{\dag} }

\newcommand{\lbsi}{{\,\lambda}^{b}_{i,\g} }
\newcommand{\lai}{{\,\lambda}^{a}_{i} }
\newcommand{\laimu}{\lambda^{a}_{i,m-1 } }
\newcommand{\lbsimu}{\lambda^{b}_{{i,\g,m-1}\,} }
\newcommand{\laim}{\lambda^{a}_{i,m } }
\newcommand{\lbsim}{\lambda^{b}_{{i,\g,m}\,} }
\newcommand{\lrarrow}{\leftrightarrow}

\newcommand{\latimu}{\Lambda^{a}_{i,m-1 } }
\newcommand{\lbtsimu}{\Lambda^{b}_{{i,\g,m-1}\,} }
\newcommand{\latim}{\Lambda^{a}_{i,m } }
\newcommand{\lbtsim}{\Lambda^{b}_{{i,\g,m}\,} }

\newcommand{\ab}[1]{{\,a}^{(#1)} }
\newcommand{\bb}[1]{{\,b}^{(#1)} }
\newcommand{\bbd}[1]{{\,b}^{(#1) 2} }
\newcommand{\bbdag}[1]{{\,b}^{(#1) \dag} }
\newcommand{\bbc}[1]{{\,b}^{(#1) \ast} }

\newcommand{\llam}[1]{{\,\lam}^{(#1)} }
\newcommand{\und}[1]{\underline{#1} }

\newcommand{\rhog}{{\,\rho}^{\g} }

\newcommand{\del}{\delta}

\newcommand{\oms}{\omega_{s} }
\newcommand{\eiod}{{\,e}^{i\oms\del} }
\newcommand{\emiod}{{\,e}^{-i\oms\del} }

\newcommand{\gamk}{{\,\gamma}(\v k)\, }
\newcommand{\gamkq}{{\,\gamma}(\v k-\v q)\, }
\newcommand{\gamkpq}{{\,\gamma}(\v k+\v q)\, }

\newcommand{\omnu}{\omega_{\nu} }
\newcommand{\eind}{{\,e}^{i\omnu\del} }
\newcommand{\emind}{{\,e}^{-i\omnu\del} }

\newcommand{\kaom}{{(\v k,\oms)\,} }
\newcommand{\kqon}{{(\v k-\v q,\oms-\omnu)\,} }
\newcommand{\qunu}{{(\v q,\omnu)\,} }
\newcommand{\mqunu}{{(-\v q,-\omnu)\,} }

\newcommand{\dsum}{\displaystyle \sum}

\newcommand{\delK}{{\,\delta}^{K} }
\newcommand{\dnuo}{\delK_{{\nu,0}\,}}

\newcommand{\XYMFO}{\und X=(\und b_0,\und 0) ,\,\und Y = (\und b_0,\und 0)}

\newcommand{\ia}{\alpha}
\newcommand{\ib}{\beta}
\newcommand{\il}{l}

\newcommand{\Gam}{\Gamma }

\newcommand{\BB}{{\cal B} }
\newcommand{\EE}{{\cal E} }
\newcommand{\CC}{{\cal C} }

\renewcommand{\evi}{\it}
\renewcommand{\beqn}{\begin{eqnarray}}
\renewcommand{\eeqn}{\end{eqnarray}}
\newcommand{\nonum}{\nonumber}

\newcommand{\typeoftext}{
\widetext    
}

 \tighten   

\begin{document}

 \preprint{Submitted to Phys. Rev. B}

\draft   

\title{
On the correct continuum limit \\ 
of the functional-integral
representation \\
 for the four-slave-boson approach to the Hubbard
model: \\
 Paramagnetic phase}

\author{E. Arrigoni
\thanks{Present address:
Institut f\"ur Theoretische Physik, Universit\"at W\"urzburg, D-97074
W\"urzburg, Germany}
}
\address{ 
Max-Planck-Institut f\"ur Physik
Complexer Systeme, D-70569 Stuttgart, Fed. Rep. Germany }

\author{ G.C. Strinati 
\thanks{Present address: 
Dipartimento di Matematica e Fisica, Universit\`a di Camerino, I-62032
Camerino, Italy}
}

\address{   Dipartimento di Fisica, III Universit\`a di Roma, 
 I-00146 Roma, Italy 
}

\maketitle

\begin{abstract}
The Hubbard model with finite on-site repulsion $U$ is studied via the
functional-integral formulation of the four-slave-boson approach by
Kotliar and Ruckenstein. It is shown that a correct treatment of the
continuum imaginary time limit (which is required by the very
definition of the functional integral) modifies the free energy when
fluctuation ($1/N$) corrections beyond  mean-field 
are considered, thus removing the inconsistencies originating from the
incorrect handling of this pathologic limit  so far 
performed in the literature. In particular, our treatment correctly 
restores the
 decrease of the average number of doubly occupied sites for
increasing $U$.  Our analysis requires us to suitably interpret the
Kotliar and Ruckenstein 
choice for the bosonic hopping 
operator and to abandon the commonly used normal-ordering prescription, 
in order to obtain
meaningful
fluctuation corrections. 
In this way
we recover the exact solution at $U=0$ 
not only at the \MF level but {\evi also} at the next  order in $1/N$. In
addition, we consider
alternative choices for the bosonic hopping operator 
 and test them numerically for a simple two-site model
for which the exact solution is readily available for any $U$. We also
discuss how the $1/N$ expansion can be formally generalized to the
four-slave-boson approach, and provide a simplified
prescription to obtain the additional terms in the free energy which
result at the order $1/N$ from  the correct continuum limit.
\end{abstract}


\typeoftext

\section{INTRODUCTION}
\label{s:intr} 
The Hubbard model (with its variations) is often adopted to represent
the essential correlations among electrons (holes) in a lattice,
especially in the intermediate- and strong-coupling regimes (relevant
 to heavy-fermion and high-Tc materials). In these regimes, the
use of conventional many-body techniques (that rely on expansions in
terms of the on-site repulsion $U$) becomes questionable and
alternative non-perturbative methods are required.\cite{fulde}
 Specifically, in the strong-coupling limit the need for {\evi projecting out}
the electronic configurations with double occupancy at any given site
acquires the role of a ``constraint'' which has to be enforced for a
correct description of correlations. Special methods have accordingly
been developed to satisfy the local constraints in the strong-coupling
limit.\cite{vollhardt}

In particular, slave-boson representations have been introduced to
allow for field-theoretical treatments of strong correlations with 
 constraints,  amounting to a projection method that employs
auxiliary (or ``slave'') bosonic particles.\cite{barnes} While in the
original version slave bosons were referring only to empty and doubly occupied
states at any given lattice site, the method was later extended by
Kotliar and Ruckenstein\cite{kotliar} (KR) who assigned slave bosons also
to the singly-occupied states. The KR four-slave-boson method
  maps the physical fermion (destruction) operator $\wti f\is$
with spin $\g$ at site $i$ onto the product of a (pseudo) fermion
$f\is$ {\evi with the same spin } and of a bosonic operator $z\is$
(see below),
and is especially suited to deal with the Hubbard model for {\evi any}
$U$.
 The KR method is also particularly appealing for the
treatment of magnetic problems, which can be approached already at the
mean-field-level owing to the presence of the single-occupancy bosons.
For these reasons the KR method has been adopted to treat several
problems, both at its mean-field level\cite{m2s3} and with the
inclusion of fluctuation corrections,\cite{lavagna} by relying on a
functional-integral formulation which is ideally suited to implement
the local constraints for slave bosons.\cite{wolfle} Specifically, it
has been found that the KR mean-field solution gives remarkable
agreement with more elaborate Monte Carlo results over a wide range of
interactions and particle densities.\cite{lilly} This success is not
entirely surprising since the KR (paramagnetic) mean-field solution
has been modeled after the Gutzwiller results and thus is identical
to the Gutzwiller approximation\cite{gutzwiller} (the latter having been
further shown recently to become exact in infinite
dimensions\cite{ref2}). The KR finding that the Gutzwiller results can
be recovered without explicit wave functions has then prompted the
hope to improve the theory systematically by the inclusion of
fluctuation corrections, using the KR saddle-point solution as a good
starting point.  

The use of the KR four-slave-boson method was for us originally
motivated by the interest in describing  the magnetic properties of
\cuo2 layers with an itinerant approach in the limit of large
repulsion $U_d$ at Cu sites. Antiferromagnetic calculations at the
mean-field level within a three-band Hubbard model with $U_d \approx
\infty$ (where the double occupancy boson does not enter) were 
actually quite promising.\cite{noiprb90} However, our attempts to include
fluctuations along the lines of Refs.\cite{lavagna} and \cite{wolfle}
met with  inconsistencies, leading to an instability
of our previous mean-field results. Further consideration of the 
single-band Hubbard model with {\evi finite} $U$ did not remove the
unpleasant features of fluctuation calculations (for instance, 
 the average number of
doubly occupied sites 
away from ``half filling'' 
was found to unphysically increase for increasing $U$
 with the inclusion of fluctuations).
 We were thus unavoidably led to conclude that
there was something {\evi systematically} wrong in the standard
procedure adopted in the literature to include fluctuation corrections
within the KR method.

About at the same time Jolicoeur and Le Guillou \cite{jolicoeur}
signaled the occurrence of additional inconsistencies 
 when including fluctuation corrections within the KR
method  at $U=0$,
 and proposed to modify the bosonic hopping operator $z$ mentioned
above at each order in the loop expansion 
(without suggesting, however, any explicit form of $z$ even at
the one-loop order).

On the other hand, it was soon clear that the inconsistencies found when including
fluctuations were not limited to the KR four-slave-boson method (and
thus  to the  presence of
the operator $z$  whose choice is to a large extent arbitrary). In
fact, Emery and coworkers \cite{zhang} have shown that inconsistencies
arise when including fluctuations even in the simpler $U=\infty$
one-slave-boson problem (where only the empty boson appears), which is
free from the arbitrariness related to the operator $z$. This result
then 
suggested that the origin of the problems was in the slave-boson
approach itself, irrespective of its particular
version.

The solution to these problems was  obtained eventually by 
deeper examination of the functional-integral formulation on which
slave-boson methods  rely. Specifically, it turned
out that the {\evi continuum} imaginary time {\evi limit} of the
functional-integral representation of the partition function had been 
incorrectly
implemented beyond the mean-field level in the slave-boson
literature, by performing the continuum limit  in
the action at the outset.
 By doing so, the bosonic commutators were not properly
represented in the functional integral, a shortcoming which 
turns out to be  crucial in the presence of (slave-) boson condensates.
The peculiarities of the continuum imaginary time limit within a
coherent-state functional-integral approach in the presence of a
slave-boson condensate have been originally demonstrated in the
context of the simpler $U=\infty$ one-slave-boson
problem.\cite{noiprep} In the present paper we concentrate on the
implications of taking the correct continuum limit with the
four-slave-boson method, which prove to be nontrivial 
due 
 to the presence  of the
bosonic operator $z$. A short preliminary version of the results
presented here can be found in Ref. \cite{noiprl}

Taking the continuum time limit  at the end of the calculation 
is actually a well-established procedure 
for  Feynman's path integrals.\cite{feynman} On the other hand, for
the functional integrals based on the coherent-state representation it
has been common practice to consider the continuum limit at the
outset, although  warnings have been given that this procedure might lead to
inconsistencies.\cite{schulman}
Taking the continuum limit at the outset considerably
simplifies the calculations, by making the expression of the action
more manageable and allowing for the use of standard Matsubara techniques
developed for the  diagrammatic Green's functions
approach.\cite{rickayzen} Conversely, keeping a finite imaginary time
mesh until the end of the calculation requires one to reconsider the
Matsubara techniques for a finite set of (fermionic or/and bosonic)
frequencies. Although this  task has been avoided in the
previous slave-boson literature,  our work \cite{noiprep,noiprl}
demonstrates unambiguously that, in the presence of bosonic
condensates, it is necessary 
to preserve the discretized form of the action until the end of the
calculation when resorting to a coherent-state functional-integral
representation of the partition function.
 In fact, proper account of the discretized nature of the
functional integral yields {\evi additional terms} for the free energy
when including fluctuations, which would otherwise be missed by taking the
continuum limit at the outset.

 Concerning further the KR
four-slave-boson method, the presence of several bosons
and of 
 the bosonic operator $z$ makes unavoidably the
derivation of the above  additional terms  more involved than for the
one-slave-boson method considered  in Ref. \cite{noiprep}.
 These additional terms turn out to be essential for both methods to
heal the inconsistencies found in the literature when taking the
continuum limit at the outset. In the present paper we  derive
in detail   these additional terms for the
four-slave-boson method with a {\evi generic} form of $z$.
In particular, we will show that the form of the bosonic operator $z$
proposed by Kotliar and Ruckenstein (which,
 for the sake of a direct mapping onto a functional-integral formulation,
has been invariably
interpreted in the literature within a normal-ordering prescription
 and which
will be referred to as $z_{KR}$ in the following) leads to
inconsistencies when including fluctuations even by taking carefully
the continuum limit at the end of the calculation.\cite{noiprl}
Nonetheless, we will also show that these inconsistencies can be
overcome by suitably relaxing the normal-ordering prescription on the
KR choice for the bosonic operator $z$. In this way, the preliminary
results presented in Ref.\cite{noiprl} are extended and improved. In
particular, this form of $z$ without the normal-ordering prescription
(which we shall refer to as $z_{SQ}$) turns out to reproduce the
correct independent-particle solution at $U=0$ not only at the \MF
level but {\evi also} with the inclusion of the $1/N$ corrections.
(Numerical results will be considered in the zero-temperature limit
only throughout this paper).
In addition,
 we shall  consider 
for comparison two alternative
 forms of $z$
(different from $z_{KR}$ and $z_{SQ}$) 
which are also free from the inconsistencies occurring for $z_{KR}$.
Specifically, we will consider a ``linearized'' 
form  $z_{LIN}$ which
preserves, by construction, the KR mean-field solution 
at ``half filling'' of the
paramagnetic 
band
 for any value of $U$ and also 
 strongly suppresses the contribution of fluctuations to the
free energy at $U=0$ (although not completely and at ``half filling'' only).
Our finding that the results obtained for the ground-state energy
alternatively with  $z_{SQ}$ and  $z_{LIN}$ do not differ appreciably
for large values of $U$ (say, for $U \cmag 4t$) even away from ``half
filling''  hints further that the results obtained with the
four-slave-boson method might not depend crucially on the choice of the
operator $z$ for practical purposes.

Numerical calculations will be mainly carried out  for a one-level
two-site model, which avoids the complications due to the spatial
structure and keeps those due to the imaginary time discretization of
the functional integral which we are mostly interested in. Restriction
to a two-site model will enable us to compare our numerical results,
obtained within alternative approximations to the functional integral,
with the exact solution which is readily available for any value of
$U$ and band filling.

The plan of the paper is the following. Section \ref{s:func} deals with
the formal problems related to the correct functional-integral
formulation of the  four-slave-boson method over a discretized
imaginary time mesh, and with the peculiar problems related to the
choice of the bosonic factor $z$. The novel terms for the free energy,
resulting at the Gaussian level from taking the continuum limit of the
functional integral only at the end of the calculation, are derived in
detail for any given form of the bosonic factor $z$ and interpreted by
the need of restoring the correct bosonic commutators. 
 Section
\ref{s:nume} presents our numerical calculations with the
four-slave-boson method for the ground-state energy of a simple
two-site model and compares them with the available exact results for
any $U$ and band filling.
The results obtained with alternative forms
 of the bosonic operator $z$ are discussed in this
context.
Section \ref{s:conc} gives our conclusions. Finally, the Appendices
contain additional mathematical arguments which are needed for making
the material presented in the text self-contained. In particular,
Appendix \ref{app:a} 
 discusses the $1/N$ expansion for the
four-slave-boson method, and Appendix \ref{app:c} recovers within the
``Cartesian'' gauge  the
additional terms for the free energy (which were derived in the text
within the ``radial'' gauge).
The problem of obtaining the additional terms (due to a correct
handling of the \CL of the \FI) also for the correlation functions
(and not only for the free energy) is briefly discussed in this context.

\section{FUNCTIONAL-INTEGRAL FORMULATION 
OF THE FOUR-SLAVE-BOSON METHOD}
\label{s:func} 
In this Section, we discuss the procedure to formulate the 
four-slave-boson method via the coherent-state functional-integral
representation of the partition function, by keeping the discretized
imaginary time mesh (which is required by the very definition of the
functional integral) until the end of the calculation.
Specifically, we will demonstrate at the Gaussian level that this  correct 
procedure yields additional terms for the free energy,
which were missed in the previous literature when the continuum limit
of the functional integral was taken incorrectly  in the
effective action at the outset.
By doing so, we extend to the four-slave-boson method the results
obtained in Ref. \cite{noiprep} for the simpler $U=\infty$
one-slave-boson problem. Mastering the formal apparatus with a
discretized time mesh, however, is now
unavoidably more involved.

For the sake of definiteness, we consider the single-band Hubbard
Hamiltonian 
\beq
\label{ham} 
 H = t \sum_{i,\nene,\g}   \nn {\wti f}{\wti f}{i,\g}{i+\nene,\g}  +
 U \sum_i \n {\wti f}{i,\up} \n {\wti f}{i,\down}
\eeq
with on-site repulsion $U$, where $\g$ is a spin label and 
$\nene$ runs over the ``star'' of
nearest neighbors to site $i$ in a two-dimensional square lattice
of $\NN$ sites.
 Extension of the following arguments to more
complex lattices with multi-band structures and to off-site repulsive
terms should, in principle, be straightforward.

As mentioned in the Introduction, an essential requirement on any
method for an approximate solution of \eqref{ham} in the
strong-coupling regime ($U \gg t$) is that double occupancy at a given
site is properly treated. To this end, suppression of double occupancy
in the limit $U/t\to\infty$ acquires the role of a ``constraint'' for
the many-body problem, whose enforcement poses no problem for systems
of 
small size but is difficult to implement for an infinite
system. In this case, the introduction of auxiliary variables in the
form of slave bosons, which keep track of the occupancy  at any
given lattice site, looks particularly convenient for enforcing the
``constraint''. 

\subsection{ THE FOUR-SLAVE-BOSON METHOD} 
For any given $U$, Kotliar and Ruckenstein \cite{kotliar} have mapped
the physical fermion operator $\wti f\is$ in Eq. \eqref{ham} onto the
product $f\is z\is$ of a (pseudo) fermion $f\is$ and of a bosonic
operator $z\is$ (at any given site). In its {\evi simplest} version
$z\is$ has the form
\beq
z\is = z\is^0 = s_{i,-\g}\dg d_i + e_i\dg s\is \;, 
\label{z1} 
\eeq
where the bosonic destruction operators $e_i$, $s\is$, and $d_i$ refer
to empty, singly, and doubly occupied states (in the order) at site
$i$. To establish a one-to-one correspondence between the original
Fock space and the enlarged one which contains also the bosonic states, the
following constraints must be satisfied:
\begin{mathletters}
\beqn
 &&
\n di + \sum_{\g} \n s{i,\g} + \n ei =1   
\label{constraintsa}
\\
 && \n f{i,\g} - \n s{i,\g} - \n di =0. 
\label{constraintsb}
\eeqn
\label{constraints} 
\end{mathletters}
However, enforcement of the constraints \eqref{constraints} allows for
alternative choices of $z\is$ different from \eqref{z1}. This
observation has been exploited by
Kotliar and Ruckenstein who have 
accordingly modified the form
\eqref{z1} 
 as follows
\beq
\label{zr} 
 z_{i,\g}=  s^{\dag}_{i,-\g} R_{i,\g} d_i + e^{\dag}_i R_{i,\g} s_{i,\g} \;,
\eeq
with the requirement that the {\evi subsidiary operator} $ R_{i,\g}$
acts as the unit operator in the subspace specified by the constraints
\eqref{constraints}. In particular, Kotliar and Ruckenstein
 have considered the
nonlinear form 
\label{rkrsq} 
\begin{mathletters}
\beqn
 R_{i,\g} &&=  R_{i,\g}^{KR} =  :R_{i,\g}^{SQ}:
\label{rkr} 
\\
  R_{i,\g}^{SQ} && = \frac{1}{\sqrt{1- \n di - \n s{i,\g}} \sqrt{1- \n ei - \n
s{i,-\g}}} \;,
\label{rsq} 
\eeqn
\end{mathletters}
where $:{\cal O}:$ denotes the normal ordering of the operator $\cal O$.
This expedient has enabled  Kotliar and Ruckenstein
 to recover the Gutzwiller solution for a
paramagnetic band in a straightforward way, by considering the
mean-field solution whereby each bosonic operator is replaced by its
average value. Other forms of  $R_{i,\g}$ different from the KR choice
\eqref{rkr} are, however, possible in principle. We shall exploit this
freedom in the following, and
explore also 
 alternative expressions 
 containing the bosonic number operators $\n
ei, \n s{i,\g}$, and $\n di$.

\subsection{FUNCTIONAL-INTEGRAL FORMULATION}
\label{func-int} 
Functional integrals provide an ideal framework to enforce constraints
like \eqref{constraints}. This is achieved by introducing a Lagrange
multiplier for each constraint (say, $\lam_i^a$ and $\lam\is^b$ for the
constraints \eqref{constraints}, in the order). The
functional-integral formulation based on coherent states
\cite{schulman} rests on breaking up the (imaginary time) interval
$(0,\bet)$ into $M$ steps (where $\bet=1/(k_BT)$ is the inverse
temperature and $M \to\infty$ eventually),\cite{kh=1} and yields the
following expression for the grand-canonical partition function:
\beqn
\label{zint}
 {\cal Z} = \lim_{M \to \infty} && 
 \int  \prod_{i} d \lam_i^a \left(\prod_{\g} d \lam\is^b\right)
 \prod_{m=0}^{M-1}  d^2 e_{i,m}  d^2 d_{i,m}   
\nonumber \\
 && \times \left(\prod_{\g}  
d^2 s_{i,\g,m}   d \> \barw f_{i,\g,m} d \> f_{i,\g,m}\right)
\exp \{-S_M\},  
\eeqn
where $m$ labels the imaginary time steps, ($e,s_{\g},d$) are complex
boson fields, $f$ and $\barw f$ are Grassmann variables, and $S_M$ is
the discretized action given by 
\beq
\label{sm} 
 S_M= \delta \sum_{m=0}^{M-1} \sum_i 
 \left( W_{i,m}^{(F)} +W_{i,m}^{(B)}+W_{i,m}^{(FB)} - \lam_i^a \right).
\eeq
Here, $\del=\bet/M$ is the elementary time step,
\beq
     W_{i,m}^{(F)} = {1 \over \delta} \sum_{\g} \wwbar f\ism \left[ 
        f\ism - \left(1-\delta\lbsi+\delta\mu\right) f\ismu 
\right]
\label{wfim} 
\eeq
($\mu$ being the chemical potential) and \cite{lambda}
\beqn
W_{i,m}^{(B)} &&=
 {1 \over \delta} e\im\cc \left[ e\im -  \left(1-\delta\lai\right) e\imu 
\right] 
\nonumber\\
 &&+ {1 \over \delta} \sum_{\g} s\ism\cc \left[ s\ism -
  \left(1-\delta\lai+\delta \lbsi\right) s\ismu \right] 
\nonumber\\
&&+{1 \over \delta} d\im\cc \left[ d\im -  \left(1-\delta\lai+\delta 
\sum_{\g}\lbsi-\delta U\right) d\imu  \right],
\label{wbim} 
 \eeqn     
are functions of the Grassmann variables and of the boson fields,
respectively, while 
\beq
\label{wfbim} 
      W_{i,m}^{(FB)} = t \sum_{\nene,\g} \wwbar f\jsm 
z\js\cc(m-1,m)  z\is(m,m-1) f\ismu 
\eeq
is the mixed fermion-boson term which derives from the kinetic term in
the Hamiltonian \eqref{ham} and thus depends explicitly on the form of the
bosonic operator $z\is$. [In the above expressions, the label $\g$
takes the two values $\pm 1$ corresponding to spin $\mez$. A
generalized label $S$ which can take $2 N$ values (with $N$ positive
integer) will be introduced in Appendix \ref{app:a} as a formal tool to
generate a $1/N$ expansion for the four-slave-boson method].

In the expression \eqref{wfbim}, $z\is(m,m-1)$ stands for the matrix
element of the operator $z\is$ between coherent states labeled by
imaginary times $\tau_m=m\delta$ (on the left) and
$\tau_{m-1}=\tau_m-\delta$ (on the right). When $z\is$ is taken of the
general form \eqref{zr}, this matrix element becomes
\beq
\label{zism} 
z\is(m,m') = \left[ e\im\cc s_{i,\g,m'} + s_{i,-\g,m}\cc d_{i,m'} \right]
R\is(m,m')
\eeq
where $R\is(m,m')$ is the matrix element of the subsidiary operator 
$R\is$ between coherent states at times $m$ and $m'$.
 In particular, when 
the
normal-ordered form \eqref{rkr} is considered, 
 the above matrix element
$R\is(m,m')$ 
acquires the simple form:
\beq
\label{rkrm} 
 R\is^{KR}(m,m')  = {1\over \sqrt{1- d\im\cc d_{i,m'} - s\ism\cc s_{i,\g,m'}}}
{1\over \sqrt{1-e\im\cc e_{i,m'} - s\imsm\cc s_{i,-\g,m'}}} .
\eeq
This form with $m=m'$ has been constantly used
 in the previous literature.\cite{schonhammer}
In the more general case when the operator $R\is$ is not explicitly in
normal-ordered form, it should be expressed as a sum of normal-ordered
terms by suitably commuting the bosonic operators. In practice, for an
operator like \eqref{rsq} this procedure can be implemented in the
context of a $1/N$ expansion, as it will be shown in subsection
\ref{s:nno}.

As  remarked in the Introduction, the functional-integral
representation of the partition function requires one to keep,
in principle,
 the
discretized imaginary time mesh  until the end of the
calculation.\cite{schulman} However, it has been common practice in
the  slave-boson literature  to violate  this requirement by
taking the continuum ($\delta\to0$) limit of the action \eqref{sm} -
\eqref{wfbim} at the outset, thus effectively transferring the
$M\to\infty$ limit in Eq. \eqref{zint} under the integral sign. This
procedure removes the distinction between the two time labels $m$ and
$m'$ in Eqs. \eqref{zism} and \eqref{rkrm}. 
These 
seemingly formal considerations 
acquire relevance from the occurrence of 
 the unphysical results
 obtained in the \CL
 already at the Gaussian level, as shown in the next
Section.

 Specifically, we argue that taking the continuum limit at the outset is
{\evi pathologic} for the slave-boson problems owing to the presence
of the slave-boson condensate, as it was already discussed in detail
in Ref. \cite{noiprep} for the $U=\infty$ one-slave-boson problem.
Accordingly, the $M\to\infty$ limit has to be properly taken in Eq.
\eqref{zint} {\evi only at the end of the calculation}. Otherwise, wrong
results for the free energy (and derived quantities) are obtained when
including fluctuation corrections beyond the saddle-point of the
functional integral. As it will be clear from the following analysis,
these problems stem from a failure of the functional integral 
to account for the bosonic
commutation rules when the continuum limit is taken at the
outset.

\subsection{ GAUSSIAN FLUCTUATIONS FOR THE FREE ENERGY WITH A
DISCRETIZED TIME MESH} 
\label{s:gauss} 
The complex boson fields at each discretized imaginary time $\tau_m$
entering the functional integral \eqref{zint} can be represented,
alternatively, either via their amplitude and phase or via their real and
imaginary parts. The two representations correspond to the so-called
``radial'' and ``Cartesian'' gauges, respectively. In this Section, we
adopt the ``radial'' gauge to evaluate on the same footing both the
conventional ``continuum'' contribution (which was regularly considered in
the previous literature) and the novel contribution resulting from a
careful handling of the continuum limit of the functional integral
(which we shall name ``\CFI'' from the range of Matsubara frequencies
it originates from). Accordingly, each stage of the calculation
reported below will be free from infrared singularities which plague
instead the ``Cartesian'' treatment. However, we shall show in
Appendix \ref{app:c} that use of the ``Cartesian'' gauge makes the
specific calculation of the ``\CFI'' somewhat simpler
than in 
the ``radial'' gauge (at least at the Gaussian level we consider).
 Comparison of
the results obtained 
for the ``\CFI''
in the two gauges 
can also serve as  a test of the 
validity of the $1/N$ expansion for the four-slave-boson method
presented in Appendix \ref{app:a}.

The boson fields in Eq. \eqref{zint} are represented via their
amplitude and phase, by setting
\beq
\label{rad}
\Biggl\{
\begin{array}{l l l}
 	e\im &=& \wti e\im \exp\{i \fiem\} \\
 	s\ism &=& \wti s\ism \exp\{i \fism\} \\
 	d\im &=& \wti d\im \exp\{i \fidm\} .
\end{array}
\eeq
It is then convenient to transform the Grassmann fields therein as follows:
\beq
\label{frad}
f\ism = \uti f\ism \exp\{i(\fiem-\fism)\} \;.
\eeq
Entering Eqs. \eqref{rad} and \eqref{frad} in the action 
 \eqref{sm} - \eqref{wfbim} makes the boson phases to appear
only in certain combinations.  It is then convenient to introduce
the following alternative variables
\beq
\label{filam}
\left\{
\begin{array}{l l l}
&& \latimu = {i\over\delta} \, (\fiem -\fiemu) + \lai \\
&&  \latimu -\lbtsimu = {i\over\delta} \, (\fism -\fismu) +\lai - \lbsi  \\
&&   g\im = i \, (  \fiem +\fidm -\sum_{\g} \fism).
\end{array}
\right .
\eeq
In this way, the three phases $\fiem$ and $\fism$ (with $\g=\pm 1$)
are eliminated in favor of the new variables $\latim$ and $\lbtsim$,
which can be considered as time-dependent Lagrange multipliers. The
need of keeping in the functional integral an additional independent
variable, like $g\im$ of Eq. \eqref{filam}, has been already 
pointed out by
Jolicoeur and Le Guillou (Ref.\cite{jolicoeur}) [although in the
continuum case], by  arguing that $\fidm$ cannot be chosen to make
$g\im$ vanishing for each $m$. For later convenience, we introduce also
the variables 

\beq
\label{filam1}
\Biggl\{
\begin{array}{l l l}
 \laimu &=& {i\over\delta} \, (\fiem -\fiemu)  \\
  \laimu -\lbsimu &=& {i\over\delta} \, (\fism -\fismu) 
\end{array}
\eeq
which, however,
 are not linearly independent because

\beq
\label{sum0}
\sum_{m=0}^{M-1} \laim = \sum_{m=0}^{M-1} \lbsim =0.
\eeq
 
To simplify the notation somewhat, we further introduce the dictionary
\beq
\label{defab} 
\begin{array}{l r l} 
\wti e \lrarrow \ab1 = \bb1 , && g \lrarrow \ab5 = \bb5 , \\
\wti s_{\up} \lrarrow \ab2 = \bb2 , &&  \lambda^b_{\up} \lrarrow \ab6 =
\llam1 , \\
\wti s_{\down} \lrarrow \ab3 = \bb3 , &&  \lambda^b_{\down} \lrarrow
\ab7 = \llam2 , \\ 
\wti d \lrarrow \ab4 = \bb4 , &&  \lambda^a \lrarrow
\ab8 = \llam3  ,
\end{array}
\eeq
with the understanding that the appropriate site and time (or wave
vector and frequency) indices (or arguments) will be indicated
whenever necessary. In addition, the notation $\und a$ will stand for
the set 
 $\{ \ab{\ia}; \ia=1,\ldots,8 \}$, 
 $ \und b$  for the set $\{ \bb{\ib}; \ib=1,\ldots,5 \}$, and 
 $\und \lam$  for the set $\{ \llam{\il}; \il=1,\ldots,3 \}$.

With the above notation, the terms \eqref{wfim}-\eqref{wfbim} of the
action \eqref{sm} can be cast in the form:

\beq
\label{wif}
     W_{i,m}^{(F)} = {1 \over \delta} \sum_{\g} \baruti f\ism \left[ 
       \uti f\ism - q^{\g}(\lam\ismu^b,\lam\is^b-\mu) \uti f\ismu \right],
\eeq

\beq
W_{i,m}^{(B)} =
 {1 \over \delta} \sum_{\ib=1}^4 \bbc{\ib}\im \left[ \bb{\ib}\im -
h^{\ib}(\und \lam\imu,\und\lam_i,g\im-g\imu) \bb{\ib}\imu \right],
\eeq
and
\beq
      W_{i,m}^{(FB)} = t \sum_{\nene,\g} \baruti f\jsm 
\wwbar\rhog(\und a\jm,\und a\jmu) \rhog(\und a\im,\und a\imu)
\uti f\ismu \quad.
\eeq
Here, $q^{\g}$, $h^{\ib}$, and $\rhog$ ($\wwbar \rhog$) are functions
of the bosonic and $\lam$ variables, defined respectively by

\beq
q^{\g}(\lam\ismu^b,\lam\is^b-\mu)=\left[1-\delta(\lam\is^b-\mu)\right]
\exp \left\{ -\del \lam\ismu^b \right\},
\label{qsigma}
\eeq

\beq
        {\,h}^{1}( \und\lam\imu,\, \und\lam_i,\, g\im-g\imu )=
        [1-\del \lam_i^a]\;\exp\left\{-\del \lam\imu^a\right\} \;,
\eeq

\beq
       {\,h}^{{5-\g \over2}}( \und\lam\imu,\, \und\lam_i,\, g\im-g\imu )=
        [1-\del (\lam_i^a -  \lam\is^b)] \; \exp\left\{-\del\,(
\lam\imu^a-\lam\ismu^b) \right\},
\eeq
\beqn
       {\,h}^{4}( \und\lam\imu,\, && \und\lam_i, \, g\im-g\imu ) =
        [1-\del (\lam_i^a - \sum_{\g}  \lam\is^b) -\del U] 
\nonum\\&&\times      \exp\left\{-\del \, ( \lam\imu^a-
\sum_{\g} \lam\ismu^b) -(g\im - g\imu) \right\}, 
\eeqn 
and
\renewcommand{\men}{\bar}
\beqn
\label{rho}
  \rhog(\und a\im ,\, \und a\imu)  &=&
 \exp\{-\del \lam\imu^a\}  
 \R( n^{(e)}_{i,m}, n^{(s_{\g})}_{i,m}, n^{(s_{\bar\g})}_{i,m},
n^{(d)}_{i,m} ) 
\nonumber\\
&\times&
 \left[ \wti e\im \wti s\ismu+  \wti s\imsm \wti d\imu \, \exp\{g\imu + \del \lam\imsmu^b\} 
\right]
\eeqn
with $\bar \g= -\g$ and
\beq
\Biggl\{ \begin{array}{l l l}
  n^{(e)}\im  &=& \wti e\im \wti e\imu \exp\{-\del \lam\imu^a\} \\
  n^{(s_{\g})}\im &=& \wti s\ism \wti s\ismu \exp\{-\del\,( \lam\imu^a-
\lam\ismu^b)\} 
\\
 n^{(d)}\im &=& \wti d\im \wti d\imu \exp\{-\del\,( \lam\imu^a- 
\sum_{\g}\lam\ismu^b ) -(g\im - g\imu)\} \quad .
\end{array}
\label{numop} 
\eeq
[ $\wwbar \rhog$ is obtained from $\rhog$ by interchanging
$\bb{\ib}\im \lrarrow \bb{\ib}\imu \quad (\ib =1,\cdots,5) $ and then
letting $g \to -g$ ]. In Eq.\eqref{rho} it is understood that 
the {\evi subsidiary function}
$\R$ is formally obtained from the subsidiary operator $R\is$ of
Eq.\eqref{zr} by replacing each bosonic occupation number therein with
the corresponding c-number argument \eqref{numop}.
This prescription 
holds provided  the subsidiary operator  $R$
is written as the 
normal ordering of a function of the bosonic occupation numbers.

Equations \eqref{wif}-\eqref{numop} are still exact. We now proceed and
expand the action in powers of the fluctuating fields about a chosen
mean-field solution, by setting for the bosonic and {\evi static} $\lam$
variables
\beq
\Biggl\{
\begin{array}{l l l}
& \bb{\ib}\im =  \bb{\ib}_0 + \uti b_{i,m}^{(\ib)}
\quad\quad &(\ib=1,\cdots5),
\\&
\llam{\il}_i =  \llam{\il}_0 + \uti{\lam}_i^{(\il)}
\quad\quad & (\il=1,\ldots3).
\end{array}
\label{fluc}
\eeq
 The choice \eqref{fluc} entails a {\evi homogeneous} (i.e.,
site independent) mean-field solution, which will restrict us in the
following to the paramagnetic (or else, to the ferromagnetic) case. It
also implies that the mean-field solution is {\evi static} (i.e.,
time-independent).\cite{allvariables} 
Accordingly, the dynamic variables \eqref{filam1} have, by their definition,
vanishing mean-field contribution.\cite{g0=0}

We  next introduce the space and time Fourier transforms of the
fluctuating fields, by setting 
\beq
\uti b_{i,m}^{(\ib)}=  \sum_{\v q}^{BZ}\sum_{\nu=0}^{M-1} e^{i( \v q \cdot
\v R_i - \omnu\tau_m)}
 \bb{\ib}\qunu \quad\quad (\ib=1,\cdots,5),
\label{bbim} 
\eeq
\beq
\uti{\lam}^{(\il)}\im \equiv \llam{\il}\im =
  \sum_{\v q}^{BZ}\sum_{\nu=1}^{M-1} e^{i( \v q \cdot
\v R_i - \omnu\tau_m)}
 \llam{\il}\qunu \quad\quad  (\il=1,\cdots,3),
\eeq
and

\beq
\uti{\lam}_i^{(\il)} =
  \sum_{\v q}^{BZ} e^{i \v q \cdot
\v R_i }
 \llam{\il}(\v q,0)  \quad\quad  (\il=1,\cdots,3).
\eeq
Here, $\v R_i$ is the lattice vector associated to site $i$, $\v q$ is
a wave vector restricted to the first Brillouin zone (BZ),
$\tau_m=m\delta$ $(m=0,\cdots,M-1)$, $\omnu=2 \pi \nu /\bet$ is a
bosonic Matsubara frequency, and $\llam{\il}(\v q,0) = \llam{\il}(\v
q,\omnu=0)$ by our definition. A similar transformation holds for the
Grassmann fields:
\beq
\uti{f}\ism=  \sum_{\v k}^{BZ}\sum_{s=0}^{M-1} e^{i( \v k \cdot
\v R_i - \oms\tau_m)}
 f_{\g}\kaom 
\label{fism} 
\eeq
where now $\oms= 2 \pi (s+1/2) /\bet$ is a fermionic Matsubara
frequency. Note that the {\evi range} of the Matsubara frequencies
remains {\evi bounded} while keeping the discretized time
mesh.\cite{transformations} 

Expanding the action up to quadratic order in the fluctuating
fields requires one to consider the first and second derivatives of the
functions \eqref{qsigma}-\eqref{rho} with respect to their arguments.
By our convention, each argument may represent either a single
variable or a set of variables (in the latter case we shall underline
the corresponding symbol). To keep the notation compact, we then
introduce the following short-hand notation for the required
derivatives. We set:
\beqn
&&
 q^{\g}_{0;0} = { q^{\g}(x,y) } \Bigl|_{x=0,y=\lam_{\g,0}^b-\mu},
\nonum\\
&&
 q^{\g}_{1;0} = {\de q^{\g}(x,y) \over \de x} \Bigl|_{x=0,y=\lam_{\g,0}^b-\mu},
\nonum\\&&
 q^{\g}_{0;1} = {\de q^{\g}(x,y) \over \de y} \Bigl|_{x=0,y=\lam_{\g,0}^b-\mu},
\nonum\\&&
 q^{\g}_{(1,1);0} = {\de^2 q^{\g}(x,y) \over \de x^2} \Bigl|_{x=0,y=\lam_{\g,0}^b-\mu},
\nonum\\&&
 q^{\g}_{0;(1,1)} = {\de^2 q^{\g}(x,y) \over \de y^2} \Bigl|_{x=0,y=\lam_{\g,0}^b-\mu},
\nonum\\&&
 q^{\g}_{1;1} = {\de^2 q^{\g}(x,y) \over \de x \de y} \Bigl|_{x=0,y=\lam_{\g,0}^b-\mu};
\label{derq} 
\eeqn

\beqn
&&
 {\,h}^{\ib}_{0;0;0} = {\,h}^{\ib}(\und X,\, \und Y , Z) \Bigl|_{\und X= \und 0,\,\und Y= \und \lam_0,\, Z=0},
\nonum\\&&
 {\,h}^{\ib}_{\il;0;0} = {\de {\,h}^{\ib}(\und X,\, \und Y , Z) \over \de X^{(\il)}} 
 \Bigl|_{\und X= \und 0,\,\und Y= \und \lam_0,\, Z=0},
\nonum\\&&
 {\,h}^{\ib}_{0;0;1} = {\de {\,h}^{\ib}(\und X,\, \und Y , Z) \over
\de Z} \Bigl|_{\und X= \und 0,\,\und Y= \und \lam_0,\, Z=0},
\nonum\\&&
 {\,h}^{\ib}_{\il;\il';0} = {\de^2 {\,h}^{\ib}(\und X,\, \und Y , Z) \over
 \de X^{(\il)} \de Y^{(\il')} } \Bigl|_{\und X= \und 0,\,\und Y= \und \lam_0,\, Z=0}, 
\nonum\\&&
 {\,h}^{\ib}_{\il;0;1} = {\de^2 {\,h}^{\ib}(\und X,\, \und Y , Z) \over
 \de X^{(\il)} \de Z} \Bigl|_{\und X= \und 0,\,\und Y= \und \lam_0,\, Z=0}, 
\nonum\\&&
 {\,h}^{\ib}_{(\il,\il');0;0} = {\de^2 {\,h}^{\ib}(\und X,\, \und Y , Z) \over
 \de X^{(\il)} \de X^{(\il')} } \Bigl|_{\und X= \und 0,\,\und Y= \und \lam_0,\, Z=0}, 
\nonum\\&&
 {\,h}^{\ib}_{0;0;(1,1)} = {\de^2 {\,h}^{\ib}(\und X,\, \und
Y , Z) \over \de Z^2} \Bigl|_{\und X= \und 0,\,\und Y= \und \lam_0,\, Z=0},  \;
\label{derh} 
\eeqn
and so on;

\beqn
&& \rhog_{0;0} = \rhog(\und X,\, \und Y)
\bigr|_{\XYMFO} \;,
\nonum\\&&
 \rhog_{\ia;0} = {\de \rhog(\und X,\, \und Y) \over \de X^{(\ia)}} 
\Bigl|_{\XYMFO} \;,
\nonum\\&&
 \rhog_{\ia;\ia'} = {\de^2 \rhog(\und X,\, \und Y) \over 
\de X^{(\ia)} \de Y^{(\ia')} } 
\Bigl|_{\XYMFO} \; ,
\nonum\\&&
 \rhog_{(\ia,\ia');0} = {\de^2 \rhog(\und X,\, \und Y)
\over \de X^{(\ia)} \de X^{(\ia')} }
\Bigl|_{\XYMFO} \;,
\label{derrho} 
\eeqn
and so on.
Similar conventions hold for $\wwbar \rhog$.
In particular, the suffices 
$\und X=(\und b_0,\und 0)$ and  $\und Y = (\und b_0,\und 0)$ in
Eqs.\eqref{derrho} signify that each bosonic field $\bb{\ib}\im$
$(\ib=1,\cdots,5)$ is replaced by the associated mean-field value 
$\bb{\ib}_0$, while each {\evi dynamic} $\llam{\il}\im$ field 
$(\il=1,\cdots,3)$ is replaced by zero.

Expanding at this point the action \eqref{sm} up to quadratic order in
the fluctuating bosonic fields, we obtain the following contributions:
\beq
S_M= \beta \NN[ S^{(0)}+S_F^{(0)}+S_F^{(1)}+S^{(2)}+S_F^{(2)} ].
\label{sm012} 
\eeq
Here:
\beq
S^{(0)}=
\lam_0^a (\sum_{\ib=1}^4 \bbd{\ib}_0 -1 ) -\sum_{\g} \lam_{\g ,\, 0}^b (
\bbd4_0 + \bbd{{5-\g \over 2}}_0 ) + U \bbd4_0 
\label{s0} 
\eeq
is the constant (mean-field) value of the bosonic part \eqref{wbim}
of the action;
\beq
S_F^{(0)} = 
\dsum_{\v k}^{BZ}\dsum_{\g}\dsum_{s=0}^{M-1}
   \wwbar f_{\g} \kaom 
\eiod 
 G_0^{\g}\kaom^{-1} 
f_{\g}\kaom
\eeq
where
\beq
 G_0^{\g}\kaom^{-1} =  {(\emiod-1) \over \del} +  \E_{\g}(\v k)
\label{g0fer} 
\eeq
is the single-particle mean-field fermionic Green's function
corresponding to the band eigenvalue
\beq
 \E_{\g}(\v k) = \lam_{\g ,\,0}^b -\mu + t \rhog_{0;0} \wwbar
\rhog_{0;0} \gamk
\label{epsk} 
\eeq
with
\beq
\label{gam} 
\gamk = \sum_{\nene} \exp\{-i \v k\cdot \nene\} \; ;
\eeq
\beqn
 S_F^{(1)}  =
\dsum_{\v q}^{BZ}\dsum_{\nu=0}^{M-1}\dsum_{\ia=1}^8
\dsum_{\v k}^{BZ}\dsum_{\g}\dsum_{s=0}^{M-1} &&
      \wwbar f_{\g}\kaom \eiod
\EE^{(1)}(\v q,\omnu ; \v k,\g | \ia)
 \nonum\\&&\times   \ab{\ia}\qunu  f_{\g}\kqon 
\eeqn
where
\beqn
\EE^{(1)}&&(\v q,\omnu ; \v k,\g | \ia) =
( 1- \del_{\nu,0}^K \sum_{l=1}^3 \del_{\ia,l+5}^K )
\nonum\\&&\times   t \Bigl[ \gamk \wwbar \rhog_{0;0} ( \rhog_{0;\ia} + \emind \rhog_{\ia;0} ) 
\nonum\\&&     +  \gamkq  \rhog_{0;0} ( \wwbar \rhog_{0;\ia} + \emind
\, \wwbar
\rhog_{\ia;0} ) \Bigr]
\nonum\\&&    -(  \del_{\ia,6}^K  \del_{\g,+1}^K + \del_{\ia,7}^K  \del_{\g,-1}^K )
\undel \Bigl[(1-\del_{\nu,0}^K) q^{\g}_{1;0} + \del_{\nu,0}^K  q^{\g}_{0;1}
\Bigr] \;,
\label{e1} 
\eeqn
$\del_{i,j}^K$ ($=1$ when $i=j$ and $0$ otherwise ) being the
Kronecker delta function;
\beq
S^{(2)} =  
\dsum_{\v q}^{BZ}\dsum_{\nu=0}^{M-1}\dsum_{\ia,\ia'=1}^8
\ab{\ia}\qunu \BB(\v q,\omnu| \ia,\ia') \ab{\ia'}\mqunu
\label{s2} 
\eeq
where

\beqn
\BB && (\v q,\omnu| \ia,\ia') = 
\del_{\ia,\ia'}^K
\sum_{\ib=1}^4 \del_{\ia,\ib}^K \undel (1-\emind h^{\ia}_{0;0;0} )
\nonum\\&&
- \mez {(1+\emind) \over \delta } \sum_{\ib=1}^4 \del_{\ia,\ib}^K
\bb{\ia}_0  \Bigl[ \del_{\ia',5}^K (\eind -1) h^{\ia}_{0;0;1}
\nonum\\&&
+  \sum_{\il'=1}^3 \del_{\ia',5+\il'}^K
 ( (1-\dnuo) h^{\ia}_{\il';0;0} +
\dnuo h^{\ia}_{0;\il';0} ) \Bigr]
\nonum\\&&
- \mez{ (1+\eind) \over \delta } \sum_{\ib'=1}^4
\del_{\ia',\ib'}^K  \bb{\ia'}_0 \Bigl[ \del_{\ia,5}^K (\emind -1)
h^{\ia'}_{0;0;1}
\nonum\\&&
+  \sum_{\il=1}^3 \del_{\ia,5+\il}^K
 ( (1-\dnuo) h^{\ia'}_{\il;0;0} +
\dnuo h^{\ia'}_{0;\il;0} ) \Bigr]
\nonum\\&&
- {1\over 2\delta} 
\sum_{\ib=1}^4  \bbd{\ib}_0 \Biggl\{  \sum_{\il=1}^3 \del_{\ia,5+\il}^K
 \sum_{\il'=1}^3 \del_{\ia',5+\il'}^K
\nonum\\&&\times
\left[   (1-\dnuo) h^{\ib}_{(\il,\il');0;0} +
\dnuo h^{\ib}_{0;(\il,\il');0}  \right]
\nonum\\&&
+ \sum_{\il=1}^3 \del_{\ia,5+\il}^K \del_{\ia',5}^K (\eind-1)
h^{\ib}_{\il;0;1}
\nonum\\&&
+ \del_{\ia,5}^K \sum_{\il'=1}^3 \del_{\ia',5+\il'}^K (\emind-1)
h^{\ib}_{\il';0;1}
\nonum\\&&  + \del_{\ia,5}^K   \del_{\ia',5}^K 
(\eind-1)(\emind-1) h^{\ib}_{0;0;(1,1)} \Biggr\} \;;
\label{bb} 
\eeqn

\beqn
S_F^{(2)} = &&
\dsum_{\v q}^{BZ}\dsum_{\nu=0}^{M-1}\dsum_{\ia,\ia'=1}^8
 \ab{\ia}\qunu 
\Biggl[ 
\dsum_{\v k}^{BZ}\dsum_{\g}\dsum_{s=0}^{M-1}
\wwbar f_{\g}\kaom \eiod 
\nonum\\&&\times  \EE^{(2)}(\v q,\omnu;\v k,\g | \ia,\ia') f_{\g}\kaom \Biggr]
 \ab{\ia'}\mqunu 
\label{s2f} 
\eeqn
where

\beqn
\EE^{(2)}(\v q, && \omnu;\v k,\g | \ia,\ia') =
 - \delK_{\ia,\ia'} ( 
\delK_{\ia,6} \delK_{\g,+1} + \delK_{\ia,7} \delK_{\g,-1} )
\nonum\\&& 
\times {1\over 2\delta} \Bigl( (1-\dnuo) q^{\g}_{(1,1);0} +  \dnuo
q^{\g}_{0;(1,1)}
\Bigr)
\nonum\\&&
+ (1-\dnuo \sum_{\il=1}^3 \delK_{\ia,5+\il})
 (1-\dnuo \sum_{\il'=1}^3 \delK_{\ia',5+\il'})
\nonum\\&&\times
\frac{t}{2} 
\Biggl[  
\gamk \rhog_{0;0}( \wwbar  \rhog_{(\ia,\ia');0} +
\wwbar  \rhog_{0;(\ia,\ia')} + \wwbar  \rhog_{\ia;\ia'}\emind+
 \wwbar  \rhog_{\ia';\ia}\eind )
\nonum\\&&
+
\gamk \wwbar \rhog_{0;0} (   \rhog_{(\ia,\ia');0} +
  \rhog_{0;(\ia,\ia')} +   \rhog_{\ia;\ia'}\emind+
   \rhog_{\ia';\ia}\eind )
\nonum\\&&
+
\gamkq  ( \wwbar  \rhog_{\ia;0}  \rhog_{\ia';0} +
          \wwbar  \rhog_{0;\ia}  \rhog_{0;\ia'} +
         \wwbar  \rhog_{\ia;0}  \rhog_{0;\ia'} \emind +
         \wwbar  \rhog_{0;\ia}  \rhog_{\ia';0} \eind )
\nonum\\&&
+
\gamkpq  ( \wwbar  \rhog_{\ia';0}  \rhog_{\ia;0} +
          \wwbar  \rhog_{0;\ia'}  \rhog_{0;\ia} +
         \wwbar  \rhog_{\ia';0}  \rhog_{0;\ia} \eind +
         \wwbar  \rhog_{0;\ia'}  \rhog_{\ia;0} \emind )
  \Biggr] \;.
\label{e2} 
\eeqn
Note that in the action \eqref{sm012}
we have not considered the linear term in the bosonic field (which
originates from the purely bosonic contribution \eqref{wbim}), since
this term  cancels at
self-consistency by definition.\cite{allvariables}
In addition, only pairs of fermionic (Grassmann) variables with the
{\evi same} $\v k$ {\evi and} $\oms$ have been retained in the contribution
\eqref{s2f}, because all other pairs would give a vanishing
contribution to the following Gaussian calculation.

The fermionic variables are  eliminated at this point by performing the
functional integration over the action $S^{(0)}_F+S^{(1)}_F+S^{(2)}_F$
which is
quadratic in the fermionic variables. Exponentiating  the
resulting expression and expanding the exponent up to second order in
the fluctuating bosonic fields, we eventually obtain the effective action
\beq
S_{eff}= \bet\NN [ S^{(0)} + \wwbar S^{(0)} + S^{(2)} + \wwbar S^{(2)}
]
\label{seff} 
\eeq
where $S^{(0)}$ and $ S^{(2)}$ are given by \eqref{s0} and \eqref{s2},
respectively, and
\beqn
&& \wwbar  S^{(0)} = -\frac{1}{\NN} 
\dsum_{\v k}^{BZ}\dsum_{\g} \frac{1}{\beta} \dsum_{s=0}^{M-1}
\ln [\delta \eiod G_0^{\g}\kaom^{-1} ],
\label{sbar0} 
\\&&
\wwbar  S^{(2)} = 
\dsum_{\v q}^{BZ}\dsum_{\nu=0}^{M-1}\dsum_{\ia,\ia'=1}^8
\ab{\ia}\qunu \CC(\v q,\omnu|\ia,\ia') \ab{\ia'}\mqunu,
\label{sbar2} 
\eeqn
with
\beqn
 \CC(\v q,\omnu|\ia,\ia') &&= \frac{1}{\NN} \dsum_{\v k}^{BZ}\dsum_{\g}
\Bigl\{ f_M(\E_{\g}(\v k)) \EE^{(2)}(\v q,\omnu;\v k,\g|\ia,\ia')
\nonum\\&&
-\mez  \EE^{(1)}(\v q,\omnu;\v k,\g|\ia) 
\Pi_M \left(\E_{\g}(\v k),\E_{\g}(\v k-\v q);\nu\right)
\nonum\\&&\times
 \EE^{(1)}(-\v q,-\omnu;\v k-\v q,\g|\ia') 
\Bigr\}
\label{cc} 
\eeqn
[cf. Eqs. \eqref{g0fer}, \eqref{e1}, and \eqref{e2}]. In 
expression \eqref{cc} we have introduced the Fermi function $f_M(\E)$
and the fermionic polarization function $\Pi_M(\E,\E';\nu)$, which
generalize to the case of the discretized time mesh considered in this
paper the familiar functions of the Matsubara Green's functions
theory, the latter being recovered in the continuum ($\delta\to 0$)
limit. The expressions of $f_M(\E)$ and $\Pi_M(\E,\E';\nu)$ are
reported in Appendix \ref{app:b}.

The free energy is obtained by performing the Gaussian integration
over the bosonic variables $\{\ab{\ia}\qunu; \ia=1,\cdots,8\}$ with
action $S_{eff}$ given by Eq. \eqref{seff}. One obtains (per lattice
site):
\beq
F=F_0+F_1
\label{f01} 
\eeq
with
\beq
\label{f0} 
F_0=S^{(0)}+\wwbar S^{(0)}
\eeq
and
\beqn
\label{f1} 
F_1= \frac{1}{\NN} 
\dsum_{\v q}^{BZ}
\frac{1}{2\beta}
\dsum_{\nu=-\frac{M-1}{2}}^{\frac{M-1}{2}} \Bigl\{
&&
 \ln\det \Gam\qunu -
\sum_{\ib=1}^4 \ln(4 \bbd{\ib}_0 )
\nonum\\&&
+(1-\dnuo) \ln\left[ \frac{\delta^2}{(1-\eind)(1-\emind)} \right]
\Bigr\},
\eeqn
where $M$ has been taken, for convenience,
 to be an odd integer and the matrix
$\Gam\qunu$ has elements [cf. Eqs. \eqref{bb} and \eqref{cc}]
\beqn
\label{gamq} 
\Gam\qunu_{\ia,\ia'} &=&
 \BB(\v q,\omnu|\ia,\ia') +  \CC(\v q,\omnu|\ia,\ia')
\nonum\\&&
+  \BB(-\v q,-\omnu|\ia',\ia) +  \CC(-\v q,-\omnu|\ia',\ia).
\eeqn
The first term within braces in Eq. \eqref{f1}
has been obtained by the method discussed in Appendix \ref{app:e}, which
is required since the relevant Gaussian matrix  is
not Hermitian. The remaining terms within
braces in Eq. \eqref{f1} originate instead from Jacobian
contributions which are specific to the ``radial'' gauge and are
included here for convenience of the following calculation. Note also
that the frequency sum in Eq. \eqref{f1} has been symmetrized between
positive and negative values, in order to extract directly its
``continuum limit'' (see below) according to a theorem reported in
Appendix \ref{app:b}.

 The mean-field contribution $F_0$ to the
(site) free energy  given by Eq. \eqref{f0} does not
present any pathology in the $\delta\to0$ limit. One obtains the
familiar expression for an (effective) system of independent particles
\beq
\lim_{M\to\infty} \beta \wwbar S^{(0)} = - \frac{1}{\NN} 
\dsum_{\v k}^{BZ}\dsum_{\g} \ln\left(1+ e^{-\beta \E_{\g}(\v k)} \right)
\label{lims0} 
\eeq
 [cf. Eq. \eqref{sbar0}],
as shown in Appendix \ref{app:b}. The fluctuation contribution $F_1$
to the (site) free energy, on the other hand, has 
pathologic behavior in the $\delta\to0$ limit owing to the presence
of the bosonic frequency sum in Eq. \eqref{f1}.
It turns out, in fact, that performing the Matsubara frequency sum
with the discretized time mesh (i. e., keeping $|\nu|\leq (M-1)/2$ and
taking the ${M\to\infty}$ limit only {\evi after} having evaluated the
sum) gives a different result from the one obtained by interchanging the
continuum limit with the frequency sum. The latter procedure 
requires standard  mathematical techniques and has invariably been
considered in the previous slave-boson literature.

The pathologic behavior of a bosonic frequency sum of the type
\eqref{f1} has been discussed in detail in Ref. \cite{noiprep} in the
context of the simpler $U=\infty$ one-slave-boson problem.
Specifically, it has been proved that this frequency sum can be
suitably partitioned into a ``continuum limit'' contribution (where
the  $\delta\to0$ limit is taken {\evi at the outset} \ in each term of
the sum while the Matsubara index $\nu$ is extended from $-\infty$ to 
 $+\infty$) plus a ``\CFI'' which accounts for what is left out by
the \CL. For the reader's convenience, we report in Appendix
\ref{app:b} the statement of a theorem which shows how the ``\CFI''
can be extracted for a class of bosonic frequency sums that satisfy
certain requirements. For a complete proof of the theorem we
refer instead to Appendix B of Ref. \cite{noiprep}.

 It has been also demonstrated in Ref. \cite{noiprep} in the
context of the $U=\infty$ one-slave-boson problem that taking into
account the ``\CFI'' (to the bosonic frequency sum entering the
Gaussian correction to the free energy) is not only required at a
formal level, but it is also {\evi essential} in practice for
validating the $1/N$ expansion for the free energy. In the next
Section we will show that taking properly into account the ``\CFI'' to
the sum \eqref{f1} is essential for obtaining meaningful results at
the Gaussian level even for the four-slave-boson case.  This will hold in
spite of the nontrivial additional complications related to the
presence of the subsidiary operator $R$ in Eq. \eqref{zr}.

In essence, the reason why taking the \CL at the outset
 in the frequency sum of Eq.
\eqref{f1} leads to incorrect results, is that the
bosonic frequency $\omnu$ enters the relevant fluctuation matrix \eqref{gamq}
{\evi only} through the phase factor $\exp\{i \omnu \delta\}$. By
taking the \CL, one effectively 
regards the argument of this phase factor to be
small (i. e., $|\omnu| \delta\ll 1$) and accordingly replaces 
$\exp\{i \omnu \delta\} \to 1 + i\omnu \delta $. The flaw of this
procedure is that, even in the limit $\delta\to0$, $|\omnu| \delta$
can be comparable to unity since, by definition, $\omnu \delta = 2\pi
\nu/M$
($\nu=-(M-1)/2,\cdots,(M-1)/2$) and $\max|\omnu\delta| \cong \pi$. The
replacement
$\exp\{i \omnu \delta\}\to 1 + i\omnu \delta $ is thus not allowed to
obtain the correct value of the bosonic frequency sum in Eq.
\eqref{f1}, and the limit $\delta\to0$ can be safely taken therein
only {\evi after} having performed the frequency sum. 

To be more specific, a bosonic frequency sum of the type \eqref{f1}
can be effectively partitioned into two contributions associated with
``small'' and ``large'' frequencies, respectively. The contribution
from ``small'' frequencies (such that $|\omnu|\delta\ll1$) corresponds
to the usual continuum limit. The contribution
from ``large'' frequencies (such that $|\omnu|\delta$ is of order
unity) requires instead expanding the argument of the sum in terms of
all $\delta$ factors not entering the combination $\exp\{i \omnu
\delta\}$. Specifically, the contribution from ``large'' frequencies
we are concerned with originates from terms $\ord(\delta)$ in this
power expansion, since the sum of a large (of order $M$) number of
terms each $\ord(\delta)$ yields a finite contribution.

According to the general procedure of Appendix \ref{app:b},
 extraction of the contribution  from ``large'' frequencies
(that we have named  ``\CFI'') amounts 
to isolating the function $g$ of Eq. \eqref{b.10} 
and 
 to determining its constant term $g_0$ of the expansion
\eqref{b.9}. 
 Whenever 
the terms in the bosonic sum have  simple expressions (as for the
$U=\infty$ one-slave-boson problem treated in Ref. \cite{noiprep}),
extraction of the ``\CFI'' $g_0$ can be done analytically with
moderate effort. In the present four-slave-boson case, however, the
expressions of the matrix elements \eqref{gamq} are exceedingly
complicated to handle the sum in Eq. \eqref{f1} 
analytically. For this reason, we have developed a computer algorithm
for symbolic calculations which (i) evaluates explicitly the
quantities \eqref{derq}-\eqref{derrho} 
in terms of the bosonic mean-field values $\ab{\ia}_0
\;\;(\ia=1,\cdots,8)$ and of the parameter $\delta$, (ii) sets up the
matrix 
elements \eqref{gamq} 
, and (iii) performs the steps indicated in Appendix \ref{app:b} to
extract the ``\CFI'' $g_0$.

In this context, we found it convenient to regularize at the outset the
matrix \eqref{gamq} for $\nu\not=0$ 
in the limit of small $\del$, by introducing an
auxiliary fluctuation matrix $\wwbar\Gam$ such that
\beq
\wwbar \Gam(\v q,\eind|\del)_{\ia,\ia'} =
\del^{1-\chi_{\ia}-\chi_{\ia'}}
\Gam\qunu_{\ia,\ia'} 
\label{trgam} 
\eeq
where $\chi_{\ia}= \sum_{\il=1}^3 \delK_{\ia,5+\il}$. 
This definition makes the matrix $\wwbar\Gam$ and its inverse regular
in $\del$ about $\del=0$ and implies that
$\det\wwbar\Gam=\del^2\det\Gam$.
Accordingly, we write (in matrix notation)
\beq
\wwbar \Gam(\v q,\zeta|\del) = \wwbar  \Gam_0(\v q,\zeta) + \del\wwbar   \Gam_1(\v q,\zeta)
+\ord(\del^2)
\eeq
with $\zeta = \exp\{i \omnu \delta\}$, and expand 
\beq
\ln\det \wwbar\Gam = \tr\ln \wwbar\Gam = \ln\det \wwbar\Gam_0 + \del
\tr \left(\wwbar\Gam_0^{-1}\wwbar\Gam_1\right) + \ord(\del^2)
\eeq
provided $\wwbar\Gam_0$ is nonsingular. For the lowest-order term we
obtain
\beq
\det\wwbar\Gam_0(\zeta) = (1-\zeta)(1-\zeta^{-1}) \prod_{\ib=1}^4 4 \bbd{\ib}_0
\,,
\eeq
which (together with the factor $\del^2$ from the transformation
\eqref{trgam})
cancels the Jacobian contributions on the right-hand side of
Eq.\eqref{f1} (except for a term $-\frac1{2\bet} \ln 4 \bbd{\ib}_0$
that will be added to the $\nu=0$ component of the fluctuation
matrix).
 We are thus left with extracting the constant term
$g_0$ from the function
\beq
g_{\v q}(\zeta)=\tr[\wwbar\Gam_0(\zeta)^{-1}\wwbar\Gam_1(\v q,\zeta)]
\eeq
for any given wave vector $\v q$, according to the prescription
discussed in Appendix \ref{app:b}.

We eventually  obtain  the following expression 
for the ``\CFI'' to the Gaussian free energy
\eqref{f1}:
\beqn
F_1^{(d)} &=& -2\lam_0^a + 2\lam_0^b - \frac{U}{2} -\frac{1}{4} 
\sum_{\ib=1}^4 \frac{\de F_0}{\de \bbd{\ib}_0}
\nonum\\ 
&+&
\F[\R;\und b_0] \frac{2t}{\NN} \dsum_{\v k}^{BZ} \gamk f_F(\E_{\v k})
\label{f1d} 
\eeqn
which holds specifically for the {\evi paramagnetic} case whereby 
$\lam_0^b=\lam_{\g,0}^b$ and $\E_{\v k} =\E_{\g}(\v k)$ are
independent of $\g$. In Eq. \eqref{f1d}, $f_F(\E)$ is the ordinary Fermi
function (which is recovered in the present context as the continuum
limit of the function $f_M(\E)$ - cf. Appendix \ref{app:b}),
 $F_0$ is given by Eq. \eqref{f0},
 and
$\F[\R;\und b_0]$ depends on the specific form of the subsidiary
operator $R$ in Eq.\eqref{zr}.
In particular, $\F[\R;\und b_0]$
 vanishes in the simplest case when $R=1$. In all
other cases, 
$\F[\R;\und {b}_0]$ depends in a nontrivial way on the subsidiary
function $\R$ introduced in Eq. \eqref{rho} 
 and on its first and second derivatives, which have to be calculated
with the arguments \eqref{numop} 
 taken at the mean-field level. Specifically, $\F[\R;\und b_0]$ can be
represented in the compact form:
\beqn
\F[\R;\und b_0] = 
-2 z_0\Biggl[&&
 \bb1_0\bb2_0 
\left( \frac{\de\R}{\de n_1}+\frac{\de\R}{\de n_2}+ \mez
\frac{\de\R}{\de n_3}+\mez \frac{\de\R}{\de n_4} \right)
\nonum\\&&
+ 
 \bb3_0\bb4_0 
\left( \mez \frac{\de\R}{\de n_1}+\mez\frac{\de\R}{\de n_2}+ 
\frac{\de\R}{\de n_3}+ \frac{\de\R}{\de n_4} \right)
\nonum\\&&
+ (\bb1_0\bb2_0+\bb3_0\bb4_0) \mez
\sum_{\ib=1}^4 \bbd{\ib}_0 \frac{\de^2 \R}{\de n_{\ib}^2}
\Biggr] .
\label{zrb} 
\eeqn
In this expression, $\R=\R(n_1,n_2,n_3,n_4)$ with
$n_{\ib}=\bbd{\ib}_0\;\;(\ib=1,\cdots,4)$,
\beq
z_0=(\bb1_0 \bb2_0+\bb3_0 \bb4_0)   \R(n_1,n_2,n_3,n_4)
\label{z0} 
\eeq
is the mean-field value  of the bosonic operator $z\is$
given by Eq. \eqref{zr}, and
$\bb2_0=\bb3_0$ for the paramagnetic case we are considering.

Equations \eqref{f1d} and \eqref{zrb} are the main results of this
subsection. It will be shown in Appendix \ref{app:c} how these results
can be obtained alternatively within the ``Cartesian'' gauge. The
reason to work here with the ``radial'' gauge is that
the resulting 
 ``\CL~'' calculation is free from infrared divergences. In particular,
the ``\CL~'' contribution to the Gaussian free energy \eqref{f1} is
given by
\beq
\label{f1c} 
F_1^{(c)} = \frac{1}{2\beta\NN} 
\dsum_{\v q}^{BZ}\dsum_{\nu=-\infty}^{+\infty} \left\{
\ln\left[ \frac{\det\Gam_c\qunu}{\dnuo+(1-\dnuo)\omnu^2} \right]
 - \dsum_{\ib=1}^4 \ln(4 \bbd{\ib}_0) \right\}
\eeq
where $\Gam_c\qunu$ is the continuum ($\delta\to0$) limit of the
matrix \eqref{gamq}.

The physical free energy (per lattice site) at the Gaussian level is
then given by [cf. Eq.\eqref{f01}]
\beq
F=F_0+F_1^{(c)}+F_1^{(d)}
\label{ftot} 
\eeq
where $F_1^{(c)}$ and $F_1^{(d)}$ are given by Eqs. \eqref{f1c} and
\eqref{f1d}, respectively, with each contribution evaluated {\evi at
self-consistency} by taking the mean-field values for the bosonic
variables $\und b_0$. In this way, each term $\de F_0/\de\bbd{\ib}_0$
in Eq.\eqref{f1d} vanishes.\cite{df0db02} 

The ``\CFI'' $F_1^{(d)}$ in Eq. \eqref{ftot} is what has been missed in
the previous literature on the four-slave-boson approach to the
Hubbard model when considering Gaussian corrections beyond mean field.
We have attributed the reason for this omission to an improper
handling of the \CL of the functional integral. We  now argue that the
occurrence of $F_1^{(d)}$ is intrinsically related to the need of
recovering the correct bosonic commutators within the functional
integral.

\subsection{ INTERPRETATION OF THE ``CONTRIBUTION FROM INFINITY''} 
\label{s:int} 
The ``\CFI'' \eqref{f1d}, when taken at 
  self-consistency, can be interpreted in an
euristic  way by the following argument. Let's consider first the
simplest case when $\R=1$ (and $\F[\R;\und b_0]=0$,
 according to Eq.\eqref{zrb}).
In this case, noncommuting bosonic operators enter the Hamiltonian
only through the number operators $\bbdag{\ib}\bb{\ib}$. We argue
that taking naively the \CL of the  action at the outset
corresponds effectively to replacing each number operator
$\bbdag{\ib}\bb{\ib}$ in the original Hamiltonian as follows 
\beq
 \bbdag{\ib}\bb{\ib} \to \mez( \bbdag{\ib}\bb{\ib}+
\bb{\ib}\bbdag{\ib}) =  \bbdag{\ib}\bb{\ib}+
  \mez[\bb{\ib},\bbdag{\ib}] \;,
\eeq
and then treating the action associated with the 
modified Hamiltonian
 correctly
(that is, by keeping the discretized time mesh until the end of the
calculation). To restore the correct original Hamiltonian, one has
thus to {\evi subtract} one half of the commutator 
$[\bb{\ib},\bbdag{\ib}]$ whenever the term $\bbdag{\ib} \bb{\ib}$
occurs in it. However, since bosonic
commutators contribute terms  of order at least $1/N$ to the action
of the functional integral (cf. Appendix \ref{app:a}, in particular
Eq. \eqref{resc} 
), one half of each commutator has to be subtracted only
when fluctuation $(1/N)$ corrections are included. Thus, to the terms
\beqn
 &&\lam_i^a (\n ei + \sum_{\g}\n s{i,\g} +\n di)
\nonum\\
-&& \sum_{\g} \lam_{i,\g}^b (\n s{i,\g} + \n di) + U \n di
\eeqn
at a given lattice site in the original Hamiltonian, there corresponds
 {\evi subtraction} of the terms
\beq
\lam_0^a \mez 4 - 2\lam_0^b \mez 2 + U \mez = 2 \lam_0^a -2 \lam_0^b +
\frac{U}{2}
\eeq
from the Gaussian contribution to the (site) free energy  calculated
in the \CL (in the paramagnetic case). In this way, we can account
for the presence of the first three terms on the right-hand side of
Eq. \eqref{f1d}.

In the general case when $\R\not=1$  (and $\F[\R;\und b_0]$ 
is nonvanishing), the
last term on the \rhs of Eq. \eqref{f1d} originates from 
 {\evi additional} noncommuting operators which are present 
in the bosonic (hopping)
operator \eqref{zr} as soon as the subsidiary operator $R\is$ is
different from unity. In this case, noncommuting bosonic operators
enter the Hamiltonian also through more general combinations than the
number operators $\bbdag{\ib} \bb{\ib}$.

Quite generally, we argue 
that taking the \CL of the action at the outset corresponds to
replacing  the original Hamiltonian operator $H$ by a different operator
$Q_c(H)$, obtained by a suitable average over the permutations of the
sequence of creation and destruction bosonic operators defining $H$.
To restore the original operator $H$, one has then to introduce the
difference $C_{\infty}(H)$ between $H$ and $Q_c(H)$ and set
\beq
\label{divp} 
H=Q_c(H)+C_{\infty}(H)\;.
\eeq
By construction, the difference $C_{\infty}(H)$ is expressed in terms
of bosonic commutators. It thus contributes to the action only 
terms  of order  at least $1/N$.
In our case this implies that $C_{\infty}(H)$ contributes to the
Gaussian correction to the free energy only its average value
$C_{\infty}^0(H)$ 
evaluated at the mean-field level.

 We can formulate the following hypotheses
 on the operator transformation
$Q_c$ acting on a generic operator $P$:

\begin{itemize}
\item[(i)] $Q_c$ is a linear transformation, in the sense that
$Q_c(P+P')=Q_c(P)+Q_c(P')$ for any given pair of operators $P$ and
$P'$;

\item[(ii)] When two operators $P$ and $P'$ refer to ``different
kinds'' of particles (i. e., to different bosons at the same or
different sites, or to the same bosons at different sites), then
$Q_c(PP')=Q_c(P)Q_c(P')$;

\item[(iii)] If $K$ is a c-number term or a fermionic operator, then 
$Q_c(KP)=KQ_c(P)$.\cite{fermionic} 
\end{itemize}

With the above hypotheses, we can restrict ourselves to considering the
transformation $Q_c$ acting on a monomial operator $P$ 
entering the expression of the Hamiltonian operator $H$, of the form
\beq
P= b^{\dag n}b^m
\label{pbnbm} 
\eeq
(with $m$ and $n$ arbitrary integers) for any given kind $b$ of bosons.
Accordingly, we might take $Q_c$ of the form 
\beq
Q_c(b^{\dag n}b^m) = \frac{1}{n_{\cal P}} \sum_{\{k\}} {\cal P}_k
(b^{\dag n}b^m) 
\label{qc} 
\eeq
where $k$ labels the $n_{\cal P}=(n+m)!/(n!m!)$ permutations 
${\cal P}_k$ of the product $b^{\dag n}b^m$. Alternatively, we could
consider the simpler form
\beq
Q'_c(b^{\dag n}b^m) = \mez (b^{\dag n}b^m+ b^m b^{\dag n}).
\label{qcp} 
\eeq
For our purposes, the two operators \eqref{qc} and \eqref{qcp} are
equivalent to each other, since the average (mean-field) values 
 $C_{\infty}^0(P)$ of the corresponding operators 
 $C_{\infty}(P)$  differ by terms  of
order at least $(1/N)^2$ with respect to the average value of $b^{\dag n}b^m$.
In the following, we shall thus consider only the simpler form
\eqref{qcp}.

To evaluate the average (mean-field) value 
$C_{\infty}^0(b^{\dag n}b^m)$, we exploit the relation
\beq
[b^{\dag},b^m] = - m b^{m-1} \;,
\eeq
from which
\beqn
b^{\dag n}b^m = b^m b^{\dag n}
 &&- \frac{1}{1!} \frac{\de x^n}{\de x}\bigr|_{x=1}
 \frac{\de y^m}{\de y}\bigr|_{y=1} b^{\dag n-1}b^{m-1}
\nonum\\&& -  \frac{1}{2!} \frac{\de^2 x^n}{\de x^2}\bigr|_{x=1}
 \frac{\de^2 y^m}{\de y^2}\bigr|_{y=1} b^{\dag n-2}b^{m-2}
 \nonum\\&& -  \frac{1}{3!} \frac{\de^3 x^n}{\de x^3}\bigr|_{x=1}
 \frac{\de^3 y^m}{\de y^3}\bigr|_{y=1} b^{\dag n-3}b^{m-3}
\nonum\\&& - \cdots \;.
\label{bnbm} 
\eeqn
Thus:
\beq
b^{\dag n}b^m = \mez \left( b^{\dag n}b^m + b^m b^{\dag n} \right) - 
\frac{nm}{2} b^{\dag n-1}b^{m-1} - \cdots
\label{bnbm1} 
\eeq
where the remaining terms originating from the commutator \eqref{bnbm}
have been omitted since they contribute to
$C_{\infty}^0(P)$  at order at least $(1/N)^2$. {\evi At the relevant
$1/N$ order}, comparison of Eq. \eqref{bnbm1} with Eqs.\eqref{divp}
(written for $P$ in the place of $H$),
\eqref{pbnbm}, and \eqref{qcp} yields eventually
\beq
C_{\infty}(b^{\dag n}b^m) = - \frac{nm}{2} b^{\dag n-1}b^{m-1} \;.
\label{cinf} 
\eeq
The corresponding average value $C^0_{\infty}$ is then obtained from
Eq. \eqref{cinf} by replacing  $b$ and $b^{\dag}$ by their
mean-field value $b_0$.

With the monomial \eqref{pbnbm} we can construct the generic
(normal-ordered) operators of interest, namely,
\beqn
 : f(b^{\dag}b): &&= \sum_{n=0}^{\infty} f_n b^{\dag n}b^n ,
\\
 : b^{\dag} f(b^{\dag}b): &&= \sum_{n=0}^{\infty} f_n b^{\dag n+1}b^n ,
\\
 : f(b^{\dag}b) b: &&= \sum_{n=0}^{\infty} f_n b^{\dag n}b^{n+1} .
\eeqn
The corresponding values of $C^0_{\infty}$ are , in the order:
\beqn
C^0_{\infty}( : f(b^{\dag}b):)  &&= - \sum_{n=1}^{\infty} f_n 
\frac{n^2}{2} b_0^{2(n-1)} 
\nonum\\ && =
-\mez \left( f'(b_0^2) + b_0^2 f''(b_0^2) \right),
\label{c0inf:f} 
\\
C^0_{\infty}( :b^{\dag} f(b^{\dag}b):)  &&= - b_0\sum_{n=1}^{\infty}
f_n \frac{(n+1)n}{2} b_0^{2(n-1)} 
\nonum\\ && =
-\frac{b_0}{2} \left( 2f'(b_0^2) + b_0^2 f''(b_0^2) \right),
\label{c0inf:f1} 
\\
C^0_{\infty}( :f(b^{\dag}b) b:) && = C^0_{\infty}( :b^{\dag} f(b^{\dag}b):)
\label{c0inf:f2} 
\eeqn
where we have assumed the series  $\sum_{n=0}^{\infty} f_n x^n$ to
represent the analytic function $f(x)$.

Thus far we have considered a single kind of bosons. For different
kinds of bosons, $C^0_{\infty}$ can be evaluated by exploiting the
following property. Given two operators $P_a$ and $P_b$ referring to
bosons $a$ and $b$, respectively, hypothesis (ii) above together with
the definition \eqref{divp} imply the identity:
\beqn
C_{\infty}(P_aP_b) &&= P_a P_b -Q_c(P_a)Q_c(P_b) 
\nonum\\
&& = P_a P_b - (P_a-C_{\infty}(P_a))(P_b-C_{\infty}(P_b))
\nonum\\
&&=  P_a C_{\infty}(P_b)+ C_{\infty}(P_a) P_b -C_{\infty}(P_a)
C_{\infty}(P_b) \;.
\label{cinfab} 
\eeqn
Taking the average value of both sides of Eq. \eqref{cinfab}, we
obtain at the leading order in $1/N$
\beq
C^0_{\infty}(P_aP_b) = <P_a>_0 C^0_{\infty}(P_b) + <P_b>_0 C^0_{\infty}(P_a) 
\label{c0infab} 
\eeq
where $<P>_0$ signifies that the average value of $P$ is evaluated at
the mean-field level.

The compound operators of interest are of the form
\beq
: f(a^{\dag},a,b^{\dag},b) : = \sum_{m,n} f_{mn} :
 a^{\dag \mu} a^{\nu}   (a^{\dag} a)^m
  (b^{\dag} b)^n b^{\dag \rho} b^{\tau} :
\label{faabb} 
\eeq
where the integers $\mu,\nu,\rho$, and $\tau$ can take the values $0$
and $1$. In particular, when   $\mu=\nu=\rho=\tau=0$ we obtain from
Eqs. \eqref{c0infab} and \eqref{cinf}:
\beqn
C^0_{\infty}(: f(a^{\dag}a,b^{\dag}b) :) &&= \sum_{m,n} f_{mn} \left(
a_0^{2m}  C^0_{\infty}( b^{\dag n} b^n ) + b_0^{2n}
 C^0_{\infty} ( a^{\dag m} a^m ) \right) 
\nonum \\&&
= -\mez \sum_{m,n}  f_{mn} \left[ 
x^m \left(\frac{\de}{\de y}+ 
y \frac{\de^2}{\de y^2} \right) y^n 
+ y^n \left(\frac{\de}{\de x}+ 
x \frac{\de^2}{\de x^2} \right) x^m \right]_{
x=a_0^2, \, y=b_0^2 
}
\nonum \\&&
= -\mez \left[  
\left(\frac{\de}{\de x}+ x \frac{\de^2}{\de x^2} \right)
f(x,b_0^2)\bigr|_{x=a_0^2}
+ \left(\frac{\de}{\de y}+ y \frac{\de^2}{\de y^2} \right)  f(a_0^2,y)\bigr|_{y=b_0^2}
\right]
\nonum\\&&
= 
C^0_{\infty}(: f(a^{\dag}a,b_0^2) :) + C^0_{\infty}(:
f(a_0^2,b^{\dag}b) :)
\eeqn
where use has been made of Eq.\eqref{c0inf:f} and of the assumption
that the series $\sum_{m,n} f_{mn} x^m y^n$ represents the analytic
function $f(x,y)$. By the same token, we obtain 
for the other cases of Eq.\eqref{faabb}:
\beq
C^0_{\infty}(: f(a^{\dag},a,b^{\dag},b) :) = 
C^0_{\infty}(:f(a_0,a_0,b^{\dag},b) :) +
C^0_{\infty}(: f(a^{\dag},a,b_0,b_0) :) \;.
\label{cinf:fafb} 
\eeq
Extension of Eq. \eqref{cinf:fafb} to several kinds of bosons is
straightforward.

We are now in a position to reproduce the last term on the \rhs of
Eq.\eqref{f1d}. This term turns out to be the $C^0_{\infty}$
contribution originating from the presence of noncommuting bosonic
operators in the kinetic part of the Hamiltonian, which occurs only
when nontrivial forms of the subsidiary operator in Eq.\eqref{zr} are
considered (i.e., when $R\is \not=1$). According to the above
prescriptions, we then obtain for the kinetic part of the Hamiltonian:
\beqn
C^0_{\infty}&&(: \frac{t}{\NN} \sum_{i,\nene,\g} f^{\dag}\is z^{\dag}\is
z\js f\js :) =
 \frac{t}{\NN}\sum_{i,\nene,\g} \left< f^{\dag}\is f\js
\right>_0 C^0_{\infty}(:z^{\dag}\is z\js:) 
\nonumber\\&&
= \frac{t}{\NN}\sum_{i,\nene,\g} \left< f^{\dag}\is f\js
\right>_0 z_0 \left( C^0_{\infty}(:z^{\dag}\is:)+ C^0_{\infty}(:z\js:)
\right)
\label{cinf:hop} 
\eeqn
where $z_0$ is given by Eq.\eqref{z0}. In deriving Eq.
\eqref{cinf:hop}
we have made use of Eq. \eqref{c0infab} and of the fact that, at the
order of the $1/N$ expansion we are considering, the fermionic average
has to be evaluated with the bosons taken at the mean-field level.
Equation \eqref{cinf:hop} further simplifies for the homogeneous and
paramagnetic mean-field solution we are restricting to. In this case
\beq
 C^0_{\infty}(:z^{\dag}\is:) =  C^0_{\infty}(:z\js:) =  
C^0_{\infty}(:z_{\up}:)
\eeq
is independent of $i$ {\evi and} $\g$. [It is here understood that the
operator $z_{\up}$ corresponds to a given reference site, say $i_0$]. We
obtain eventually:
\beqn
C^0_{\infty}(: \frac{t}{\NN} \sum_{i,\nene,\g} f^{\dag}\is z^{\dag}\is
z\js f\js :) &&=
2 z_0 C^0_{\infty}(:z_{\up}:)\frac{t}{\NN}  \sum_{i,\nene,\g} \left< 
f^{\dag}\is f\js \right>_0 
\nonum\\&& =
2 z_0 C^0_{\infty}(:z_{\up}:)\frac{2 t}{\NN}  \sum_{\v k}^{BZ} \gamk
f_F(\E_{\v k})
\label{cinf:hop1} 
\eeqn
with $\gamk$ given by Eq.\eqref{gam}.

We are thus left with evaluating $  C^0_{\infty}(:z_{\up}:)$. Taking
$z_{\up}$ of the form [cf. Eq.\eqref{zr}]
\beq
z_{\up} = s^{\dag}_{\down} R_{\up} d + e^{\dag} R_{\up} s_{\up}
\eeq
with
\beq
R_{\g} = :\R( \n e{},\n s{\g},\n s{\bar \g},\n d{}):
\eeq
by our conventions, and using Eqs. \eqref{c0inf:f}-\eqref{c0inf:f2}
and
\eqref{cinf:fafb}, we obtain
\beq
 C^0_{\infty}(:z_{\up}:) =  C^0_{\infty}(: s^{\dag}_{\down} R_{\up} d:)
+  C^0_{\infty}(: e^{\dag} R_{\up} s_{\up}:)
\label{c0inf:z} 
\eeq
where [cf. the dictionary \eqref{defab}]
\beqn
 C^0_{\infty}  &&\left( : s^{\dag}_{\down} R_{\up} d: \right) =
 C^0_{\infty}\left( \bb3_0 :\R( \bbdag1 \bb1 , \bbd2_0, \bbd3_0, \bbd4_0):
\bb4_0\right)
\nonum\\&& 
+ C^0_{\infty}\left( \bb3_0 :\R( \bbd1_0 ,\bbdag2 \bb2 , \bbd3_0, \bbd4_0):
\bb4_0\right)
\nonum\\&& 
+ C^0_{\infty}\left( : \bbdag3 \R( \bbd1_0 , \bbd2_0, \bbdag3 \bb3 , \bbd4_0):
\bb4_0\right)
\nonum\\&&
 + C^0_{\infty}\left( \bb3_0 :\R( \bbd1_0 ,\bbd2_0 , \bbd3_0, \bbdag4 \bb4)
\bb4 :\right)
\nonum\\&&
= -\mez \bb3_0\bb4_0 \biggl\{  
\left[ \frac{\de}{\de n_1} +\bbd1_0 \frac{\de^2}{\de n_1^2} +
\frac{\de}{\de n_2} +\bbd2_0 \frac{\de^2}{\de n_2^2} \right] \R
\nonum\\&&
+\left[ 2 \frac{\de}{\de n_3} +\bbd3_0 \frac{\de^2}{\de n_3^2} +
2 \frac{\de}{\de n_4} +\bbd4_0 \frac{\de^2}{\de n_4^2} \right] \R
\biggr\}
\label{csrd} 
\eeqn
with the same notation used in Eq.\eqref{zrb}. The other contribution 
$ C^0_{\infty}(: e^{\dag} R_{\up} s_{\up}:) $ can be obtained from
Eq.\eqref{csrd} with the replacement $1 \lrarrow 4$ and  $2 \lrarrow 3$
 for the boson indices. Comparison of Eqs. \eqref{c0inf:z} and
\eqref{csrd} with  Eq.\eqref{zrb} identifies eventually
\beq
\F[\R;\und b_0] = 2 z_0  C^0_{\infty}(:z_{\up}:) \;.
\label{ident} 
\eeq
This result, in turn, implies that the last term on the \rhs of
Eq.\eqref{f1d} coincides with the $ C^0_{\infty}$ contribution
\eqref{cinf:hop1}
from the kinetic part of the Hamiltonian.

The method that we have here provided to rationalize the ``\CFI''
\eqref{f1d} might also serve as a practical prescription to avoid the
burden of keeping {\evi explicitly } the discretized nature of the
functional integral until the end of the calculation.\cite{cfi} 
This method
 may thus be useful when considering
Hamiltonians different from \eqref{ham}
 and more general  phases (like magnetic ones). However,
extension of this procedure to physical quantities other than the free
energy and its derivatives (for instance, to the electronic
correlation functions) is not  straightforward and requires a
separate study. This problem will be touched upon in Appendix
\ref{app:c} in the framework of the ``Cartesian'' gauge.

\subsection{ EXTENSION TO OPERATORS  NOT EXPLICITLY IN NORMAL-ORDERED
FORM }
\label{s:nno} 
Thus far we have considered subsidiary operators $R\is$ that are {\evi
explicitly } written in normal-ordered form. In fact, in subsection
\ref{s:gauss} the normal-ordering prescription entered the operative
definition of the subsidiary function $\R$ [cf. the comment following
Eq. \eqref{numop}], while in subsection \ref{s:int} the
normal-ordering prescription for all terms in the Hamiltonian was
assumed throughout [cf. the main result \eqref{ident} therein].
Nonetheless, it might be also relevant to consider operators, like
$R\is^{SQ}$ given by Eq.\eqref{rsq}, which are {\evi not} explicitly
written in normal-ordered form. For instance, it will result from the
numerical calculations presented in the next Section that the choice
\eqref{rsq} remedies the unphysical results obtained with its
normal-ordered version \eqref{rkr}.\cite{normal-ordered}

It is clear that, when adopting a functional-integral formalism based
on coherent states, any non normal-ordered form for $R\is$ should be
preliminarly rearranged into a sequence of normal-ordered terms. In
general, this rearrangement might result into hardly manageable
expressions (or even into divergent expressions when non polynomial
operators like \eqref{rsq} are adopted). This difficulty can be
overcome as follows in the spirit of the $1/N$ expansion considered in
this paper.

Let's consider first the rearrangement of the simple bosonic monomial
operator $(b^{\dag}b)^n$ into a sequence of normal-ordered terms:
\beq
(b^{\dag}b)^n = \sum_{k=0}^n a_{n,k} :(b^{\dag}b)^{n-k}: \;.
\label{n-no} 
\eeq
It is clear that $a_{n,0}=1$. A recursive relation can be readily
obtained for the other coefficients of the expansion \eqref{n-no}:
\beq
a_{n,k} = a_{n-1,k} + (n-k) a_{n-1,k-1}
\label{ank} 
\eeq
with the initial conditions $a_{n,n+1}= a_{n,-1}=0$ and $a_{0,0}=1$.
In particular, Eq. \eqref{ank} gives for $k=1$
\beq
a_{n,1}=\frac{n(n-1)}2 \;.
\label{an1} 
\eeq
Knowledge of coefficients with $k>1$ is not required at the order
$1/N$ we are considering in this paper.

For a general bosonic operator we then write
\beq
f\left( \frac{b^{\dag}b}N \right) = \sum_{n=0}^{\infty} f_n \;
\left( \frac{b^{\dag}b}N \right)^n = 
\sum_{n=0}^{\infty}\sum_{k=0}^{n} f_n a_{n,k} \frac 1{N^n} 
:(b^{\dag}b)^{n-k}: \;,
\label{fbb} 
\eeq
where the expansion \eqref{n-no} has been used and $b^{\dag}b$ has
been divided by $N$ according to a standard procedure of the $1/N$ expansion 
[see Appendix \ref{app:a}].
The right-hand side of Eq.\eqref{fbb} admits a direct representation
in terms of the functional integral, by mapping $:(b^{\dag}b)^q:$ onto
$(b^{*}_m b_{m-1})^q$ in the action for any (positive) integer $q$. In
the spirit of the $1/N$ expansion, we then rescale the bosonic
variables $b_m$ by setting $b_m=\sqrt{N} \tilde b_m$. The operator
\eqref{fbb} is thus mapped in the action onto the series
$\sum_{k=0}^{\infty}N^{-k} \bar f^{(k)} (\tilde b^{*}_m \tilde
b_{m-1})$, where
\beq
\bar f^{(k)}(x) = \sum_{q=0}^{\infty} f_{k+q} a_{k+q,k} x^q \;.
\label{fbark} 
\eeq
At the first order in $1/N$, only terms $ \bar f^{(k)}(x)$ with $k=0$ and
$1$ have to be retained in the action. Specifically, the term with
$k=0$ will contribute both to the leading term $F_0$ of the free
energy [cf. Eq. \eqref{a.10}] and to its $1/N$ correction $F_1$  
[cf. Eq. \eqref{f1bis}], by considering its mean-field value and its
Gaussian terms, respectively. The term with $k=1$ will instead
contribute only to $F_1$, by considering its mean-field
value.\cite{action} 
In this way, the contribution to $F_1$ of the term with $k=1$ can be
effectively reabsorbed into the ``contribution from infinity``
 to $F_1$ originating from
the term with $k=0$. Specifically, it turns out that this contribution
simply cancels the term containing the second derivative of $f$ in Eq.
\eqref{c0inf:f}.

Entering Eq. \eqref{an1} into Eq. \eqref{fbark} we, in fact, obtain
for $\bar f^{(1)}(x)$:
\beq
\bar f^{(1)}(x) = \sum_{q=0}^{\infty} f_{1+q} a_{1+q,1} x^q = \mez x
f''(x)\;, 
\eeq
so that
\beq
C^0_{\infty} (:f(b^{\dag}b):) + \bar f^{(1)}(b_0^2) = -\mez f'(b_0^2)
\label{cinftot} 
\eeq
owing to Eq. \eqref{c0inf:f}. In Eq. \eqref{cinftot} we have
consistently replaced the argument of the function  $\bar f^{(1)}(x)$
by $b_0^2$ and set $N=1$ eventually.
Generalization of the above results to the case of several bosons is
straightforward.

 In conclusion, for an operator $R\is$ not {\evi
explicitly } in normal-ordered form the above considerations have the
effect of modifying the factor $\F[\R;\und b_0]$ in Eq. \eqref{f1d},
by (i) dropping the last terms in Eq. \eqref{zrb} containing the
second derivatives of $\R$ and (ii) interpreting the function $\R$ in the
remaining terms in Eq. \eqref{zrb} (as well as in the expression for
the continuum part of the fluctuations) as being associated with
$:R\is:$, i.e., with the normal-ordered component of $R\is$.

It will be shown in the next Section that, when specified to
$R\is^{SQ}$, the above prescription succeeds in eliminating the unpleasant
 features occurring for its normal-ordered version $R\is^{KR}$ [cf. Eq.
\eqref{rkr}]. Namely, it eliminates an unphysical divergence of the
ground-state energy when the zero-occupancy limit is approached
(making the ground-state energy to  correctly vanish in this limit) and
completely suppresses the fluctuation corrections to the KR \MF
ground-state  energy for $U=0$ and  any filling value.

Finally, we remark that the right-hand side of Eq. \eqref{cinftot} could
be obtained directly by generalizing the euristic argument of
subsection \ref{s:int} to the case of non normal-ordered operators.
To this end, one suitably defines the operator $Q_c(P)$ (and
consequently $C_{\infty}(P)$) for a non normal-ordered operator $P$ by
extending the ``specular'' prescription \eqref{qcp}, and evaluates 
$C^0_{\infty}$ accordingly. By this procedure, one is led to interpret
the right-hand side of Eq. \eqref{cinftot} as the ``\CFI'' originating
from the non normal-ordered operator $f(b^{\dag}b)$.

\section{NUMERICAL RESULTS FOR A ONE-LEVEL TWO-SITE MODEL}
\label{s:nume} 
In the previous Section we have shown that the additional
contributions \eqref{f1d} to the Gaussian free energy originate from a
careful handling of the continuum time limit of the functional
integral. The question naturally arises whether these additional
contributions might substantially affect the numerical value of the
free energy (and of related physical quantities) in the practical cases
of interest, or even modify its behavior in a qualitative way.
Specifically, we may inquire whether the additional contributions
\eqref{f1d} could serve to overcome the inconsistencies 
encountered when including Gaussian fluctuations 
by treating  the \FI in  the \CL~. In this respect,
we have found that the {\evi wrong curvature} of the free energy
versus $U$ results with the conventional \CL treatment, thus
producing an unphysical {\evi increase} of the (average) number of
doubly occupied sites for in creasing $U$. Additional inconsistencies
have been pointed out for
 the noninteracting ($U=0$) case in Ref. \cite{jolicoeur}. 

In this Section we will show that inclusion of the ``\CFI''
\eqref{f1d} indeed succeeds in overcoming the inconsistencies mentioned
above. 
In addition, we will show that relaxing the normal-ordering
prescription 
\eqref{rkr} 
of the KR choice  for the subsidiary operator,
 makes the $1/N$ results 
{\evi coinciding} with the exact (free-particle) solution at $U=0$
for any filling and  number of lattice sites. In this way, the
goodness of the KR \MF solution will not be spoiled by the inclusion of
fluctuations. 

 We shall illustrate the above effects
 by numerical calculations at a generic value of $U$ for a
one-level two-site model, which avoids the complications due to the
spatial structure while dealing 
explicitly
with the imaginary time discretization.
 We expect, however, these effects
 to hold  for a more general multi-site system as well,
since the time \CL does not depend on the spatial structure
[In this respect, we will also present some numerical results for the
lattice case when $U=0$]. The
restriction to a two-site model will also enable us to compare the
numerical results, obtained within alternative approximations to the
\FI, with the available exact solution discussed in Appendix
\ref{app:d}.\cite{periodic} 
In addition, to get a practical insight into the  usefulness of a given
choice for the bosonic hopping operator $z$, we compare the results obtained
with several alternative choices of $z$.

\subsection{MEAN-FIELD SOLUTION}
\label{mfsol} 
At the \MFL, the free energy (per lattice site) of the paramagnetic
solution is given
by [cf. Eqs.  \eqref{f0}, \eqref{s0}, and \eqref{lims0}]:
\beq
F_0= -\frac{1}{\bet} \sum_{p=1}^2 \ln \left( 1+e^{-\bet\E_p} \right) +
F_0^{(B)}
\eeq
where
\beq
F_0^{(B)} = \lam_0^a (\bbd1_0 +2\bbd2_0 + \bbd4_0-1) - 2 \lam_0^b (
\bbd2_0 + \bbd4_0) + U \bbd4_0 
\eeq
and [cf. Eq. \eqref{epsk}]
\beq
\Biggl\{ \begin{array}{l}
      \E_1=\lam_0^b-\mu -2 t \rho_0^2
\\       \E_2=\lam_0^b-\mu +2 t \rho_0^2
\end{array}
\label{e1e2} 
\eeq
with the notation 
\beq
\rho_0= \bb2_0(\bb1_0+\bb4_0) \R(n_1,n_2,n_2,n_4)
\label{rho0} 
\eeq
in the place of $z_0$ [cf. Eq. \eqref{z0}]. We have here specified to
the two-site model the general expressions reported in Section \ref{s:func}
for the lattice model, and made use of the paramagnetic {\evi ansatz}
$\bb2_0=\bb3_0$.

The parameters $\bb1_0,\bb2_0,\bb4_0,\lam_0^a,\lam_0^b$, and $\mu$ are
determined by minimizing $F_0+\mu n$ (where $n$ stands for the
particle density per lattice site). The resulting mean-field equations are:
\beqn
&&
\frac{\de F_0}{\de \bb{\ib}_0} = 4 t (f_F(\E_2)-f_F(\E_1)) \rho_0
\frac{\de \rho_0}{\de  \bb{\ib}_0} + \frac{\de F_0^{(B)}}{\de
\bb{\ib}_0}=0 \quad (\ib=1,2,4) ,
\label{eqbb} 
\\&&
\frac{\de F_0}{\de \lam_0^a} = \bbd1_0+ 2 \bbd2_0 +  \bbd4_0-1=0,
\label{mfcons1} 
\\&&
\frac{\de F_0}{\de \lam_0^b} = f_F(\E_1)+f_F(\E_2) - 2 (\bbd2_0+
\bbd4_0)=0 \;,
\label{mfcons2} 
\\ 
-&& 
\frac{\de F_0}{\de \mu} = f_F(\E_1)+f_F(\E_2) =n \;.
\label{eqmu} 
\eeqn
In particular, {\evi in the zero-temperature limit} Eq.\eqref{eqmu}
implies 
\beq
f_F(\E_1)=1 \quad\quad, \quad f_F(\E_2)=n-1 \;,
\label{ffsc} 
\eeq
whenever $n\geq1$ and $\rho_0$ is nonvanishing, with the chemical
potential jumping from $\E_1$ to $\E_2$ across $n=1$.

At ``half filling'' (i. e., when $n=1$) there exists a critical value
$U_c$ such that $\bb1_0=\bb4_0=0$ and $\bb2_0=1/\sqrt{2}$ for $U\geq U_c$
(and the associated Helmholtz free energy density $F_0+\mu n$ vanishes
in the zero-temperature limit). It can be readily shown from Eqs.
\eqref{eqbb}-\eqref{eqmu} that $U_c$ is given by\cite{lattice} 
\beq
U_c= 4 t \R^2(0,\mez,\mez,0) \;.
\label{rc} 
\eeq
In deriving Eq.\eqref{rc} we have exploited the symmetry properties
\beq
\R(n_1,n_2,n_2,n_4)=\R(n_4,n_2,n_2,n_1)=\R(n_2,n_1,n_4,n_2)
\label{rsimm} 
\eeq
which are satisfied by the choices we shall consider for $\R$. This
transition at $U=U_c$ makes the effective hopping in Eq.\eqref{e1e2}
 vanishing, and  it is referred to as a Mott-Hubbard
transition.\cite{vollhardt} 
For the two-site model we are considering this transition is clearly
an artifact of the slave-boson approach at the \MFL, which will be (at
least partly) cured by the inclusion of (Gaussian) fluctuations.

For comparison, we also give the results of the conventional
Hartree-Fock decoupling. For the one-level two-site model of interest,
the Hartree-Fock ground-state energy (per lattice site) in the
paramagnetic phase is given by
\beq
E^{HF} = 2 t ( f_F(\E^{HF}_2) - f_F(\E^{HF}_1) ) + (\frac{n}{2})^2 U
\label{ehf} 
\eeq
where
\beq
\Biggl\{ \begin{array}{l}
      \E^{HF}_1= -2 t + \frac{n}{2} U - \mu
\\       \E^{HF}_2= 2 t + \frac{n}{2} U - \mu
\end{array}
\label{e1e2hf} 
\eeq
with $f_F(\E_1)$ and $f_F(\E_2)$ given by Eq. \eqref{ffsc}
(in the zero-temperature limit for $n\geq1$). Since 
 $\E^{HF}_p$ ($p=1,2$) are independent of $U$ (at fixed particle
density), the energy \eqref{ehf}
grows linearly without bound for increasing $U$. Equation \eqref{ehf}
also shows that, within the (paramagnetic) Hartree-Fock decoupling, the
site probability of double occupancy is given by $(n/2)^2$,
irrespective of $U$. One thus concludes that the (paramagnetic)
Hartree-Fock decoupling can be approximately correct  only
for $U\cmin t$ (being, in particular, exact when $U=0$). Numerical
comparison of the mean-field ground-state (site) energy $E=F_0+\mu n$
(in the zero-temperature limit) and of the associated $1/N$ 
fluctuation results
with the Hartree-Fock  $E^{HF}$ and with the exact solution 
 will be presented in the following.

\subsection{ OVERCOMING THE PROBLEMS ASSOCIATED WITH THE CONTINUUM 
FLUCTUATIONS BY THE DISCRETIZED FLUCTUATIONS} 
A naive handling of the \FI (whereby the \CL is taken at the outset) yields,
at the Gaussian level, the (site) free energy \eqref{ftot} with the
term $F_1^{(d)}$ omitted and $F_1^{(c)}$ given by Eq.\eqref{f1c}. In
that expression, the continuum fluctuation matrix $\Gam_c\qunu$ is
obtained by performing the $\delta\to0$ limit of the matrix
\eqref{gamq} (with $\BB$ and $\CC$ given by Eqs.\eqref{bb} and
\eqref{cc}, in the order), i. e., by letting  $\delta\to0$ {\evi
everywhere} $\delta$ appears.

In particular, the \CL of the matrix $\BB$ given by Eq.\eqref{bb} has elements
\beqn
&&
\BB_c(\v q,\omnu|1,1)= i \omnu + \lam_0^a ,
\nonum \\&&
\BB_c(\v q,\omnu|1,2)= \cdots =  \BB_c(\v q,\omnu|1,7) =0,
  \nonum \\&&
\BB_c(\v q,\omnu|1,8)= \BB_c(\v q,\omnu|8,1)=  \bb1_0,
\eeqn
and so on; the \CL of the factors $\EE^{(1)}$ entering the expression
\eqref{cc} for the matrix $\CC$ is given by
\beqn
&&
\EE^{(1)}_c (\v q,\omnu;\v k,\up|\ib) = t z_0 (\gamk + \gamkq) \frac{\de
z_0}{\de \bb{\ib}_0}, \quad\quad (\ib=1,\cdots,4)
\nonum \\&&
\EE^{(1)}_c (\v q,\omnu;\v k,\up|5) = t z_0 \R (\gamk - \gamkq)
\bb2_0\bb4_0,
\eeqn
and so on (with $z_0$ given by Eq. \eqref{z0}); the \CL of the factors
$\EE^{(2)}$ entering the expression \eqref{cc} for the matrix $\CC$ is
given by
\beq
\mez \sum_{\g} \EE^{(2)}_c (\v q,\omnu;\v k,\g|1,1) = \frac{t}{2} \left[
 2z_0 \frac{\de^2 z_0}{\de \bbd1_0} \gamk + \left(\frac{\de
z_0}{\de\bb1_0} \right)^2 (\gamkq+\gamkpq) \right],
\eeq
and so on.\cite{matrices} 
Finally, the \CL of the Fermi function $f_M(\E)$ and of the
polarization function $\Pi_M(\E,\E';\nu)$ entering the expression of
the matrix $\CC$ are given in Appendix \ref{app:b}.

The above expressions hold for a general lattice model. The two-site
model that we consider for numerical calculations is recovered by
restricting the wave vectors to the two values $0$ and
$\pi/|\nene|$.\cite{periodic}

Proper handling of the functional integral (with the \CL taken only
at the end of the calculation) adds to the \CL (site) free energy
$F_0+F_1^{(c)}$ the ``\CFI'' $F_1^{(d)}$ given by Eq.
\eqref{f1d} with the appropriate form \eqref{zrb} for $\F[\R;\und
b_0]$, with or
without the terms containing the second derivatives
 depending on the ordering prescription (cf. subsection \ref{s:nno}). 
 Our purpose here is to demonstrate to what
extent the contribution $F_1^{(d)}$ modifies the free energy in a
quantitative and even in a qualitative way, using the two-site model
as a simple prototype system. In this way, the contribution  $F_1^{(d)}$
will prove to be important  also from a practical point of view.

To proceed further, we have to select an {\evi explicit} form for the
subsidiary operator $R\is$ introduced in Eq.\eqref{zr}.
The choice  commonly adopted in the four-slave-boson
literature is the original KR choice \eqref{rkr}. The associated
 subsidiary function introduced in
Eq.\eqref{rho} takes the form  
\beq
\R^{KR}(n_e,n_{\g},n_{\bar\g},n_d) = \frac{1}{\sqrt{1-n_e-n_{\bar\g}}
\sqrt{1-n_d-n_{\g}}},
\label{rkrfun} 
\eeq
which recovers at the \MF level the (paramagnetic) Gutzwiller
approximation for the lattice Hubbard Hamiltonian at any band
filling.\cite{kotliar} 
However, it will turn out that the  normal-ordering prescription 
leads to 
results which markedly depart from the free-particle
solution at $U=0$ {\evi even} when the ``\CFI'' is properly included.
In addition, this prescription leads to a divergent free
energy in the zero-occupancy limit.
These shortcomings can be remedied  by
 suitably relaxing the normal-ordering
prescription, as shown in the next subsection.

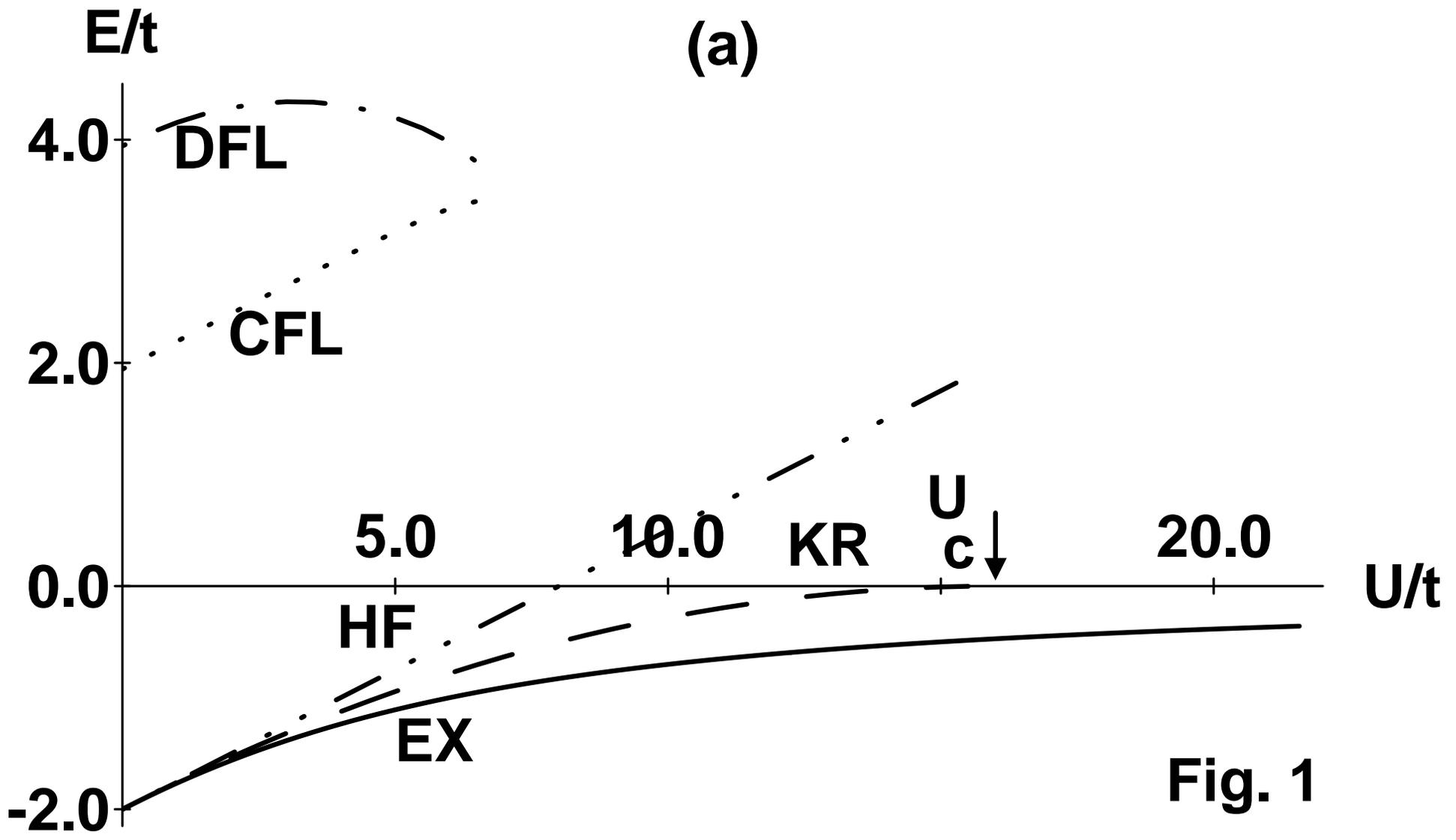
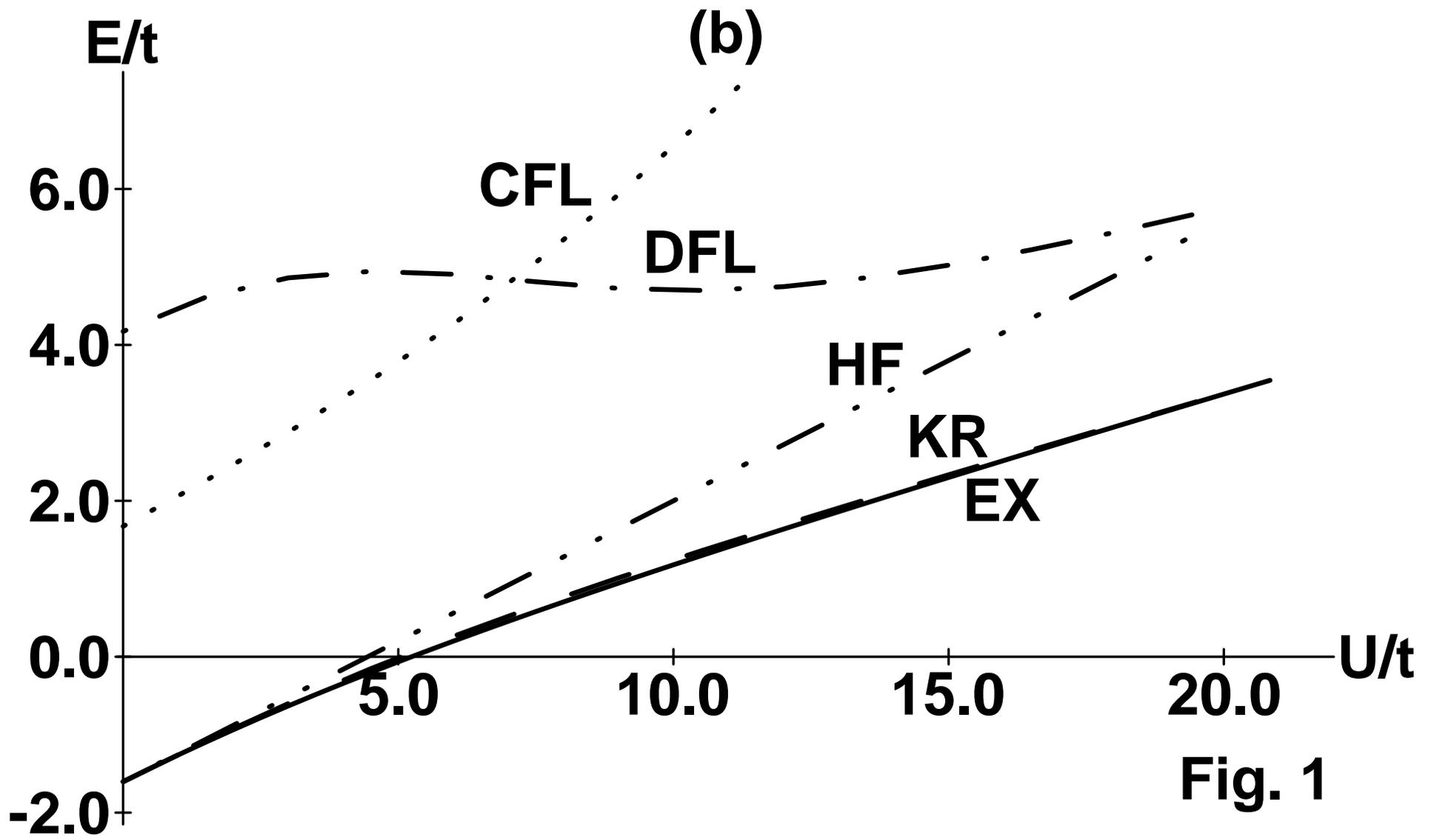
\begin{figure}
\caption{ Ground-state energy (in units of $t$) for the one-level
two-site model versus $U/t$ when (a) $n=1.0$ and (b) $n=1.2$, obtained
by the exact solution and with the KR choice (\figref{rkr}), as explained
in the text.}
\label{fig1}
\end{figure}

Figure \ref{fig1} compares the ground-state energy per lattice site
(in units of $t$) for the one-level two-site model versus $U/t$ at
``half filling'' ($n=1.0$) and at $n=1.2$, as obtained from the
exact solution of Appendix \ref{app:d} and from alternative
(paramagnetic) approximations to the four-slave-boson method with the
KR form \eqref{rkr} for $R\is$.\cite{numerical} 
The Hartree-Fock result \eqref{ehf} is also shown for comparison (HF).
The KR mean-field solution (KR) is seen to be in  good agreement
with the exact solution (EX) for both doping values, while the
fluctuation results considerably worsen this agreement {\evi
regardless} they are obtained by the incorrect continuum (CFL) or by
the correct discretized (DFL) time limiting procedure in the \FI (i.
e., without and with the inclusion of the term $F_1^{(d)}$ of Eq.
\eqref{ftot}). The HF result, on the other hand, is in agreement with
the exact solution only for small values of $U$ (say,
when $U\cmin 5t$). Note from Fig.\ref{fig1}a that at ``half filling''
both the CFL and DFL curves terminate at a critical value of $U$ (
$\cong 6.8 t$) where an \af c instability develops at the \MFL,
thus preventing the inclusion of paramagnetic fluctuations.
 Note
also the arrow which locates the critical value $U_c/t$
corresponding to the Mott-Hubbard transition for the KR mean-field solution
(the choice \eqref{rkrfun} for $\R$ gives $U_c/t=16$
according to Eq.\eqref{rc}).

We conclude from Fig.\ref{fig1} that, with the KR choice \eqref{rkr}
for $R\is$, the DFL results can be even worse than the CFL results, as
they depart  from the exact solution more than
 the mean-field results.
This finding could  appear  surprising, since one would
expect  a more complete treatment of the
\FI to produce better results.
That this is not the case with the KR choice \eqref{rkr}
can definitely be concluded by looking at the DFL free energy in the
 zero-occupancy limit, which diverges for this choice in this limit.

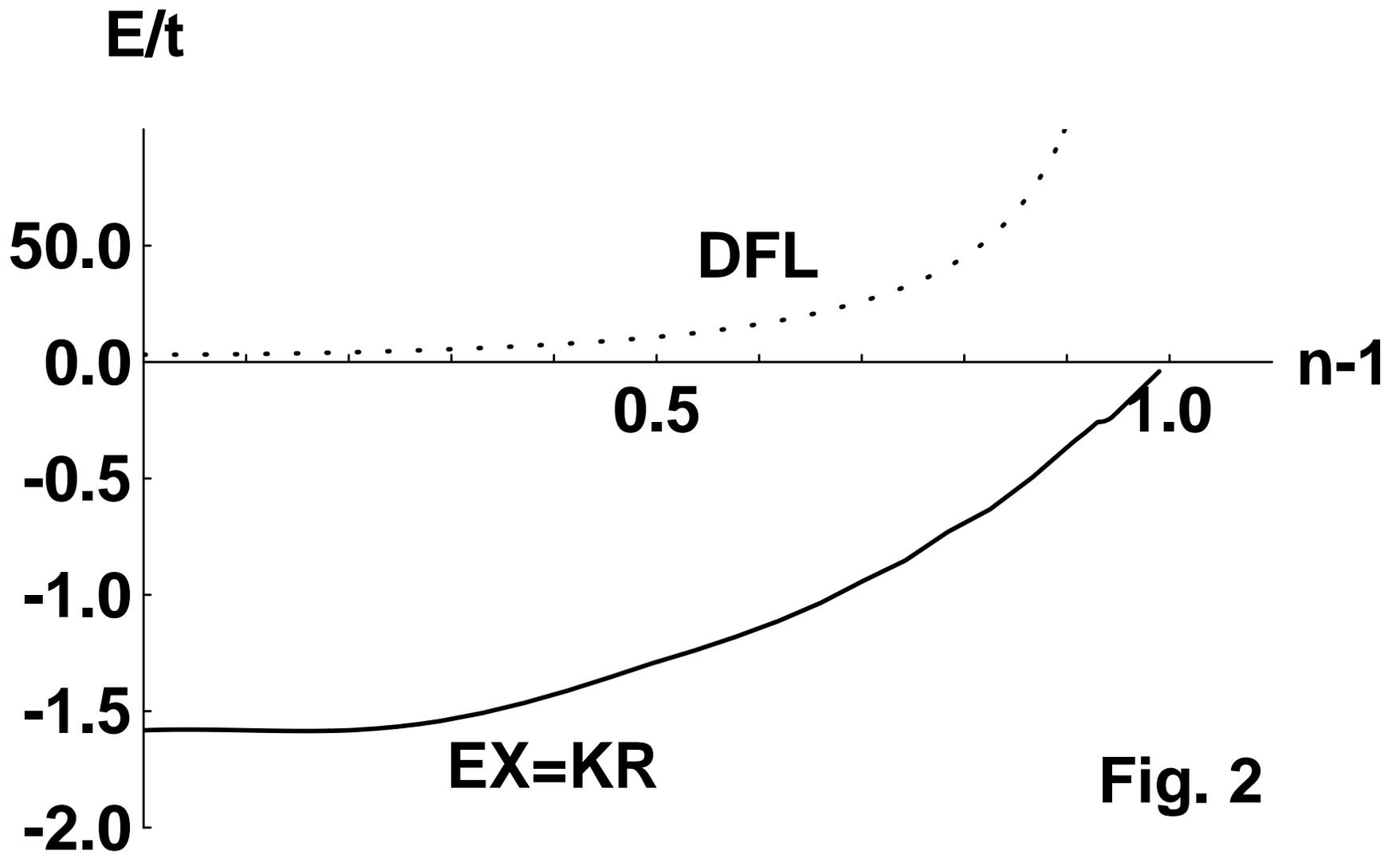
\begin{figure}
\caption{ Ground-state energy (in units of $t$) per lattice site for
the lattice model versus doping $n-1$ at $U=0$, obtained with the
KR choice (\figref{rkr}). Conventions are explained in the text.
Note the change of scale between positive and negative values of $E$.
}
\label{fig2}
\end{figure}

Before searching for  alternative prescriptions for the subsidiary
operator $R\is$ in Eq.\eqref{zr}, it is relevant to
verify whether the above unpleasant results were peculiar to the finite-size
model.
 To this end, we have performed the $U=0$ calculations 
also for the lattice case 
versus the doping value $n-1$ (since in the lattice  case the \af c instability
occurs already at infinitesimal $U$ due to the perfect nesting of the
Fermi surface). 
The results
shown in Fig.\ref{fig2} confirm our conclusion that, regardless of its
successes at the \MFL, the KR form of $R\is$ is not suited for the
inclusion of fluctuations.
Note, in particular, the divergence of
the DFL results when $n=2$, which was mentioned above for the two-site model.

This conclusion calls for  alternative expressions
of the subsidiary operator.
One could initially resort to 
 the {\evi simplest} possible expression 
\eqref{z1} for $z\is$, 
for which 
 $R\is=R\is^0=1$ and
$\R=1$.
In this case
 the factor $\F[\R;\und b_0]$ of Eq.\eqref{f1d} vanishes identically [cf.
Eq.\eqref{zrb}]. The corresponding results for the ground-state energy
(per lattice site) of the one-level two-site model are shown in
Fig.\ref{fig3}, with the same notations of Fig.\ref{fig1} (except for
the labeling of the mean-field curve, which is now MF). Note that the
critical value for the Mott-Hubbard transition to occur at
``half filling'' is now $U_c=4 t$, according to Eq.\eqref{rc}. We see from
this figure that the DFL results now improve the agreement with the
exact solution with respect to the MF results, while the CFL results
considerably worsen the agreement. This finding
thus gives us support for
treating correctly the time \CL of the functional integral
 within the four-slave-boson
approach.

 In addition, we note the
following {\evi shortcomings of the CFL results:}
\begin{itemize}
\item[(i)] At finite doping ($n>1$) the CFL result yield the {\evi
wrong curvature} of $E/t$ versus $U/t$, thus producing an {\evi unphysical
increase} of the (average) number of doubly occupied sites for
increasing $U/t$.\cite{curvature} 
This shortcoming is remedied by the DFL results.

\item[(ii)] As mentioned above, in the zero-occupancy ($n=0$) limit, the ground-state
energy has to vanish irrespective of the value of $U$. The CFL results
 are instead 
finite in that limit. (This
shortcoming is also shared by the previous KR choice for $R\is$). The mean-field
results MF and HF behave instead correctly in the limit. Once more, the
DFL calculation remedies the shortcoming (with the exception of the
previous KR
choice for $R\is$ which leads to divergent DFL results in the $n=0$
limit). The $n=0$ limit thus provides us with a common {\evi reference
level} for comparing alternative results for the ground-state energy.

\end{itemize}

Although the choice $R\is^0=1$ cures formally all flaws occurring 
with the  KR choice
when fluctuation corrections are included, it is seen from
Fig.\ref{fig3} that the mean-field solution associated with  $R\is^0=1$
 constitutes a rather poor starting point for the inclusion of
fluctuations. In this respect, even the HF mean field  is closer to
the exact solution for $U/t \cmin8$. In particular, the choice $R\is^0=1$
does not reproduce at the \MFL the independent-particle result when
$U=0$. By contrast, the  KR form for $R\is$ has been tuned to reproduce
{\evi exactly} the $U=0$ result at the \MFL, by introducing the
nontrivial factor $\R^{KR}$ just as a ``normalization'' factor.

It is thus clear from the above discussion that a 
more sophisticated prescription than $R^0\is=1$ is required
 for the subsidiary operator, which would preserve 
  the
nice features associated with $R\is^{KR}$ at the \MFL and, at the same
time, would avoid the unphysical behavior resulting from 
$R\is^{KR}$ 
when fluctuations are included.

\subsection{ALTERNATIVE SELECTIONS OF THE SUBSIDIARY OPERATOR $R$} 
An appealing feature of the KR (paramagnetic)
mean-field solution is that (besides recovering the  $U=0$
limit) it reproduces the Gutzwiller approximation for the single-band
(lattice) Hubbard Hamiltonian (at any filling).\cite{kotliar}
 This approximation is,
in turn, expected to capture the essential electronic correlations for
the Hubbard model by reducing the site double occupancy at large 
$U$.\cite{ref2} 
One would then like to select the form of the operator $R\is$ in such a
way that the Gutzwiller approximation is still recovered at the \MFL
(at least for some particular filling). Recall in this context that  
 the subsidiary operator $R_{i,\g}$ of
Eq.\eqref{zr} has to be equivalent to the unit operator in the relevant
Fock subspace.

An alternative selection for the operator $R\is$  might rest on 
the following
criterion which exploits the mean-field constraints
\eqref{mfcons1}-\eqref{eqmu} in the zero-temperature limit. In
particular, {\evi at half filling} ($n=1$) the relation
$\bbd1_0+\bbd2_0=\bbd2_0+\bbd4_0=1/2$ holds for any value of $U$,
 so that any function $\R$ of
$\bbd1_0+\bbd2_0$ and $\bbd2_0+\bbd4_0$ will also be independent of
$U$. This property is satisfied by the KR form \eqref{rkrfun}. For
any other form of $\R$  satisfying this property, it is then enough
to ``normalize'' $\R$ in such a way that it coincides with $\R^{KR}$
when 
$\bbd1_0+\bbd2_0=\bbd2_0+\bbd4_0=1/2$.
This criterion guarantees that this new $\R$ and $\R^{KR}$ give the same
mean-field results.
Note that the \MF free energy $F_0$ is
also independent of $\R$ (whenever the property \eqref{rsimm} holds)
since $\lam_0^a$ does not enter $F_0$ at self-consistency.

A  form of the function $\R$, which
 reproduces the KR \MF results at
half-filling, is obtained by suitably ``linearizing''  the expression 
\eqref{rkrfun}, namely, by setting \cite{noiprl}
\beq
\R^{LIN}(n_e,n_{\g},n_{\bar \g},n_d)= [1+x(n_e+n_{\bar\g})][1+x(n_{\g}+n_d)]
\label{rlin} 
\eeq
with $x=x_0$ such that (in the paramagnetic case)
\beq
\left( 1+\frac{x_0}{2} \right)^2 = \frac{1}{(1-\mez)} =2 \;.
\label{x0} 
\eeq
This gives $x_0=2(\sqrt{2}-1)\cong 0.828$ for the positive solution.
Although this particular value of $x$ has been selected at ``half
filling'', in the following we shall use the same value of $x$ for any
filling. Note that the subsidiary function \eqref{rlin} corresponds to
the ``linearized'' subsidiary operator 
$R\is^{LIN}=[1+x(\n ei+\n s{i,-\g})] [1+x(\n di+\n s{i,\g})]$ for
which the normal ordering is irrelevant. This simple choice is free
from the nonanalyticity problems affecting instead
$R\is^{KR}$.\cite{normal-ordered}

The  results for the ground-state energy (per lattice site) of the two-site
model corresponding to \eqref{rlin} 
are shown in Fig.\ref{fig4} for filling values $n=1.0$,
$n=1.2$, and $n=1.5$. Curves are labeled with the conventions used in
Fig.\ref{fig3}. Note that the DFL results at ``half filling'' are now
quite close to the exact solution for $U/t\cmin8$. In particular,
 the DFL fluctuation corrections do not
change appreciably (i. e., within less than $1\%$) the exact solution at
$U=0$. (We have verified that this finding
remains true  also for an infinite system). The agreement between DFL
and EX, however, is not so good for $U$ close to the critical value
$U_c$ ($=16 t$) for the Mott-Hubbard transition to occur in the \MF
solution. Note also the presence of a ``cusp'' about $U_c$ in the
fluctuation results. This ``cusp'' is of no special concern since it
is a characteristic feature of the Gaussian corrections near the
transition point where higher-order fluctuations are expected to be
important. This ``cusp'', in fact, quickly disappears as soon as one
moves away from half filling (cf. Figs.\ref{fig4}b and  \ref{fig4}c).
Note finally that the exact solution for the ground-state energy
approaches a linear trend with respect to $U$ for increasing doping.
As a consequence, the HF approximation becomes progressively more
reliable for increasing doping.

The behavior of the ground-state energy (per lattice site) versus doping $n-1$ 
is shown in Fig. \ref{fig5} for three characteristic values of $U/t$
with the same ``linearized'' $\R^{LIN}$ used for Fig. \ref{fig4}. The CFL
results, however, have been excluded from Fig. \ref{fig5} as they
depart markedly from the other results. Even in this case, the correct
fluctuation contribution (DFL) give quite good overall results.
Note that the piecewise linear behavior of the exact
solution versus doping is  due to the finite size of
the system (as well as to the zero-temperature limit), since the exact
solution for this system is {\evi constructed} by interpolating
between the solutions with $n=1.0$ and $n=1.5$ and 
with $n=1.5$ and $n=2.0$.
Note, however, from Fig.\ref{fig5}a that the free-particle results at $U=0$
 are reproduced by the \MF solution and by the fluctuation corrections
 at ``half filling'' only (and when $n=0.0$ or $2.0$),
 while departures from the correct results become
more pronounced at finite doping. (Recall that the KR mean field
provides instead - by construction - the correct $U=0$ results at any doping
and thus the correct value for the compressibility). 

 The above shortcoming is obviously due to the fact that 
 the parameter $x$ of the ``linearized'' 
 $\R^{LIN}$ has been selected
at ``half filling'' according to Eq.\eqref{x0}. Although one might fix
$x$ at any selected doping, it is certainly not satisfactory to adjust
$x$ to be
doping dependent  in order to reproduce the
independent-particle results at any doping. For this reason the
``linearized'' form \eqref{rlin} proves to be  not completely
satisfactory.

 The fact that  the KR mean field 
reproduces the ($U=0$) free-particle results for {\evi any} doping is
due to the 
perfect cancellation of the doping dependence of the factor $\bb2_0
(\bb1_0 +\bb4_0 )$  in Eq.\eqref{rho0} with the self-consistent 
doping dependence of
the $\R^{KR}$ factor  \eqref{rkrfun}. It is also evident that no
``polynomial'' 
generalization of 
 \eqref{rlin} can yield this perfect cancellation for 
all doping values.

To fully preserve the nice features of the KR mean field, we finally
 consider the ``square root'' form $R\is^{SQ}$ 
\eqref{rsq} for the subsidiary
operator, whereby the KR normal-ordering prescription 
in \eqref{rkr} has been removed.
 Following
 the discussion of subsection \ref{s:nno}, this can readily be achieved
within the $1/N$ expansion by 
dropping the last terms in Eq.\eqref{zrb} containing the second
derivatives and using $\R^{KR}$ in the rest of the calculation.

The results obtained for the ground-state energy (per lattice site) 
of the two-site model
with the choice $R\is=R\is^{SQ}$ are shown in Fig.\ref{fig6} versus
$U$ for $n=1.0$ and $n=1.2$, where the \MF (KR) and the complete $1/N$
(SQ) calculations are compared with the exact solution (like in
Fig.\ref{fig1}, an \af c instability develops 
at ``half filling''
in SQ when $U\cong 6.8 t$).
 The corresponding results versus doping $n-1$ are shown in
Fig.\ref{fig7} for three characteristic values of $U$. Note, in
particular, from Fig.\ref{fig7}a that {\evi the $1/N$ fluctuation
corrections are now completely suppressed at $U=0$ for any doping}. We are
thus able to recover exactly the $U=0$ solution not only at the \MF
level but {\evi also} with the inclusion of fluctuations, without
having to adjust any parameter (like, for instance, the $x$ parameter
in Eq. \eqref{rlin}). In addition, this finding is not limited to the
two-site model but applies also to the lattice case, as shown in
Fig.\ref{fig8}. In this way, the non normal-ordering $SQ$ prescription
\eqref{rsq} overcomes the difficulties signaled in
Ref.\cite{jolicoeur}, without the need of changing the functional form
of the bosonic operator $z$ at the leading $1/N$ order beyond mean
field. Note, finally, from Figs.\ref{fig6} and \ref{fig7} that the SQ
prescription gives good agreement with the exact solution for the
two-site model at any doping {\evi even} at finite $U$. In particular,
for the filled band ($n=2$) the SQ prescription recovers the exact
result $E=U$ for any $U$ [or $E=0$ for the empty band ($n=0$)], in
contrast with the KR normal-ordering prescription that gives divergent
results as $n\to 2$ (or $n\to0$). This divergence resides in the second
derivative terms of Eq. \eqref{zrb}, which are appropriately
eliminated by the non normal-ordering SQ prescription.

\section{CONCLUDING REMARKS}
\label{s:conc} 

In this paper we have demonstrated the importance of taking correctly
the \CL of the functional integral for the four-slave-boson method by
Kotliar and Ruckenstein, by deriving in detail the additional
(ultraviolet) contributions to the free energy 
associated with a proper handling of the
vanishing of the imaginary time step. Although these additional
contributions to the free energy have been already derived for the
one-slave-boson case in Ref.\cite{noiprep}, the presence of several
bosons and of a nontrivial bosonic hopping factor $z$ (which is
peculiar to the KR method) results into {\evi additional noncommuting
terms} in the slave-boson Hamiltonian, which, in turn, make the proper
handling of the \CL of the functional integral quite more involved.
Nonetheless, the rather elaborate mathematical derivations that
resulted proved  necessary to implement correctly the KR method
and make it working properly  in practice.  We have
also provided an ``effective'' rule to obtain the additional terms in
the free energy directly from the original Hamiltonian, without the
need of going explicitly through the limiting discretization procedure
of the functional integral. The fact that this careful procedure has
anyhow produced unphysical results at the  $1/N$ order beyond mean
field when adopting the conventional KR choice for the bosonic hopping
factor $z$, has further required us to search for alternative forms
(or prescriptions) for $z$. Specifically, we have shown that removing
the KR normal-ordering prescription for $z$ enables us to overcome the
above difficulties. In particular, we have verified numerically that
the non normal-ordering prescription succeeds in suppressing
completely the $1/N$ fluctuation corrections to the ground-state
energy at $U=0$ and for any band filling. We regard this finding to be
rather compelling to remove the doubts which have been raised on
the capability of the KR four-slave-boson method to yield meaningful
fluctuation corrections (see, in particular, Ref.\cite{jolicoeur}).

In this context, some additional features still need, however, to be
verified. For instance, it should be verified that removing the
normal-ordering prescription from the KR choice of $z$ suppresses the
$1/N$ corrections at $U=0$ also for dynamical quantities (like the
fermionic Green's functions).
In this respect, we have developed preliminary arguments which show 
 that the non normal-ordered subsidiary operator \eqref{rsq}
reproduces the exact $U=0$ results (at zero temperature) {\evi for any
value of $N$}, thus suppressing the fluctuation corrections to the
free energy and to dynamical quantities {\evi exactly} when $U=0$.
\cite{unpub}

Although the formal results obtained in this paper are valid even at
finite temperature, numerical results have been restricted to the
zero-temperature limit throughout. Extension to finite temperature is
then in order. Besides, one may also consider multi-band Hubbard-type
Hamiltonians (like the Emery model for a $CuO_2$ layer), for which
using our ``effective'' rule to obtain the additional terms in the
free energy should be straightforward.

The numerical results presented in this paper have primarily concerned
the two-site model, although results have also been presented for the
infinite lattice when $U=0$. Restriction to $U=0$ is connected with
the paramagnetic $\evi ansatz$ we have considered throughout, since
the matrix of Gaussian fluctuations develops instabilities for $U=0^+$
and $n=1$. Consideration of magnetic phases is thus necessary, at
least near ``half filling''. We mention in this context that the KR
four-slave-boson method, which is not manifestly
spin-rotation-invariant, might not properly account for transverse
magnetic fluctuations at the order $1/N$. To this end, it would be
necessary to include additional ``angular'' variables associated with
 transverse fluctuations. This could be achieved by the
spin-rotation-invariant method of Ref.\cite{wolfle} where two
additional ``s'' bosons are introduced at the outset. 
 Resorting to the method of Ref.\cite{wolfle}, in turn, requires 
 one to assess preliminarly the effects of the proper handling
of the time \CL of the functional integral at the order $1/N$ that
have  been discussed in the present paper. Work along these lines is in
progress. 

\section*{ Acknowledgments }

We are indebted to C. Castellani and R. Raimondi for discussions.
One of us (E. A.) would also like to thank P. Fulde for his support
and interest in this work.

\appendix

\section{ $1/N$ Expansion with the Four-Slave-Boson Method }
\label{app:a} 
In this Appendix we show how a suitable $1/N$ expansion can be
constructed for the four-slave-boson problem. The need to organize the
theory by means of a $1/N$ expansion originates from the lack of an
intrinsic small parameter, in powers of which a more conventional
perturbation theory could be defined. In addition, the introduction of
the parameter $1/N$ enables one to establish a meaningful comparison
of the results obtained via different representations of the
integration variables in the functional integral (i.e., by using
either the ``radial'' or the ``Cartesian'' gauges).

As for the one-slave-boson problem, \cite{noiprep} 
also in the present context the
integer $2N$ represents the number of spin components of the fermionic
field coupled to the slave bosons. Also in this case, it turns out
that the \MF solution becomes exact in the $N\to \infty$ limit and
that the leading $1/N$ corrections to the free energy are associated
with the Gaussian fluctuations. For physical quantities different from
the free energy, however, the
leading $1/N$ corrections require also the inclusion of higher-order
fluctuations (which is equivalent to a $1/N$ shift of the mean-field
parameters). 

Since the physical case of interest corresponds to the value $N=1$,
one may wonder whether including only the $1/N$ corrections to the
\MF solution might be sufficient to obtain in practice sensible
results when $N=1$. In this respect, the numerical results presented
in Section \ref{s:nume} for a simple model system give us confidence
that extrapolating the $1/N$ results down to $N=1$ can work rather
accurately for practical purposes (provided the form of the subsidiary
operator is chosen appropriately).\cite{comparison} 

A meaningful $1/N$ expansion for the four-slave-boson problem can be
introduced in the following way:\cite{similar}
\begin{itemize}
\item[(i)]
 To begin with, the fermionic spin label (which in the
original action \eqref{sm}-\eqref{wfbim} was restricted to the two
values $\pm1$) is formally extended to run over the $2N$ values $S=\g
n$ with $\g=\pm 1$ and $n=1,2,\cdots,N$.
\item[(ii)] 
In order not to proliferate the number of slave bosons in the action 
\eqref{sm}-\eqref{wfbim} (thus keeping consistently the number of \MF
equations independent of $N$), we retain only two ``magnetic'' bosons 
$s_{\g=+1}$ and $s_{\g=-1}$, where now $s_{\g=+1}$ is associated with
 fermions with positive ($S>0$) spin projection and 
$s_{\g=-1}$ with  fermions with negative ($S<0$) spin
projection. This corresponds to generalizing the constraints
\eqref{constraints} as follows:
\beqn
 &&
\n di + \sum_{\g=\pm 1} \n s{i,\g} + \n ei =N \;,   
\label{constraintsaN} 
\\
 && \sum_{S=\g}^{\g N} \n f{i,S} - \n s{i,\g} - \n di =0 \quad\quad
(\g=\pm1)\; . 
\label{constraintsbN} 
\eeqn
In this way, the ``spin'' label $\g$ of the bosonic variables
$\lam^b\is$ and $z\is$ in Eqs. \eqref{wfim}-\eqref{wfbim} is
maintained. Note that the replacement $1\to N$ on the right-hand-side of
Eq. \eqref{constraintsaN} is necessary since the mean value of the
fermionic term in Eq. \eqref{constraintsbN} is of order $N$. By the
same token, the unity in the square roots in Eq. \eqref{rsq} (and
consequently in its ``linearized'' version discussed in Section
\ref{s:nume}) has also to be replaced by $N$.
\item[(iii)] 
To keep the \MF bosonic parameters independent of $N$, we rescale the
bosonic fields by letting
\beq
e_{i,m} \longrightarrow \sqrt{N} e_{i,m} \;,
\label{resc} 
\eeq
and similarly for $s_{i,\g,m}$ ($\g=\pm1$) and $d_{i,m}$.
\item[(iv)] 
Requiring the kinetic term \eqref{wfbim} to be homogeneous 
in $N$ of the same order as the other two terms 
\eqref{wfim} and \eqref{wbim} of the action, 
leads to suitably rescaling the hopping integral $t$
 depending on the
choice of the subsidiary factor $R\is$ in Eq. \eqref{zism}. In
particular, $t\to t/N^2$ when $R\is=1$; $t\to t$ with the KR choice
\eqref{rkrm}; $t\to t/N^6$ with the ``linearized'' choice corresponding
to \eqref{rlin}.
\end{itemize}

With the above prescriptions, the action $S_M$ is generalized as
follows:
\beq
S^{(N)}_M = \sum_{m=0}^{M-1}\sum_{n=1}^{N}\sum_{\g=\pm1}\sum_{<i,j>}
 \bar f_{i,\g n,m} {\cal G}_{i,m;j,m-1}[\und b;\g]  f_{j,\g n,m-1}
+ N {\cal B}[\und b]
\label{sgen} 
\eeq
where $<i,j>$ limits the lattice sum to nearest-neighbor sites. $\cal
G$ and $\cal B$ in Eq. \eqref{sgen} are functionals of the bosonic
fields $\und b$ (including now the Lagrange multipliers) which, by our
assumptions, are {\evi independent of $N$}.
[For the non normal-ordering prescription for $R\is$, however, ${\cal
G}$ is itself written as a power series in $1/N$ - see subsection
\ref{s:nno} and footnote \protect{\onlinecite{action}}].
 When $N=1$ the action
\eqref{sgen} consistently reduces to the form \eqref{sm}-\eqref{wfbim}
 considered in the text.

It is important to emphasize that the $N$-component generalization of
the slave-boson Hamiltonian (which has led to the action \eqref{sgen})
is equivalent to the $N$-component generalization of the {\evi
original} fermionic Hamiltonian \eqref{ham} {\evi only for $N=1$}. The
resulting $1/N$ expansion acquires thus physical meaning only when $N$
is eventually set equal to $1$. However, this should not be considered
to be an inconvenience of the present approach since the $N$-component
slave-boson Hamiltonian has not been conceived to study a physical
system with truly large spin; rather, the resulting $1/N$ expansion
must be considered as a purely formal tool to control the
approximations systematically.

The lack of correspondence of the two (purely fermionic and
fermionic-bosonic) Hamiltonian problems when $N>1$ results from (i)
the constraints \eqref{constraintsaN} and \eqref{constraintsbN} no
longer providing (in general) a one-to-one correspondence between the
Fock spaces on which the two Hamiltonians respectively act, and (ii)
the matrix elements of a given operator being different even for pairs
of corresponding states. As an example, let's consider the case $N=2$.
In this case, to the original configuration 
$\tilde f^{\dag}_{i,+3/2} \tilde f^{\dag}_{i,-1/2}|0>$,
 for instance, the constraints \eqref{constraintsaN} and
\eqref{constraintsbN} associate the two distinct configurations 
$d_i^{\dag} e_i^{\dag}  f^{\dag}_{i,+3/2}  f^{\dag}_{i,-1/2}|0>$ and
$s_{i,+1}^{\dag} s_{i,-1}^{\dag}  f^{\dag}_{i,+3/2}
f^{\dag}_{i,-1/2}|0>$.
On the other hand, to the original configuration 
$\tilde f^{\dag}_{i,+3/2} \tilde f^{\dag}_{i,+1/2}|0>$
there corresponds the only configuration
$(2)^{-1/2} (s_{i,+1}^{\dag})^2
f^{\dag}_{i,+3/2}f^{\dag}_{i,+1/2}|0>$.
[Note that the mapping between the two Fock spaces (i.e., with and
without slave bosons) preserves the fermionic configurations (specified
by the operators $ f^{\dag}$ and $\tilde f^{\dag}$, in the order) and
adjusts the bosonic configurations to satisfy the constraints]. For
the latter state the average value of the interaction term $U\n di$
vanishes, while for the corresponding original configuration the
average value of the original interaction term $U \sum_{S>S'} 
 \tilde f^{\dag}_{i,S}  \tilde f^{\dag}_{i,S'}  \tilde
f_{i,S'} \tilde f_{i,S}$ 
equals $U$ (with $S$ and $S'$ taking the $2N$ values
$-N,\cdots,-1,+1,\cdots,+N$). Obviously, when $N=1$ both the
one-to-one correspondence between states and the equality of the
matrix elements is fully preserved.

Returning to the action \eqref{sgen}, we carry out the associated $1/N$
expansion in the usual way by first integrating out the fermion
(Grassmann) fields. The action \eqref{sgen} is then replaced by
\beq
S^{(N)}_{eff} = -N(\tr\log {\cal G}[\und b;\g=+1] + \tr\log{\cal
G}[\und b;\g=-1]) + N{\cal B}[\und b] \equiv N S[\und b]
\label{sgeff} 
\eeq
where the trace is taken over the indices $(i,j,m)$ of Eq.
\eqref{sgen} and $S[\und b]$ is independent of $N$. Note that 
{\evi $S^{(N)}_{eff}$ depends on $N$ only through the explicit factor
$N$}
on the right-hand side of Eq. \eqref{sgeff}, as a consequence of the
assumptions (i)-(iv) spelled out above. 
[Apart from the non normal-ordering prescription for $R\is$, for which
$S[\und b]$ of Eq.\eqref{sgeff} is itself written as a power series in
$1/N$ - see footnote   \protect{\onlinecite{action}}].   
By this remark, the following
procedure is completely analogous to the standard method for the
one-slave-boson case (cf. Ref. \cite{noiprep}).

We then expand schematically $S[\und b]$ in Eq. \eqref{sgeff} in
powers of the deviation $\tilde{\und b}$ of the bosonic field $\und b$
from its \MF value $\und b_0$ as follows:
\beqn
S[\und b] && =  S[\und b_0] + 
\sum_{r_1} \Gam^{(1)}_{r_1}[\und b_0] \tilde{b}_{r_1} 
+ 
\sum_{r_1r_2} \Gam^{(2)}_{r_1r_2}[\und b_0] \tilde{b}_{r_1} \tilde{b}_{r_2} 
\nonum \\&&
+
\sum_{r_1r_2r_3} \Gam^{(3)}_{r_1r_2r_3}[\und b_0] \tilde{b}_{r_1} \tilde{b}_{r_2} \tilde{b}_{r_3} 
\nonum \\&&
+
\sum_{r_1r_2r_3r_4} \Gam^{(4)}_{r_1r_2r_3r_4}[\und b_0]
\tilde{b}_{r_1} \tilde{b}_{r_2} \tilde{b}_{r_3}  \tilde{b}_{r_4} 
+ \cdots \;,
\label{exps} 
\eeqn
where
\beq
\Gam^{(k)}_{r_1\cdots r_k}[\und b_0]=
 \frac 1{k!} \frac{\de^k S}{\de b_{r_1} \cdots \de
b_{r_k}}\Bigr|_{\und b_0}
\label{gamk} 
\eeq
and ${\und b_0}$ is chosen as usual according to the condition 
\beq
 \Gam^{(1)}_{r}[\und b_0] = \frac {\de S}{\de b_r}\Bigr|_{\und b_0} = 0
\label{conds} 
\eeq
for all values of $r$. For convenience, in the above expressions the
indices $r_1,\cdots,r_k$ label the whole set of bosonic variables
(that is, including also the static and homogeneous variables) to be
integrated over in the functional integral. For the partition function
we obtain accordingly
\beqn
\ZZ && = e^{-N S[\und b_0]} \int \prod_r (\sqrt{N} d \tilde b_r) \exp\left\{-N
\sum_{k=2}^{\infty} \sum_{r_1\cdots r_k} 
\Gam^{(k)}_{r_1\cdots r_k}[\und b_0]
\tilde b_{r_1} \cdots \tilde b_{r_k} \right\}
\nonum \\&&
=
e^{-N S[\und b_0]} \int \prod_r d x_r \exp\left\{-
\sum_{k=2}^{\infty} N^{\frac{2-k}2}
\sum_{r_1\cdots r_k} 
\Gam^{(k)}_{r_1\cdots r_k}[\und b_0]
x_{r_1} \cdots x_{r_k} \right\}
\label{zn} 
\eeqn
where the factor $\sqrt{N}$ in front of $d \tilde b_r$ originates from
the rescaling \eqref{resc} and we have defined $x=\sqrt{N} \tilde b$
(that affects the fluctuating part only). It is then evident from Eq.
\eqref{zn} that only the term with $k=2$ gives the leading $1/N$
correction to the \MF free energy, namely,
\beq
F_0 = N \frac{S[\und b_0]}{\bet} \;,
\label{a.10} 
\eeq
which justifies keeping only the Gaussian fluctuations to calculate
the $1/N$ corrections to the free energy.

Besides the free energy, let's consider a physical quantity (or
correlation function) which can be expressed as
\beqn
\phi &&= \frac{\int\prod_r d \tilde b_r \phi(\und b) e^{-N S[\und b]}}
      {\int\prod_r d \tilde b_r  e^{-N S[\und b]}}
\nonum \\&&
= \phi(\und b_0) + \frac1{\ZZ} \int\prod_r d x_r \biggl[
\frac1{\sqrt{N}} \sum_{r_1} \frac{\de \phi}{\de b_{r_1}}\Bigr|_{\und
b_0} x_{r_1} 
\nonum \\&&
+ \frac1{2 N} \sum_{r_1r_2} \frac{\de^2 \phi}{\de b_{r_1}\de
b_{r_2}}\Bigr|_{\und b_0} x_{r_1} x_{r_2}
+\cdots \biggr]  e^{-N S[\und b]} \; .
\label{phi} 
\eeqn
Since $ \phi(\und b_0)$ is of order $(1/N)^0$, evaluation of its $1/N$
corrections requires one to keep also the cubic terms in the expansion
\eqref{exps}. In particular, the $1/N$ corrections to $\phi$ derive
from (i) the quadratic term in the expansion \eqref{phi} (which is
itself proportional to $1/N$) integrated only with the Gaussian term
in the expansion \eqref{exps}, and (ii) the linear term in the
expansion \eqref{phi} (which is proportional to $1/\sqrt{N}$)
integrated by keeping also the cubic term in the expansion
\eqref{exps}. In the latter case, one needs to expand the exponential
of the cubic term in Eq. \eqref{exps} and keep only the leading term
in  $1/\sqrt{N}$ of this expansion. [The normalization factor $\ZZ$ of
Eq. \eqref{phi} should be consistently taken only up to the Gaussian
order]. Correction (i) to $\phi(\und b_0)$ is just the ordinary
Gaussian contribution, while correction (ii) is equivalent to
considering the $1/N$ shift of the mean-field parameters $\und b_0$ in 
$\phi(\und b_0)$. To show this, we rewrite correction (ii) in the
following way:
\beqn
&&
-\frac1N \sum_{r_1\cdots r_4} \Gam^{(3)}_{r_1r_2r_3}[\und b_0]
\frac{\de \phi}{\de b_{r_4}}\Bigr|_{\und b_0}
 \frac{
 \int \prod_r d x_r \, (x_{r_1}x_{r_2}x_{r_3}x_{r_4})
\exp\left\{-\sum_{r_5r_6}
\Gam^{(2)}_{r_5r_6}[\und b_0] x_{r_5}x_{r_6}
\right\}
}
{
 \int \prod_r d x_r \, 
\exp\left\{-\sum_{r_5r_6}
\Gam^{(2)}_{r_5r_6}[\und b_0] x_{r_5}x_{r_6}
\right\}
}
\nonum \\&&
= -\frac1N 
 \sum_{r_1\cdots r_4} \Gam^{(3)}_{r_1r_2r_3}[\und b_0]
\frac{\de \phi}{\de b_{r_4}}\Bigr|_{\und b_0}
\frac34  \left( \Gam^{(2)}[\und b_0]^{-1}\right)_{r_1r_2}
 \left( \Gam^{(2)}[\und b_0]^{-1}\right)_{r_3r_4}
\nonum \\&&
= -\frac1N 
 \sum_{r_1\cdots r_4} \left(\frac{\de}{\de b_{r_3}}
\Gam^{(2)}_{r_2r_1}[\und b]\right)_{\und b_0} 
\frac{\de \phi}{\de b_{r_4}}\Bigr|_{\und b_0}
\frac14  \left( \Gam^{(2)}[\und b_0]^{-1}\right)_{r_1r_2}
 \left( \Gam^{(2)}[\und b_0]^{-1}\right)_{r_3r_4}
\nonum \\&&
=
 -\frac1N 
 \sum_{r_3r_4} \left(\frac{\de}{\de b_{r_3}} \mez
\tr\ln
\Gam^{(2)}[\und b]\right)_{\und b_0} 
\frac{\de \phi}{\de b_{r_4}}\Bigr|_{\und b_0}
\frac12  \left( \Gam^{(2)}[\und b_0]^{-1}\right)_{r_3r_4}
\nonum \\&&
=
 -\frac1N 
 \sum_{r_3r_4} \left(\frac{\de}{\de b_{r_3}} 
\bet F_1[\und b]\right)_{\und b_0} 
\frac{\de \phi}{\de b_{r_4}}\Bigr|_{\und b_0}
\frac12  \left( \Gam^{(2)}[\und b_0]^{-1}\right)_{r_3r_4}
\;,
\label{1ncorr} 
\eeqn
where we have used the definition \eqref{gamk} and recalled that the
$1/N$ correction to the free energy is given by (cf. Eq.\eqref{f1})
\beq
F_1= \frac1{2\bet} \tr\ln\Gam^{(2)}[\und b_0] \;.
\label{f1bis} 
\eeq
In fact, owing to space and time translational invariance, the wave
vector and frequency labels $q(r)$ of the integration variables in
Eq. \eqref{1ncorr} are related by
\beq
q(r_1)+q(r_2)+q(r_3)=q(r_1)+q(r_2)=q(r_3)+q(r_4)=0 \;,
\eeq
which gives $q(r_3)=q(r_4)=0$. For these static and homogeneous
variables we can write
\beq
  \Gam^{(2)}[\und b_0]_{r_3r_4} = \frac1N \frac{\bet}2 
\frac{\de^2 F_0[\und b]}{\de  b_{r_3}\de b_{r_4}} \Bigr|_{\und b_0}
\label{d2f0} 
\eeq
and recognize in the expression 
\beq
 b_{r_4}^{\left(1\right)} = -\frac1N \frac{\bet}2 \sum_{r_3} 
  \left( \Gam^{\left(2\right)}[\und b_0]^{-1}\right)_{r_4r_3}
 \left(\frac{\de}{\de b_{r_3}} 
 F_1[\und b]\right)_{\und b_0} 
\label{1nshift} 
\eeq
the $1/N$ shift of the \MF parameter $ b_{r_4}^{(0)}$ [cf., e.g.,
Appendix C of Ref.\cite{noiprep}]. Expression \eqref{1ncorr} thus
reduces to $\sum_{r_4} (\de \phi/\de b_{r_4})_{\und b_0}
b_{r_4}^{(1)} $
and corresponds to shifting $\und b_0 \to \und b_0 + \und b^{(1)}$ in
the argument of $\phi(\und b_0)$, as anticipated 
above.\cite{nno}

Finally, we comment briefly on how the $1/N$ expansion discussed above
can be combined with the procedure to restore the spin-rotation
invariance in the results obtained via the KR four-slave-boson method.
To this end, we will essentially follow the procedure used in Appendix
B of Ref.\cite{noiincomm} for the physical case $N=1$, albeit with
some modifications.

The problems originating from using a slave-boson method which is not
manifestly spin-rotation invariant can be evidenced when considering,
for instance, spin correlation functions of the type
\beq
\tilde\chi_{\al,\al'}(i,j;\tau) = < T_{\tau}[ \tilde S_i^{(\al)}(\tau)
\tilde S_j^{(\al')}(0)] >
\label{ss} 
\eeq
($\al,\al'=x,y,z$), where $ T_{\tau}[\cdots]$ is the time-ordering
operator for imaginary time $\tau$ and the spin operators
\beq
 \tilde S_i^{(\al)} = \mez \sum_{\g,\g'=\pm1} \tilde f\is^{\dag}
{\bfmath \g}^{(\al)}_{\g,\g'} \tilde f_{i,\g'} 
\label{sia} 
\eeq
are defined in terms of the physical spin $\mez$ fermion operators ($
{\bfmath \g}^{(\al)}$ being a Pauli matrix). Violation of
spin-rotation invariance occurs in \eqref{ss} when considering the
slave-boson replacement $ \tilde f_{i,\g} \to  f_{i,\g} z\is$. For
instance, in the paramagnetic case the $(z,z)$ element of the tensor
\eqref{ss} differs from the $(x,x)$ and $(y,y)$ elements when the \MF
approximation for the slave bosons is considered.

In Eqs. \eqref{ss} and \eqref{sia} the spin label $\g$ refers to a
given quantization axis (say $\hat z$), which is common to all sites.
Spin-rotation invariance can be restored upon averaging over all
possible quantization axes, which are identified by the spherical
angle $\Omega=(\theta,\varphi)$ {\evi common} to all lattice sites. In
this respect, we follow Ref.\cite{noiincomm} and introduce the
rotated fermion operators
\beq
\tilde g_{i,\xi} = \sum_{\g=\pm1} [\R^{\dag}(\Omega)]_{\xi,\g}
\tilde f\is 
\label{rot} 
\eeq
where $\xi=\pm1$ and
\beq
\R(\Omega) = e^{-i(\varphi/2)\g_z} e^{-i(\theta/2)\g_y}
\label{rrot} 
\eeq
is the relevant rotation operator. Introducing at this point the
slave-boson mapping $ \tilde g_{i,\xi} \to  z_{i,\xi} g_{i,\xi} $
where now the bosonic operators refer to the new quantization axis
identified by $\Omega$, the spin operator \eqref{sia} becomes 
\beqn
 \tilde S_i^{(\al)} &&= T_{\al z}(\Omega) \mez \sum_{\xi,\xi'=\pm1}
g^{\dag}_{i,\xi} {\bfmath \g}^{(z)}_{\xi,\xi'} g_{i,\xi'}
\nonum \\&&
+  \sum_{\bet=x,y}  T_{\al \bet}(\Omega) \mez 
 \sum_{\xi,\xi'=\pm1}
 z^{\dag}_{i,\xi} g^{\dag}_{i,\xi} {\bfmath \g}^{(\bet)}_{\xi,\xi'}
 g_{i,\xi'}  z_{i,\xi'}
\label{siatr} 
\eeqn 
with $  T_{\al \bet}(\Omega)$ defined by
\beq
\R^{\dag}(\Omega) {\bfmath \g}^{(\al)} \R(\Omega) = \sum_{\bet} 
 T_{\al \bet}(\Omega)  {\bfmath \g}^{(\bet)} \;.
\eeq
Note that the pair of bosonic operators at the same site with equal
spin labels $(\xi=\xi')$ have been consistently replaced by unity in
Eq. \eqref{siatr}. Note also that, within the \MF approximation for
the slave bosons, the remaining bosonic operators in Eq. \eqref{siatr}
can be replaced by the spin-independent factor $z$ in the paramagnetic
phase we are considering.

To evaluate the average over $\Omega$, it is necessary to transform
back to the old $\hat z$ quantization axis, by using the
transformation
\beq
 g_{i,\xi}  = \sum_{\g=\pm1} [ \R^{\dag}(\Omega)]_{\xi, \g} f_{i,\g}
\eeq
for the pseudo fermion operators. The operator \eqref{siatr} then
reads
\beqn
 \tilde S_i^{(\al)} &&= (1-z^2)
T_{\al z}(\Omega) 
\sum_{\bet} \left(
\mez \sum_{\g,\g'=\pm1}
f^{\dag}_{i,\g} {\bfmath \g}^{(\bet)}_{\g,\g'} f_{i,\g'}
\right) T_{\bet z}(\Omega)
\nonum \\&&
+ z^2 \mez
 \sum_{\g,\g'=\pm1}
 f^{\dag}_{i,\g} {\bfmath \g}^{(\al)}_{\g,\g'}
 f_{i,\g'} \;.
\label{siatr1} 
\eeqn 
Extension to arbitrary $N\,(>1)$ is performed by adding an additional
label $n \; (=1,2,\cdots,N)$ to the fermionic operators and
considering the action in the form \eqref{sgeff} (which does not
depend on $\Omega$). Accordingly, $ \tilde S_i^{(\al)} $ of
Eq.\eqref{siatr1} is replaced by
\beq
 \tilde S_i^{(\al)}(N) = (1-z^2)
T_{\al z}(\Omega) 
\sum_{\bet}
S_i^{(\bet)}(N)
T_{\bet z}(\Omega)
+ z^2 S_i^{(\al)}(N)
\label{tsian}
\eeq
with the notation
\beq
S_i^{(\al)}(N)= \frac1{2N} \sum_{n=1}^N  \sum_{\g,\g'=\pm1}
 f^{\dag}_{i,\g,n} {\bfmath \g}^{(\al)}_{\g,\g'}
 f_{i,\g',n} \;.
\label{sian}
\eeq
Entering the expression \eqref{tsian} into the correlation function
\eqref{ss} in the place of $S_i^{(\al)}$ and averaging over the
spherical angle $\Omega$, we obtain for the averaged spin correlation
function:
\beqn
&& \int \frac{d \Omega}{4\pi} 
< T_{\tau}[ \tilde S_i^{(\al)}(N;\tau)
\tilde S_j^{(\al')}(N;0)] >
\nonum \\&&
= (1-z^2)^2 \sum_{\bet \bet'}
\chi^{(N)}_{\bet \bet'}(i,j;\tau) 
 \int \frac{d \Omega}{4\pi} 
T_{\al z}(\Omega) T_{\al' z}(\Omega) T_{\bet z}(\Omega) T_{\bet' z}(\Omega) 
\nonum \\&&
+ (1-z^2)z^2 \sum_{\bet} 
\chi^{(N)}_{\bet \al'}(i,j;\tau) 
 \int \frac{d \Omega}{4\pi} 
T_{\al z}(\Omega) T_{\bet z}(\Omega)
\nonum \\&&
+  z^2(1-z^2) \sum_{\bet'} 
\chi^{(N)}_{\al \bet'}(i,j;\tau) 
 \int \frac{d \Omega}{4\pi} 
T_{\al' z}(\Omega) T_{\bet' z}(\Omega)
\nonum \\&&
+ z^4 \chi^{(N)}_{\al \al'}(i,j;\tau) 
\label{avss} 
\eeqn
where
\beq
\chi^{(N)}_{\al \al'}(i,j;\tau) \equiv
< T_{\tau}[  S_i^{(\al)}(N;\tau)
S_j^{(\al')}(N;0)] >
\label{xnaa} 
\eeq
is a diagonal tensor with equal diagonal elements (at the \MF level
for the slave bosons).
Equation \eqref{avss} then reduces to
\beqn
&&
\int \frac{d \Omega}{4\pi} 
< T_{\tau}[ \tilde S_i^{(\al)}(N;\tau)
\tilde S_j^{(\al')}(N;0)] >
\nonum \\&&
= \del_{\al,\al'} \Biggl\{
 (1-z^2)^2 \sum_{\bet}
\chi^{(N)}_{\bet \bet}(i,j;\tau) 
 \int \frac{d \Omega}{4\pi} 
T_{\al z}(\Omega)^2 T_{\bet z}(\Omega)^2
\nonum \\&&
+ \left[\frac23 z^2(1-z^2) + z^4\right]
\chi^{(N)}_{\al \al}(i,j;\tau) 
\Biggr\}
\nonum \\&&
=\del_{\al,\al'} \frac{ \left[(1-z^2)^2 + 2 z^2(1-z^2) + 3 z^4\right]}3 
\chi^{(N)}_{\al \al}(i,j;\tau)
= 
\del_{\al,\al'} \left(\frac{1+2z^4}3\right)
\chi^{(N)}_{\al \al}(i,j;\tau) \;.
 \label{avss1} 
\eeqn
In this way spin-rotation invariance for the spin correlation
functions has been restored at arbitrary values of $N$.\cite{averaging}

\section{ Summary of the Relevant Mathematical Expressions Involving
Matsubara Frequency sums with a Discretized Time Mesh}
\label{app:b} 
The Fermi function and the fermionic polarization function introduced
in Section \ref{s:func} for the case of a discretized (imaginary) time
mesh are defined, respectively, as follows:
\beqn
&& f_M(E)=-\frac{1}{\beta}\displaystyle{\sum_{s=0}^{M-1}}
\frac{1}{E-(1-e^{-i\omega_s\delta})/{\delta}} \;,
\label{b.1}
\\&&
\Pi_M( E, E'; \nu )=\Pi_M( E', E; -\nu )
\nonum \\&&
= -\displaystyle\frac{1}{\beta}\displaystyle{\sum_{s =0}^{M-1}}
\frac{1}{E-\left(1-e^{-i\omega_s\delta}\right)/{\delta}} \;
 \displaystyle\frac{1}{ E' -\left(1-e^{-i(\omega_s-\omega_{\nu})
\delta}\right)/{\delta}} 
\label{b.2} 
\eeqn
with $\omega_s= 2\pi (s+1/2)/\bet$ and $\omega_{\nu}= 2 \pi \nu/\bet$
($\nu=0,1,\cdots,M-1$). The fermionic frequency sums in Eqs.
\eqref{b.1} and \eqref{b.2} can be evaluated according to the method
developed in Appendix B of Ref.\cite{noiprep}. For the reader's
convenience we report here the results:
\beqn
f_M(E) && =\displaystyle\frac{1}{1-(\beta
E)/{M}}\;\displaystyle\frac{1}{1+\left(1-(\beta E)/{M}\right)^{-M}}
\nonum
\\&&
\undersymbolsub{M\to\infty}{\longrightarrow} \displaystyle{1\over{1+e^{\beta
E}}}=f_F(E) \;,
\label{b.3}
\eeqn
\beqn
\Pi_M( E,E'; \nu ) &&=\displaystyle{{\beta}\over{M}}\;\displaystyle{{\eind
f_M(E')-f_M( E)}\over {\eind \beta E/M-\beta E'
/M+1-\eind}}
\nonum \\&&
\undersymbolsub{M\to\infty}{\longrightarrow}
\displaystyle{{f_F(E')-f_F(E)}\over{E- E' - i\omega_{\nu}}} \;\;,
\label{b.4}
\eeqn
where the limiting expressions for $M\to \infty$ recover the standard
finite-temperature Matsubara results.

Similarly, extraction of the ``\CFI'' to a given bosonic frequency sum
can be obtained according to a theorem proved in Appendix B of
Ref.\cite{noiprep}. Since that procedure constitutes an essential step
to obtain the novel results of Section \ref{s:func} (and of Appendix
\ref{app:c} below), we summarize it here for convenience. According to
this theorem, a bosonic frequency sum of the type
\beq
S_M = {1\over\beta} \displaystyle{\sum^{(M-1)/2}_{\nu=-{(M-1)/2}}} h\,(\eind,
\delta)
\label{b.5}
\eeq
can be evaluated as the sum of its ``\CL'' $S_c$ plus a ``\CFI''
$g_0$:
\beq
S_M = S_c +g_0 \,.
\label{b.6}
\eeq
In this expression
\beq
S_c = {1\over\beta} \displaystyle{\sum^{+\infty}_{\nu=-\infty}} h_c (\omega_\nu)
\label{b.7}
\eeq
is an ordinary Matsubara bosonic frequency sum, where
\beq
h_c(z) =\lim_{\delta\rightarrow 0} h\,(e^{iz\delta},\delta)
\label{b.8}
\eeq
is the continuum limit of the function $h$ of  Eq.\eqref{b.5},
 and $g_0$ is the
constant term of the function
\beq
g(\zeta ) = \displaystyle{\sum^{-1}_{s=-S}} g_s\zeta^s + g_0 +
\displaystyle{\sum^S_{s=1}} g_s\zeta^s 
\label{b.9}
\eeq
(with $\zeta = \exp\,\{i z\delta\}$)
 which enters the expansion
\beq
h\,(\zeta, \delta) = \delta g\,(\zeta) + \ord(\delta^2)
\;.
\label{b.10}
\eeq
The result \eqref{b.6} holds under some restrictions on the function
$h$, which have been spelled out in detail in Appendix B of
Ref.\cite{noiprep}. These restrictions are definitely satisfied by the
functions of interest considered in Section \ref{s:func} (and in
Appendix \ref{app:c}). It is important to recall here that the
function $h\,( e^{i \omega_{\nu}\delta}, \delta)$ in Eq.
\eqref{b.5} has to be taken to be  even in $\omnu$ (at last, by
suitably symmetrizing it), so that the associated \CL function
$h_c(\omnu)$ vanishes at least like $|\omnu|^{-2}$ when
$|\omnu| \to \infty$.

We conclude this Appendix by remarking that the result \eqref{b.3}
can also be used to evaluate the contribution from the level $E$ to
the fermionic free energy, namely,
\beq
F_M(E) = -\frac1{\bet} \sum_{s=0}^{M-1}
\ln \left( 1 - e^{i \oms \delta} (1-\delta E) \right) \,.
\label{b.11}
\eeq
In fact, differentiating both sides of \eqref{b.11} with respect to
$E$ we obtain
\beq
\frac{\de F_M(E)}{\de E} =  f_M(E)
\label{b.12}
\eeq
according to the definition \eqref{b.1}.
On the other hand, the result \eqref{b.3} can be rewritten in the form
\beq
f_M(E)= - {1\over\bet} {d  \over d \, E }
  \ln\left(1+(1-\delta E)^M \right)
\label{b.13}
\eeq
with $\delta=\bet/M$. Comparison of \eqref{b.12} with \eqref{b.13}
defines $F_M(E)$ up to a constant, which can be determined by noting
from \eqref{b.11} that $F_M(1/\delta)=0$ for any $M$. We thus obtain:
\beq
F_M(E)=  -\frac1{\bet}\ln\left(1+(1-\delta E)^M \right)
\label{b.14}
\eeq
which recovers the known result
\beq
F_{\infty}(E) =  -\frac1{\bet}\ln\left( 1+e^{- \bet E} \right)
\label{b.15}
\eeq
in the continuum ($M\to \infty$) limit. This result has been used to
derive Eq.\eqref{lims0} of the text.

\newcommand{\GGam}{{\bfmath\Gamma} }
\newcommand{\AAA}{{\bfmath A} }
\newcommand{\BBB}{{\bfmath B} }
\newcommand{\UU}{{\bfmath 1} }

\section{ Obtaining the Contribution from Infinity in the Cartesian Gauge }
\label{app:c} 
The ``radial'' gauge was used in the text to represent the bosonic
fields in the \FI. This choice has enabled us to deal with expressions
which are free from infrared divergences in the continuum
($\delta\to0$) limit. The price we had to pay was that the calculation
of the ``\CFI'' to the free energy turned out to be rather involved in
the ``radial'' gauge (cf. Section \ref{s:func}). On the other hand, we
expect that the ``\CFI'' can be obtained equivalently in any other gauge at
a given order in $1/N$, provided a separate $1/N$ expansion holds in both
discretized and continuum versions of the \FI~.

We shall show in this Appendix that the ``\CFI'' to the free energy
\eqref{f1d}-\eqref{z0} taken at self-consistency can alternatively be
obtained at the same $1/N$ order in the ``Cartesian'' gauge, when one
takes the real and imaginary parts of the complex bosonic fields as
integration variables.\cite{polar} It will turn out that the derivation
of the ``\CFI'' in the ``Cartesian'' gauge is considerably simpler
than in the ``radial'' gauge, a result which contrasts the situation
for the \CL expressions that are plagued by infrared divergences in
the ``Cartesian'' gauge. This result gives support to the validity of
the $1/N$ expansion discussed in Appendix \ref{app:a}, and suggests at
the same time that, at least at the order $1/N$ we are explicitly
considering, it would be convenient to evaluate a given physical
quantity by representing alternatively its \CL in the ``radial'' gauge
and the associated ``\CFI'' in the ``Cartesian'' gauge. 
To this end, besides recovering the ``\CFI'' \eqref{f1d}-\eqref{z0}
to the free energy using the ``Cartesian'' gauge, in the following we
shall also indicate how the calculation can be extended to other
physical quantities (such as correlation functions) by working out
explicitly a simple example.

We begin by rewriting the discretized action \eqref{sm} in the compact
form:
\beqn
S_M 
= \sum_{m=0}^{M-1} \sum_i && \Bigl\{ \sum_{\g} \bar f\ism [ f\ism
-(1-\delta q\is) f\ismu] 
\nonum \\ &&
+ \sum_{\ib=1}^4 \bbc{\ib}\im [ \bb{\ib}\im - (1-\del h_i^{(\ib)})
\bb{\ib}\imu ]
\nonum \\&&
+ \del t \sum_{\nene,\g} \bar f\jsm z_{i+\nene,\g}^{\ast}(m-1,m) 
z_{i,\g}(m,m-1) f\ismu - \del \lam_i^a \Bigr\}
\label{c.1}
\eeqn
where the site-dependent coefficients $q\is$ and $   h_i^{(\ib)}$
can be readily read off Eqs. \eqref{wfim} and \eqref{wbim}. In the
``Cartesian'' gauge we set:\cite{polar}
\beq
\bb{\ib}\im = \bb{\ib}_0 + \uti b^{(\ib)}\im  \quad,\quad
\bbc{\ib}\im = \bb{\ib}_0 + \uti b^{(\ib)\ast}\im
\label{c.2}
\eeq
$(\ib=1,\cdots,4)$, and expand the action \eqref{c.1} up to quadratic
order in the ``small'' quantities $ \uti b^{(\ib)}\im $ . Introducing
the (normalized) Fourier transform
\beq
 \uti b^{(\ib)}\im = \frac1{\sqrt{M\NN}} \sum_{\v q}^{BZ} \sum_{\nu=1}^{M-1}
 e^{i(\v q \cdot \v R_i - \omnu \tau_m)} \bb{\ib}\qunu
\label{c.3}
\eeq
[cf. expression \eqref{bbim} of the text] and a similar transformation
for the Grassmann fields, the relevant part of the action \eqref{c.1}
becomes:
\beqn
S_M && =\bet \NN S^{(0)}
+  \sum_{\v k}^{BZ} \sum_{\g} \sum_{s=0}^{M-1}
\bar f_{\g}\kaom [1-\eiod(1-\del \E_{\g}(\v k))] f_{\g}\kaom
\nonum \\&&
+  \sum_{\v q}^{BZ} \sum_{\ib=1}^4 \sum_{\nu=1}^{M-1}
\bb{\ib}\qunu^{\ast} [ 1-\eind (1-\del h_0^{(\ib)})] \bb{\ib}\qunu
\nonum \\&&
+ \del \frac1{\sqrt{M \NN}}  \sum_{\v k, \v q}^{BZ}  \sum_{\g}
\sum_{\ib=1}^4  \sum_{s=0}^{M-1} \sum_{\nu=1}^{M-1}
\eiod 
\nonum \\&&
\times \Bigl\{
\bar f_{\g}\kaom \E_1(\v q;\v k,\g |\ib) \bb{\ib}\qunu 
 f_{\g}\kqon
\nonum \\&&
+ \bar f_{\g}\kqon \bar  \E_1(\v q;\v k,\g |\ib)
\bb{\ib}\qunu^{\ast} f_{\g}\kaom \Bigr\}
\nonum \\&&
+ \del   \frac1{M \NN} 
 \sum_{\v k, \v q}^{BZ}  \sum_{\g}
\sum_{\ib,\ib'=1}^4  \sum_{s=0}^{M-1} \sum_{\nu=1}^{M-1}
\bar f_{\g}\kaom \eiod  f_{\g}\kaom 
\nonum \\&&
\times
\Bigl\{ 
\E_{11}(\v q;\v k,\g|\ib,\ib') \eind \bb{\ib}\qunu^{\ast}
\bb{\ib'}\qunu 
\nonum \\&&
+ \E_{12}(\v q;\v k,\g|\ib,\ib')  \bb{\ib}\qunu^{\ast}
\bb{\ib'}\mqunu^{\ast} 
\nonum \\&&
+ \E_{21}(\v q;\v k,\g|\ib,\ib')  \bb{\ib}\qunu
\bb{\ib'}\mqunu
\Bigr\} \;.
\label{c.4}
\eeqn
In this expression, $S^{(0)}$ is given by Eq.\eqref{s0}, the
single-particle eigenvalue $\E_{\g}(\v k)$ is given by
Eq.\eqref{epsk}, $h_0^{(\ib)}$ are \MF values (which are
spin-independent in the {\evi paramagnetic} phase we are 
considering),\cite{zerofreq} 
$(\E_1,\bar\E_1)$ contain the first
derivatives of the bosonic hopping factor while $(\E_{11},\E_{12},\E_{21})$
contain the second derivatives. Note that, similarly to the expression
\eqref{s2f} in the ``radial'' gauge, in the last term on the
right-hand side of Eq.\eqref{c.4} only pairs of fermionic variables
with the same $\v k$ and $\oms$ have been retained (owing to space and
time translational invariance).

 After integration of the Grassmann variables, the
terms in Eq.\eqref{c.4} containing  $(\E_1,\bar\E_1)$ will be
$\ord(\del^2)$ 
due to the fact that the function $\Pi_M$ given by Eq.\eqref{b.4} is of
$\ord(\delta)$
and  can thus  be ignored. By the same
token, we can avoid reporting the expressions for $\E_{12}$ and $\E_{21}$
since they contribute at $\ord(\del^2)$. In fact, to the 
 $\ord(\del)$ we are concerned here we need  only the symmetry
properties
\beqn
&& \E_{11}(\v q;\v k,\g|\ib,\ib') = \E_{11}(\v q;\v k,\g|\ib',\ib)
\nonum \\&&
 \E_{21}(\v q;\v k,\g|\ib,\ib') = \E_{12}(\v q;\v k,\g|\ib,\ib')
\nonum \\&&
 \E_{21}(-\v q;\v k,\g|\ib,\ib') = \E_{21}(\v q;\v k,\g|\ib',\ib)
\label{c.5}
\eeqn
as well as the explicit expression of $\E_{11}$ for $\ib=\ib'$:
\beqn
 \E_{11}(\v q;\v k,\g|\ib,\ib) && = 
  t\Biggl\{ \gamk z_0 \left(
 \frac{\de^2 z^{\ast}_{j,\g}(m-1,m)}{\de \bbc{\ib}_{j,m}\de
\bb{\ib}_{j,m-1}}\Bigr|_{\und b_0} +
 \frac{\de^2 z_{j,\g}(m,m-1)}{\de \bbc{\ib}_{j,m}\de
\bb{\ib}_{j,m-1}}\Bigr|_{\und b_0}
\right)
\nonum\\&&
+\gamkpq 
 \frac{\de z^{\ast}_{j,\g}(m-1,m)}{\de
\bbc{\ib}_{j,m}}\Bigr|_{\und b_0}
 \frac{\de z_{j,\g}(m,m-1)}{\de \bb{\ib}_{j,m-1}}\Bigr|_{\und b_0}
\nonum\\&&
+\gamkq 
 \frac{\de z^{\ast}_{j,\g}(m-1,m)}{\de
\bb{\ib}_{j,m-1}}\Bigr|_{\und b_0}
 \frac{\de z_{j,\g}(m,m-1)}{\de \bbc{\ib}_{j,m}}\Bigr|_{\und b_0}
\Biggr\}
\label{c.6}
\eeqn
where $z_0$ and $\gamk$ are given by Eqs.\eqref{z0} and \eqref{gam} of
the text, respectively, and $\und b_0$ stands for the set
$\{\bb{\ib}_0;\ib=1,\cdots,4\}$ of Eq.\eqref{c.2}. These are, in
essence, the simplifying features which make it easier obtaining the
``\CFI'' to the free energy in the ``Cartesian'' gauge.

Integrating out the Grassmann variables, expanding the resulting
logarithm up to quadratic order in the bosonic fields, and keeping
only terms up to $\ord(\delta)$ (while preserving, however, the full
exponential $\exp\{i \omnu \del\}$ - cf.
 Section \ref{s:func}), we obtain the effective action:
\beqn
S_{eff} &&= \bet \NN F_0 
+ \del \sum_{\v q}^{BZ} \sum_{\nu=1}^{M-1}
\sum_{\ib,\ib'=1}^4 \Biggl\{
\bb{\ib}\qunu^{\ast} \biggl[ \del_{\ib,\ib'} \frac{1-\eind}{\del} 
\nonum\\&&
+ \eind \left(  \del_{\ib,\ib'} h_0^{(\ib)} + \frac1{\NN} 
\sum_{\v k}^{BZ} \sum_{\g} f_F(\E_{\g}(\v k)) \E_{11}(\v q;\v k,\g|\ib,\ib')
\right) \biggr] \bb{\ib'}\qunu
\nonum\\&&
+ \bb{\ib}\qunu^{\ast}
\left[ \frac1{\NN} 
\sum_{\v k}^{BZ} \sum_{\g} f_F(\E_{\g}(\v k)) \E_{12}(\v q;\v k,\g|\ib,\ib')
 \right] \bb{\ib'}\mqunu^{\ast}
\nonum\\&&
+ \bb{\ib}\qunu
\left[ \frac1{\NN} 
\sum_{\v k}^{BZ} \sum_{\g} f_F(\E_{\g}(\v k)) \E_{21}(\v q;\v k,\g|\ib,\ib')
 \right] \bb{\ib'}\mqunu \Biggr\}
\label{c.7}
\eeqn
with $F_0$ given by Eq.\eqref{f0} of the text. It is convenient to
introduce at this point the column vector
\beq
\und{a}\qunu =  \left(
\begin{array}{l}
\bb1\qunu \\
\vdots
\\
\bb4\qunu
\\
\bb1\mqunu^{\ast}
\\
\vdots
\\
\bb4\mqunu^{\ast}
\end{array}
\right)
\label{c.8}
\eeq
and its adjoint 
$\und a^{\dag}\qunu =$ 
$ (\bb1\qunu^{\ast},$
 $\cdots, $
$ \bb4\qunu^{\ast},$
 $  \bb1\mqunu, $ 
$ \cdots,$ 
$ \bb4\mqunu)$,
which allow to rewrite Eq. \eqref{c.7} in the compact form
\beq
S_{eff} = \bet \NN F_0 +\mez \sum_{\v q}^{BZ} 
\sum_{\nu=-\frac{M-1}2}^{\frac{M-1}2\;\prime} \und a^{\dag}\qunu \GGam\qunu
\und a\qunu
\label{c.9}
\eeq
(excluding the $\nu=0$ term).
The fluctuation matrix in \eqref{c.9} can be conveniently decomposed
as follow:
\beq
\GGam\qunu = \GGam_0(\omnu) + \del \GGam_1\qunu
\label{c.10}
\eeq
where
\beq
\GGam_0(\omnu)=\left(\begin{array}{c c}
(1-\eind) \UU & 0 \\
0 & (1-\emind) \UU 
\end{array}
\right)
\label{c.11}
\eeq
($\UU$ being the $4 \times 4$ unit matrix) and $\GGam_1$ has the block form
\beq
\GGam_1\qunu=\left(\begin{array}{c c}
\AAA\qunu & \BBB\qunu \\
\BBB\mqunu & \AAA\mqunu
\end{array}
\right) \;.
\label{c.12}
\eeq
In this expression:
\beq
\AAA\qunu_{\ib\ib'}=\eind \left( \del_{\ib,\ib'} h_0^{(\ib)} + \frac1{\NN}
\sum_{\v k}^{BZ} \sum_{\g} f_F(\E_{\g}(\v k)) \E_{11} (\v q;\v
k,\g|\ib,\ib')\right)
\label{c.13}
\eeq
and
\beq
\BBB\qunu_{\ib\ib'}= \frac2{\NN}
\sum_{\v k}^{BZ} \sum_{\g} f_F(\E_{\g}(\v k)) \E_{12} (\v q;\v
k,\g|\ib,\ib')
\label{c.14}
\eeq
are $4 \times 4$ matrices. Note that the block structure in \eqref{c.12} has
been obtained by exploiting the symmetry properties \eqref{c.5}.

The ``\CFI'' to the (site) free energy can now be obtained from the
effective action \eqref{c.9} in the following way. Performing the
Gaussian integration over the bosonic variables $\{\und a\qunu\}$
according to the method of Appendix \ref{app:e} and expanding the
result up to $\ord(\del)$, yields
\beqn
&&
\frac1{\NN} \sum_{\v q}^{BZ} \frac1{2\bet}
 \sum_{\nu=-\frac{M-1}2}^{\frac{M-1}2\;\prime} 
\tr\ln(\GGam_0(\omnu)+\del \GGam_1\qunu)
\nonum \\ &&
 = \frac1{\NN} \sum_{\v q}^{BZ} \frac1{2\bet}
 \sum_{\nu=-\frac{M-1}2}^{\frac{M-1}2\;\prime} 
\tr\left(\ln\GGam_0(\omnu)+\GGam_0(\omnu)^{-1}
\del \GGam_1\qunu+\ord(\del^2) \right)
\nonum \\ &&
= \frac1{2\bet}
 \sum_{\nu=-\frac{M-1}2}^{\frac{M-1}2\;\prime} 
\ln\det\GGam_0(\omnu) + 
\frac1{\NN} \sum_{\v q}^{BZ} \frac1{2\bet}
 \sum_{\nu=-\frac{M-1}2}^{\frac{M-1}2\;\prime} 
\nonum \\ && \times
\left( \frac{\del}{1-\eind} \tr\AAA\qunu +  \frac{\del}{1-\emind}
\tr\AAA\mqunu \right)  
\nonum \\ &&
= \frac4{\bet} \sum_{\nu=1}^{\frac{M-1}2}
\ln(1-\eind)(1-\emind)
\nonum \\ &&
+ \frac1{\NN} \sum_{\v q}^{BZ} \frac1{2\bet}
 \sum_{\nu=-\frac{M-1}2}^{\frac{M-1}2\;\prime}
\sum_{\ib=1}^4
\left(  \frac{\del \eind}{1-\eind} E_{\ib}(\v q) +
 \frac{\del \emind}{1-\emind} E_{\ib}(-\v q)
\right)
\nonum \\ &&
=\frac4{\bet} \ln M -\mez \sum_{\ib=1}^4 \frac1{\NN}
\sum_{\v q}^{BZ} E_{\ib}(\v q)
\label{c.15}
\eeqn
where the ``quasi-particle'' energies $ E_{\ib}(\v q)$ are given by
the diagonal elements of the matrix
\eqref{c.13} (apart from the factor $\exp\{i \omnu \del\}$):
\beq
 E_{\ib}(\v q) = h^{(\ib)}_0 +  \frac1{\NN}
\sum_{\v k}^{BZ} \sum_{\g} f_F(\E_{\g}(\v k)) \E_{11}(\v q;\v
k,\g|\ib,\ib) \;.
\label{c.16}
\eeq
The first term on the right-hand side of Eq.\eqref{c.15} can be
absorbed into the overall normalization factor of the \FI, while the
second term gives the desired ``\CFI'' $F^{(d)}_1$ to the free energy.
Upon recalling the expression \eqref{c.6} for $\E_{11}$, as well as
the property
\beq
 \sum_{\v q}^{BZ} \gamma(\v k \pm \v q)=0
\label{c.17}
\eeq
[cf. Eq.\eqref{gam}], we obtain eventually:
\beqn
F_1^{(d)}  =-\mez \sum_{\ib=1}^4 h^{(\ib)}_0
&& -\mez t z_0 \sum_{\ib=1}^4 \sum_{\g} \left( 
 \frac{\de^2 z^{\ast}_{j,\g}(m-1,m)}{\de \bbc{\ib}_{j,m}\de
\bb{\ib}_{j,m-1}}\Bigr|_{\und b_0} +
 \frac{\de^2 z_{j,\g}(m,m-1)}{\de \bbc{\ib}_{j,m}\de
\bb{\ib}_{j,m-1}}\Bigr|_{\und b_0}
\right)
\nonum \\ && \times
\frac1{\NN} 
\sum_{\v k}^{BZ}  f_F(\E_{\g}(\v k)) \gamk \;.
\label{c.18}
\eeqn

There remains to verify that the result \eqref{c.18} obtained
within the ``Cartesian'' gauge coincides {\evi at self-consistency}
(i.e., when $\de F_0/\de \bbd{\ib}_0=0$ for each $\ib$) with the
result \eqref{f1d} obtained within the ``radial'' gauge. To this end,
we get by comparing Eqs. \eqref{c.1} and \eqref{wbim}
\beq
-\mez \sum_{\ib=1}^4 h_0^{(\ib)} = -2\lam_0^a + 2 \lam_0^b -\frac{U}2 
\label{c.19}
\eeq
(in the paramagnetic phase), which indeed coincides with the result
\eqref{f1d} evaluated at self-consistency for the choice $\R=1$ (so
that $\F[\R;\und b_0]$ vanishes according to Eq. \eqref{zrb}). To
verify the complete equivalence of the expressions \eqref{c.18} and
\eqref{f1d} at self-consistency for {\evi any} choice of $\R$, it is
thus enough to show that 
\beqn
\F[\R;\und b_0] &&= -\frac{z_0}4 \sum_{\ib=1}^4 \sum_{\g} 
\left(
 \frac{\de^2 z^{\ast}_{j,\g}(m-1,m)}{\de \bbc{\ib}_{j,m}\de
\bb{\ib}_{j,m-1}}\Bigr|_{\und b_0} +
 \frac{\de^2 z_{j,\g}(m,m-1)}{\de \bbc{\ib}_{j,m}\de
\bb{\ib}_{j,m-1}}\Bigr|_{\und b_0}
\right)
\nonum \\ && =
-z_0  \sum_{\ib=1}^4 
 \frac{\de^2 z_{\up}}{\de \bbc{\ib}\de
\bb{\ib}}\Bigr|_{\und b_0} 
\label{c.20}
\eeqn
for the paramagnetic phase, where
\beq
z_{\up}=\left(\bbc1 \bb2+\bbc3 \bb4\right) \R\left(\bbc1 \bb1,\bbc2 \bb2,\bbc3
\bb3,\bbc4 \bb4\right) \;. 
\label{c.21}
\eeq
For instance, when $\ib=1$ we obtain 
\beqn
 \frac{\de^2 z_{\up}}{\de \bbc{1}\de
\bb{1}}\Bigr|_{\und b_0}
&&=
\left(2 \bb1_0\bb2_0+\bb3_0\bb4_0\right) \frac{\de\R}{\de n_1} 
\nonum \\ &&
+ \left( \bb1_0\bb2_0+\bb3_0\bb4_0\right) \bbd1_0 \frac{\de^2\R}{\de n_1^2} 
\;,
\label{c.22}
\eeqn
with similar expressions for $\ib=2,3,4$. Comparison with
Eq.\eqref{zrb} thus enables us to verify that Eq.\eqref{c.20} holds.
This completes the proof. [Recall that, as for Eq.\eqref{zrb}, the
above derivation holds for a subsidiary operator $R\is$ in
Eq.\eqref{zr} which is explicitly in normal-ordered form].

Although we have focused thus far in the calculation of the free energy (and
derived quantities),
this is not the only physical quantity
of interest. In particular,
it will be relevant to extend
 the methods
developed in this paper to the calculation of correlation functions.
Similarly to what we have shown for the free energy, we expect that
taking the \CL at the outset in the \FI might miss
important contributions also for the correlation functions. In
principle, one would have to carry out the calculation of correlation
functions via the \FI formulation with the discretized time mesh and
to take consistently the \CL only at the end of the calculation.
In practice, one could exploit the experience that has been developed
for the calculation of the free energy and try to formulate simpler
alternative methods for the calculation of the correlation functions.
Specifically, developing a suitable diagrammatic theory for the
correlation functions could enable one to focus directly on those diagrams
which admit ``contributions from infinity''.
 In this respect, there exist significant
differences between the ``Cartesian'' and the ``radial'' gauge, since
in the latter keeping the discretized time mesh results in additional
vertices of the diagrammatic theory which vanish identically in the
\CL. However, the results obtained in the two gauges should coincide
order by order in the $1/N$ expansion.
 For this reason, we expect the calculation of the ``\CFI'' for
the correlation functions to be less involved in the ``Cartesian''
than in the ``radial'' gauge, in analogy to what we have just shown
for the free energy. One might thus envisage calculating
diagrammatically the ``\CFI'' to a given correlation function (at
least at the order $1/N$) in the ``Cartesian'' gauge and the
corresponding continuum contribution in the ``radial'' gauge since
this is not plagued by infrared singularities.
To this end, it will be necessary to determine, for each elementary
propagator and vertex of the diagrammatic structure, the associated
order in the time step $\del$, in order to select directly those
diagrams which have a nonvanishing ``\CFI''.\cite{consistently}

A rigorous classification of the diagrammatic theory associated with
correlation functions is beyond the scope of this paper. Nonetheless,
we illustrate the above arguments by
working out in detail one example for a physical quantity which can
also be derived alternatively from the free energy. This example will
explicitly show that a suitable handling of the diagrammatic structure
in the ``Cartesian'' gauge can provide the correct ``\CFI'' at the
order $1/N$.

Let's consider the average number $<d^{\dag}d>$ of doubly occupied
sites, defined by
\beq
<d^{\dag}d> = \frac1\NN \sum_i \frac1M \sum_{m=0}^{M-1} <d\im^{\ast}
d\imu>
\label{c.23}
\eeq
(in the limit $M\to \infty$)
where the average $<\cdots>$ is evaluated with the action \eqref{sm}.
Equation \eqref{c.23} can be interpreted as the $\v q=0$ and $\omnu=0$
component of a correlation function; alternatively, it can be obtained
as the (total) derivative of the free energy (per lattice site):
\beq
<d^{\dag}d> = \frac{d \,F}{d\,U} \;.
\label{c.24}
\eeq
We are specifically interested in the ``\CFI'' to $<d^{\dag}d>$ at
the order $1/N$. We may therefore write
\beq
<d^{\dag}d>_{\infty} = \frac{d \,F_1^{(d)}}{d\,U} 
\label{c.25}
\eeq
with $F_1^{(d)}$ given by Eq. \eqref{c.18}, the derivative being
evaluated {\evi at self-consistency} for the \MF parameters.

With the simplest choice $\R=1$, $F_1^{(d)}$ is given by Eq.
\eqref{c.19}. We obtain in this case
\beq
<d^{\dag}d>_{\infty} = -\mez \sum_{\ib=1}^4
\frac{d\,h_0^{(\ib)}}{d\,U}
= -\mez \left(\sum_{\il=1}^3 v_l \frac{d\,\lam^{(\il)}}{d\,U} +1 \right)
\label{c.26}
\eeq
with the use of the dictionary \eqref{defab} for the Lagrange
multipliers, where $v_1=v_2=-2$ and $v_3=4$. [The magnitude of each
$v_l$ represents the number of vertices in the action \eqref{sm}
containing the associated $\lam^{(l)}$ and any bosonic number operator
$b^{\dag}b$, while the sign of a given $v_l$ corresponds to the sign
of those vertices in that action]. Connection of Eq.
\eqref{c.26} with the diagrammatic structure is then obtained by
expressing the derivatives $d\,\lam^{(\il)}/d\,U$ therein as follows:
\beq
\frac{d\,\lam^{(\il)}}{d\,U} = \sum_{\ia=1}^8 G_B(\v
q=0,\omnu=0)_{5+l,\ia} \frac{\de^2F_0}{\de a_0^{(\ia)} \de U} 
= 2 d_0  G_B(\v q=0,\omnu=0)_{5+l,4}
 \label{c.27}
\eeq
where $d_0=a_0^{(4)}$ and the $\v q=0$ and $\omnu=0$ bosonic
propagator matrix $G_B$ is defined in terms of the inverse of the
matrix of the second derivatives of $F_0$ [cf. Eq. \eqref{d2f0}].
Equation \eqref{c.26} thus becomes:
\beq
<d^{\dag}d>_{\infty} = -\mez 
\left(2 d_0
\sum_{\il=1}^3 v_l 
G_B(\v q=0,\omnu=0)_{5+l,4}
\;\; + \;\; 1 \right) \; .
\label{c.28}
\eeq
This result can be obtained directly from the diagrammatic structure,
with the provision of associating with {\evi each} bosonic loop (of
the type $b^{\dag}b$) a
``\CFI'' equal to $-1/2$. To identify the bosonic loops associated
with $<d^{\dag}d>$ at the order $1/N$, we rewrite the definition
\eqref{c.23} in terms of Eqs.\eqref{c.2} and \eqref{c.3}:
\beqn
<d^{\dag}d> &&= d_0^2 + 2 d_0 \frac1{\NN} \sum_i \frac1M \sum_{m=0}^{M-1}
<\uti{d}\im> 
+  \frac1{\NN} \sum_i \frac1M \sum_{m=0}^{M-1}
<\uti{d}\im^{\ast} \uti{d}\imu > 
\nonum \\ &&
= d_0^2 + 2 d_0 d^{(1)} +  \frac1{\NN} \sum_{\v q}^{BZ} 
 \frac1M \sum_{\nu=0}^{M-1} \eind 
<d\qunu^{\ast} d\qunu > 
\label{c.29}
\eeqn
where $d^{(1)}$ is the $1/N$ shift of the \MF parameter $d_0$ (cf. 
 Appendix \ref{app:a}).
According to our prescription, the bosonic loop given by the last term
of Eq.\eqref{c.29} supplies the term $-1/2$ to $<d^{\dag}d>_{\infty}$.
Additional bosonic loops, however, are hidden in the definition of $d^{(1)}$.
We can, in fact, write [cf. Eq. \eqref{1nshift}]
\beq
d^{(1)} = \sum_{\ia=1}^8 
G_B(\v q=0,\omnu=0)_{4,\ia}
\frac{\de F_1}{\de a^{(\ia)}}
\label{c.30}
\eeq
and identify $|v_l|$ loops in $\de F_1/\de a^{(\ia)}$ for $\ia=5+l$
associated with a vertex $\lam^{(\il)} b^{\ast}b$. Taking also into
account the sign of these vertices in the action, we obtain for the
``\CFI'' to $d^{(1)}$ the expression
\beq
d^{(1)}_{\infty} =\sum_{\il=1}^3 
G_B(\v q=0,\omnu=0)_{4,5+\il} v_l \left(-\mez\right) 
\label{c.31}
\eeq
according to our prescription. We have thus verified that the ``\CFI''
obtained from the diagrammatic structure of the correlation function
\eqref{c.29} coincides with the result \eqref{c.28}, at least for the
simplest choice $\R=1$ of the subsidiary function.

When $\R$ is not unity, $F_1^{(d)}$ is given by \eqref{c.18} and not
simply by \eqref{c.19}. Besides the terms \eqref{c.26}, there exists
thus the following additional contribution to 
$<d^{\dag}d>_{\infty}$:
\beqn
&&
\frac{d}{d U} \left(\F[\R;\und b_0] \frac{2t}{\NN} \sum_{\v k}^{BZ}
\gamk f_F(\E_{\v k}) \right) =
\nonum \\ &&
= 2 d_0 \sum_{\ia=1}^8 \frac{\de}{\de a_0^{(\ia)}} \left(
\F[\R;\und b_0] \frac{2t}{\NN} \sum_{\v k}^{BZ}
\gamk f_F(\E_{\v k}) \right) G_B(\v q=0,\omnu=0)_{\ia,4}
\label{c.32}
\eeqn
since $\F[\R;\und b_0]$ and $\E_{\v k}$ depend on $U$ only through
$\und a_0$ (with $\F$ given by Eq. \eqref{c.20}). The derivatives on
the right-hand side of Eq.\eqref{c.32}, which act alternatively on the
bosonic or the fermionic factor, are associated with two different
classes of diagrams each containing the relevant (i.e., $b^{\ast}b$)
bosonic loops. It is clear that these diagrams represent additional
contributions to the $1/N$ shift $d^{(1)}$ [cf. Eq. \eqref{c.30}], and
that the relevant bosonic loops therein arise from the ``vertex''
\eqref{c.20} and its derivatives. One can readily verify that the
result \eqref{c.32} is eventually recovered by following again our
prescription of assigning a ``\CFI'' equal to $-\mez$ to each bosonic
loop of the type $b^{\dag} b$ 
identified in the diagrammatic structure.\cite{prescription}

\section{ Exact Solution for a One-Level Two-Site Model }
\label{app:d} 
The one-level two-site model Hamiltonian, which was considered in
Section \ref{s:nume} for numerical calculations with the \FI approach,
can readily be diagonalized. For the physical case of spin
$1/2$, the energy eigenvalue $\E_l \;(l=1,2,\cdots,10)$ and the
associated degeneracy factor $g_l$ for the $l$-th level with particle
number $N_l \; (=0,1,\cdots,4)$ [or, equivalently, with particle 
density per lattice site $n_l(=0,1/2,1,3/2,2)$]
are reported in Table I, where
\beq
\E_{\pm} = \frac U2 \pm \sqrt{ \left(\frac U2 \right)^2 + (4t)^2 } \;.
\label{d.1}
\eeq
States with $g_l=1,2,3$ correspond to spin singlet, doublet, and
triplet in the order. Note that, due to the use of periodic boundary
condition, the ``effective'' hopping between the two sites is $2t$ and
not $t$.\cite{periodic}

Once the set $(N_l,\E_l,g_l)$ is known, the grand-canonical partition
function can be obtained from its definition 
\beq
\ZZ= \sum_l g_l \exp\{-\bet(\E_l -\mu N_l)\}
\label{d.2}
\eeq
for given $\bet$ and $\mu$. The chemical potential can be then
eliminated in favor of the mean particle density $n$ by setting
\beq
2n= \frac1{\ZZ} \sum_l g_l N_l \exp\{-\bet(\E_l -\mu N_l)\}
\;.
\label{d.3}
\eeq
In this way, the canonical (site) free energy $\bar F$ 
(and thus the site ground-state
energy E in the zero-temperature limit) can be obtained as
\beq
2 \bar F = - \frac1{\bet} \ln \ZZ + 2 n \mu 
\undersymbolsub{\bet\to\infty}{\longrightarrow}
2 E 
\label{d.4}
\eeq
for arbitrary (even noninteger) values of $2 n$. [We refer the reader
to Ref.\cite{noiprep} for a discussion about the equivalence between
the grand-canonical and canonical ensembles for a finite-size system
in the limit $\bet\to\infty$].

The result \eqref{d.4} obtained by exact diagonalization for the
finite-size system can be compared quantitatively with the approximate
results obtained for the same system by the slave-boson approach
discussed in the text. This comparison is shown in Section \ref{s:nume}.

\section{ Gaussian Integration with a non Hermitian Matrix }
\label{app:e} 
The Gaussian matrices, which we have dealt with by
considering the functional-integral representation of the partition
function, are not Hermitian (and not even normal). Therefore,
integration of the associated quadratic form cannot be performed in a
straightforward way by direct matrix diagonalization. Nonetheless, it
is possible to readily generalize the procedure for Gaussian
integration to the cases of interest, for which the action can be cast
in the form
\beq
S(\und x,\und y) = \und x^T M \und x + 2 i  \und x^T Q \und y +
 \und y^T N \und y =
 \left( \und x^T , \und y^T \right)
\left( \begin{array}{l l} M & iQ \\ iQ^T & N \end{array} \right)
\left( \begin{array}{l} \und x \\ \und y \end{array} \right)
\; .
\label{e.1}
\eeq
In this expression, $\und x$ and  $\und y$ are sets of $n$ real
variables each, $M$ and $N$ are real and symmetric matrices, and $Q$
is a real (but not necessarily symmetric) matrix.\cite{general}
Note that the presence of the factor $i$ (instead of $-i$) in front of
$Q^T$ makes the
Gaussian matrix in \eqref{e.1} non Hermitian.

The Gaussian integral
\beq
I = \int d \, \und x \, d \, \und y e^{-S(\und x,\und y)}
\label{e.2}
\eeq
with action \eqref{e.1} can be evaluated by iteration, performing the
integration over, say, the variables $\und y$ first. To this end, we
assume that all eigenvalues of the matrix $N$ are positive 
\cite{stability} 
 and exploit the identity
\beq
\int d \, \und y \exp\{ - \und y^T N \und y - 2 i \und x^T Q \und y \}
= [\det(N/\pi)]^{-1/2} \exp\{-\und x^T Q N^{-1} Q^T \und x\}
\label{e.3}
\eeq
which can be readily proved by diagonalizing the matrix $N$. The effective action
in the variables $\und x$ then reads
\beq
S_{eff}(\und x)= \und x^T(M+Q N^{-1} Q^T ) \und x
\label{e.4}
\eeq
with an effective Gaussian matrix which is now real and symmetric.
Assuming the eigenvalues of this matrix to be all
positive,\cite{stability}  we obtain
eventually
\beq
I = [\det(N/\pi)]^{-1/2} \cdot [\det(M+Q N^{-1} Q^T)/\pi]^{-1/2}
\;.
\label{e.5}
\eeq

There remains to show how the product of the two determinants in
\eqref{e.5} is related to the determinant of the original Gaussian
matrix in \eqref{e.1}. To this end, we introduce the notation
\beq
P=M+Q  N^{-1} Q^T
\label{e.6}
\eeq
and use the familiar properties of the determinants, to obtain:
\beqn
&&
\det \left( \begin{array}{l l} M & iQ \\ iQ^T & N \end{array} \right)
=
\det \left( \begin{array}{l l} P-Q  N^{-1} Q^T & iQ \\ iQ^T & N
\end{array} \right) 
\nonum \\&&
= \det 
 \left( \begin{array}{l l} P & iQ  N^{-1} \\ 0 & \UU \end{array}
\right)
\cdot
\det \left( \begin{array}{l l} \UU & 0  \\ i Q^T & N \end{array} \right)
\nonum \\&&
= 
\det P \cdot \det N = \det(M+Q  N^{-1} Q^T)\cdot \det N \;.
\label{e.7}
\eeqn
In conclusion, we obtain for the Gaussian integral 
\eqref{e.2}:
\beq
I = \left[ \det \frac1{\pi}
  \left( \begin{array}{l l} M & iQ \\ iQ^T & N \end{array} \right)
\right]^{-1/2} \;.
\label{e.8}
\eeq
Note that the same result would have been formally obtained if the original
Gaussian matrix were Hermitian. By the same token, it can be proved
that Wick's theorem for pairwise contractions holds also for a non
Hermitian Gaussian matrix of the form \eqref{e.1}.

There remains to show that the Gaussian actions considered in this
paper can be cast in the form \eqref{e.1}. We consider the ``radial''
and ``Cartesian'' gauges separately.

\subsection{ RADIAL GAUGE }
\label{sapp:e.a} 
Quite generally, the Gaussian part of the action in the ``radial''
gauge reads
\beq
R = \sum_q\nolimits'
 \left( \und r^T(q) , \und \lam^T(q) \right)
\left( \begin{array}{l l} \Gam_{rr}^g(q) & i  \Gam_{r\lam}^g(q)\\ 
i \Gam_{\lam r}^g(q)  &  -\Gam_{\lam \lam}^g(q) \end{array} \right)
\left( \begin{array}{l} \und r(-q) \\ \und \lam(-q) \end{array} \right)
\;,
\label{e.9}
\eeq
where $q=\qunu$ is restricted to half  the available values (say,
$\omnu>0$), $\und r(q) $ and $ \und \lam(q)$ are four-component
vectors (corresponding to the components $\ia=1,\cdots,4$ and 
$\ia=5,\cdots,8$, respectively, of the vector $\und a(q)$ in Eqs.
\eqref{s2} and \eqref{sbar2}), and the $4 \times 4$ block matrices 
$\Gam_{rr}^g,  \Gam_{r\lam}^g, \Gam_{\lam r}^g$, and $\Gam_{\lam
\lam}^g$ are defined in terms of the $8 \times 8$ matrix 
\beq
\Gam^g(q)_{\ia,\ia'} = \Gam(q)_{\ia,\ia'} + \Gam^T(-q)_{\ia,\ia'}
\label{e.10}
\eeq
with $T$ standing for the transposed matrix and with
\beq
 \Gam(q)_{\ia,\ia'} = \BB(q|\ia,\ia') +\CC(q|\ia,\ia') 
\label{e.11}
\eeq
[cf. Eq. \eqref{gamq}]. The matrix \eqref{e.11} (and thus the matrix
\eqref{e.10}, too) satisfies the symmetry property
\beq
\Gam\qunu_{\ia,\ia'} = \Gam(-\v q,\omnu)_{\ia,\ia'} = 
\Gam(-\v q,-\omnu)^{\ast}_{\ia,\ia'}
\label{e.12}
\eeq
for a lattice with inversion symmetry. [The modes with $\omnu=0$ and
any $\v q$ can be treated separately in a straightforward way].

Note that in Eq.\eqref{e.9} we have restored the imaginary unit $i$
whenever necessary to identify $\und\lam(q)$ as the Fourier transform
of $\evi real$ variables.\cite{lambda}
Writing
\beqn
&&
\und r(q) = \und x(q)+ i\und y(q) \quad ,\quad 
\und r(-q) = \und r(q)^{\ast} =  \und x(q) - i\und y(q)
\nonum \\ &&
\und \lam(q) = \und \xi(q)+ i\und \eta(q) \quad ,\quad 
\und \lam(-q) = \und \lam(q)^{\ast} =  \und \xi(q) - i\und \eta(q)
\label{e.13}
\eeqn
where
$\und x(q), \und y(q), \und \xi(q)$, and $\und \eta(q)$ are
independent real variables, the Gaussian action \eqref{e.9} reads:
\beq
R = \sum_q\nolimits'
 \left( \und x^T(q) ,\und y^T(q) ,\und \xi^T(q) , 
\und \eta^T(q) \right)
\left( \begin{array}{l l} M(q) & i  Q(q)\\ 
i Q^T(q)  &  N(q) \end{array} \right)
\left( \begin{array}{l} \und x(q) \\  \und y(q) \\ \und \xi(q) \\ \und
\eta(q)  \end{array} \right) 
\;.
\label{e.14}
\eeq
In this expression,
\beqn
&&
M(q) = U 
\left( \begin{array}{c c} \Gam_{rr}^g(q) & {\bfmath 0}\\ 
 {\bfmath 0}   &   \Gam_{rr}^g(q)^T \end{array} \right)
 U^{\dag}
\\ &&
Q(q) = U 
\left( \begin{array}{c c} \Gam_{r\lam}^g(q) & {\bfmath 0} \\ 
 {\bfmath 0}   &   \Gam_{\lam r}^g(q)^T \end{array} \right)
 U^{\dag}
\\ &&
N(q) = U 
\left( \begin{array}{c c} -\Gam_{\lam \lam}^g(q) & {\bfmath 0}\\ 
 {\bfmath 0}   &   -\Gam_{\lam \lam}^g(q)^T \end{array} \right)
 U^{\dag} 
\eeqn
are $8\times 8$ matrices\cite{simplified} with
\beq
U=\frac1{\sqrt2} 
\left( \begin{array}{r r} \UU & \UU\\ 
 i \UU   &   -i \UU \end{array} \right) \;,
\label{e.18}
\eeq
$\UU$ being the $4 \times 4$ unit matrix. The matrices $M(q)$ and $N(q)$ are
real and symmetric, while the matrix $Q(q)$ is real but not symmetric.
The quadratic form \eqref{e.14} is thus of the type \eqref{e.1}. The
associated Gaussian integral then becomes: 
\beq
I_R= \prod_q^{\prime} \left[ \det \frac1{\pi} 
\left( \begin{array}{l l} M(q) & i  Q(q)\\ 
i Q^T(q)  &  N(q) \end{array} \right)
\right]^{-1/2}
=
\prod_q^{\prime} \left[\det \frac1{\pi} 
\left( \begin{array}{l l} \Gam_{rr}^g(q) & i  \Gam_{r\lam}^g(q)\\ 
i \Gam_{\lam r}^g(q)  &  -\Gam_{\lam \lam}^g(q) \end{array} \right)
\right]^{-1}
\label{e.19}
\eeq
where \eqref{e.7} has been used to derive the last expression. Note
that the same result would have been formally obtained by considering naively
the determinant of the original Gaussian matrix in Eq.\eqref{e.9}
twice.

\subsection{ CARTESIAN GAUGE }
\label{sapp:e.b} 
Similarly, the Gaussian part of the action in the ``Cartesian'' gauge
can be cast in the general form [cf. Eqs. \eqref{c.8} and \eqref{c.9}]
\beq
C = \sum_q\nolimits'
 \left( \und b^T(q)^{\ast} , \und b^T(-q) \right)
\left( \begin{array}{l l} A(q) & B(q)\\ 
B(-q)  &  A(-q) \end{array} \right)
\left( \begin{array}{l} \und b(q) \\ \und b(-q)^{\ast} \end{array} \right)
\;,
\label{e.20}
\eeq
where $ \und b(q)$ and $\und b(-q)$ are four-component vectors and the
$4\times4$ block matrices satisfy the symmetry properties
\beqn
&&
A(q)=A^T(q) = A(-q)^{\ast}
\label{e.21}
\\ &&
B(q)=B^T(q) = B(-q) = B(q)^{\ast}
\eeqn
for a lattice with inversion symmetry. Introducing the unitary
transformation
\beq
V =\frac1{\sqrt2} 
\left( \begin{array}{r r} \UU & -\UU\\ 
 \UU   &   \UU \end{array} \right) \;,
\label{e.23}
\eeq
we write
\beq
V 
\left( \begin{array}{l} \und b(q) \\ \und b(-q)^{\ast} \end{array} \right)
=
\left( \begin{array}{l} \und X(q) + i \und W(q) \\ 
\und Y(q) + i \und Z(q) \end{array} \right)
\label{e.24}
\eeq
where $\und X(q), \und Y(q),\und W(q)$, and $\und Z(q)$ are
independent real variables, and set
\beq
V
 \left( \begin{array}{l l} A(q) & B(q)\\ 
B(-q)  &  A(-q) \end{array} \right)
V^{\dag}
= 
\left( \begin{array}{l l} M(q) & i Q(q)\\ 
i Q^T(q)  &  N(q) \end{array} \right)
\;.
\label{e.26}
\eeq
In this expression
\beqn
&&
M(q) = \mez (A(q)+A(q)^{\ast}) - B(q)
\\&&
N(q) = \mez (A(q)+A(q)^{\ast}) + B(q)
\\&&
Q(q) = \frac1{2i}  (A(q)-A(q)^{\ast})
\eeqn
are now all real and symmetric $4\times4$ matrices. The Gaussian action
\eqref{e.20} thus reads:
\beq
C = 
 \sum_q\nolimits'
 \left( \und X^T(q) ,\und Y^T(q) ,\und W^T(q) , 
\und Z^T(q) \right)
\left( \begin{array}{c c} 
 \begin{array}{l l} M(q) & i Q(q) \\ i Q^T(q) & N(q) \end{array}
 & {\bfmath 0} \\  {\bfmath 0}   & 
 \begin{array}{l l} M(q) & i Q(q) \\ i Q^T(q) & N(q) \end{array}
\end{array} \right)
\left( \begin{array}{l} \und X(q) \\  \und Y(q) \\ \und W(q) \\ \und
Z(q)  \end{array} \right) 
\label{e.29}
\eeq
which is just the sum of two quadratic forms of the type \eqref{e.1}.
The associated Gaussian integral then becomes
\beq
I_C = 
\prod_q^{\prime} \left[ \det \frac1{\pi} 
\left( \begin{array}{l l} M(q) & i  Q(q)\\ 
i Q^T(q)  &  N(q) \end{array} \right)
\right]^{-1}
=
\prod_q^{\prime} \left[ \det \frac1{\pi} 
\left( \begin{array}{l l} A(q) & B(q)\\ 
B(-q)  &  A(-q) \end{array} \right)
\right]^{-1} \;.
\label{e.30}
\eeq
Again, this result would have been formally obtained by considering simply the
determinant of the original Gaussian matrix in Eq. \eqref{e.20} twice.

\def\prl{Phys. Rev. Lett. }
\def\prb{Phys. Rev. B }
\def\ut#1{{\bf #1}}


\begin{table}

\medskip
\[
\begin{array}{|c||c|c|c|c|c|c|c|c|c|c|} \hline
 l      & 1     & 2     & 3     & 4     & 5     & 6     & 7 & 8 & 9 & 10 \\
N_l  & 0 & 1 & 1 & 2 & 2 & 2 & 2 & 3 & 3 & 4  \\
\E_l & 0 & -2t & 2t & \E_{-} & 0 & U & \E_{+} & U-2t & U+2t& 2U \\
g_l & 1& 2& 2& 1& 3 & 1 & 1& 2 & 2 & 1 \\ \hline
\end{array}
\]
\caption{ Energy value $\E_l$ and degeneracy factor $g_l$ for the $l$-th
level of the one-level two-site model Hamiltonian corresponding to
particle number $N_l$.
$\E_{\pm}$ are given by Eq. (\figref{d.1}). For given $N_l$, the levels
are ordered for increasing energy. The ``empty'' ($N_l=0$) state has
been taken as the reference level with $\E_l=0$.
}
\end{table}

\begin{figure}
\caption{ Same as in Fig. \figref{fig1}, with the choice $R\is=1$.
}
\label{fig3}
\end{figure}

\begin{figure}
\caption{ Ground-state energy (in units of $t$) for the one-level
two-site model versus $U/t$ when (a) $n=1.0$, (b) $n=1.2$, and (c)
$n=1.5$, with the ``linearized'' $\R^{LIN}$ given by (\figref{rlin}) with
$x=x_0$. Conventions are explained in the text. Note the change of
scale and the consequent step in the HF solution in (a).}
\label{fig4}
\end{figure}

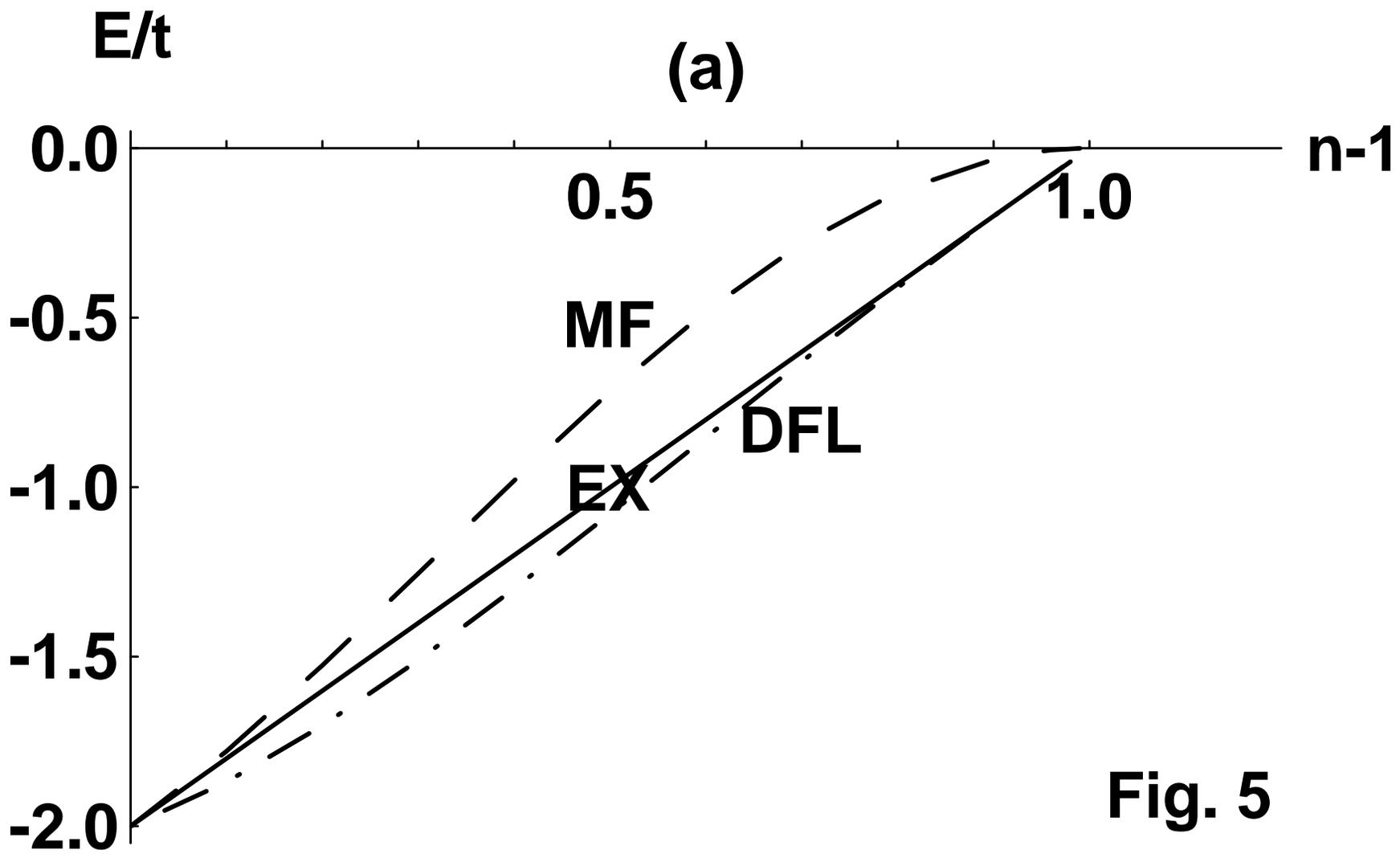
\begin{figure}
\caption{ Ground-state energy (in units of $t$) for the one-level
two-site model versus doping $n-1$ when
 (a) $U/t=0.0$  (b) $U/t=2.0$, and (c) $U/t=5.0$, 
 with the choice  $\R^{LIN}$.
}
\label{fig5}
\end{figure}

\begin{figure}
\caption{ 
Same as in Fig.\figref{fig1}, with the non normal-ordering
prescription (\figref{rsq}). Conventions are explained in the text.
}
\label{fig6}
\end{figure}

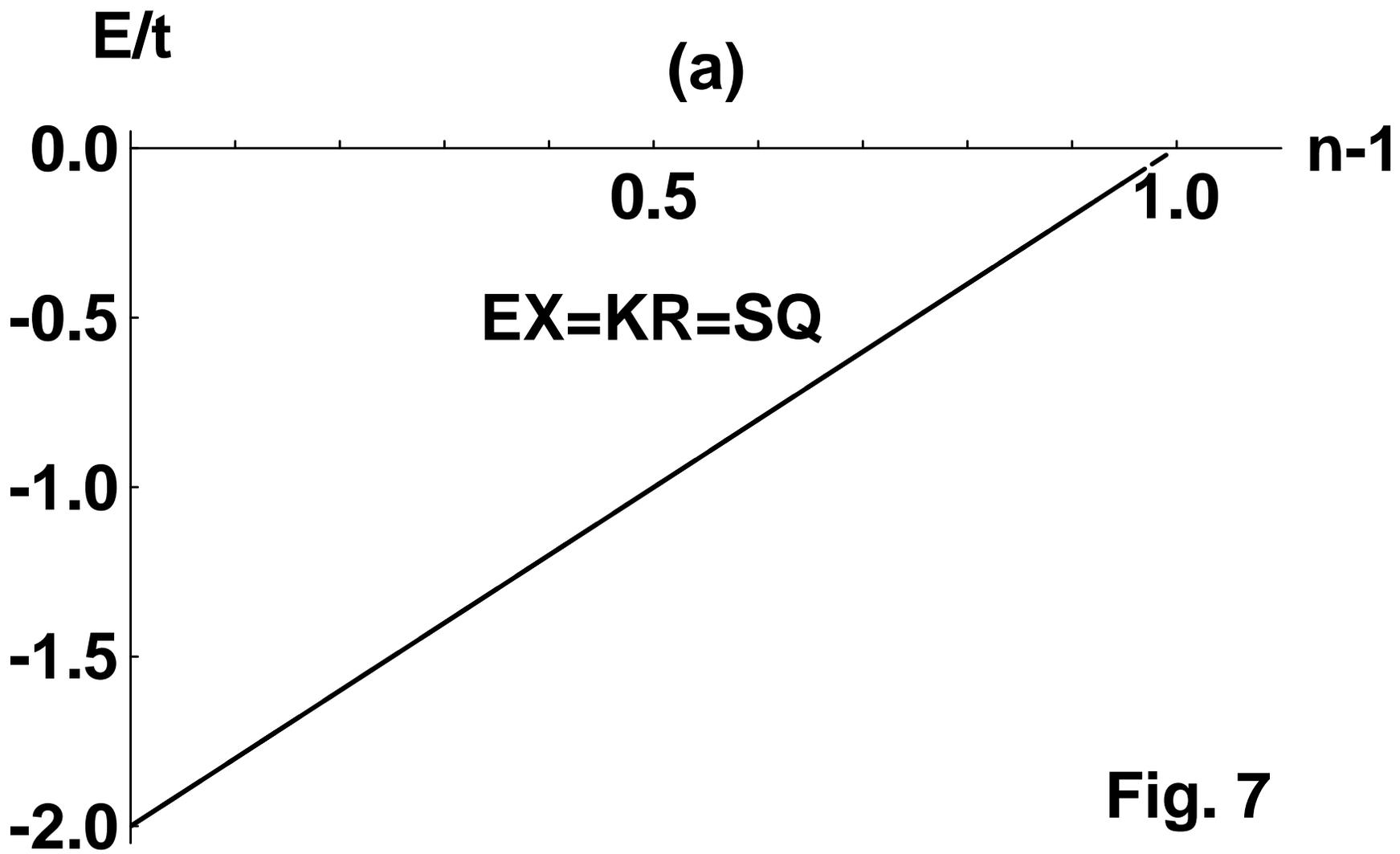
\begin{figure}
\caption{ 
Same as in Fig.\figref{fig5}, with the non normal-ordering
prescription (\figref{rsq}). Conventions are explained in the text.
}
\label{fig7}
\end{figure}

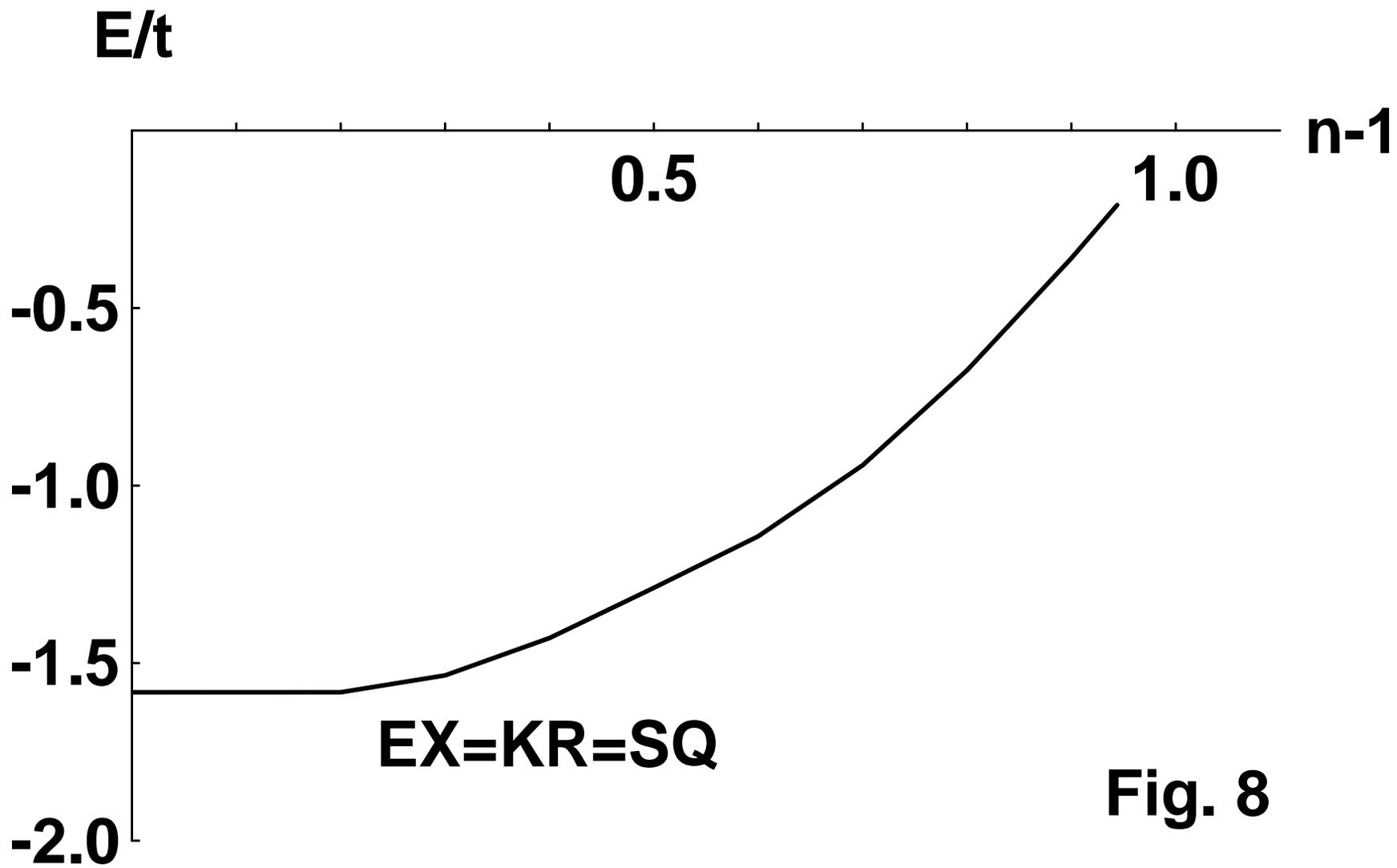
\begin{figure}
\caption{ 
Same as in Fig.\figref{fig2}, with the non normal-ordering
prescription (\figref{rsq}). Conventions are explained in the text.
}
\label{fig8}
\end{figure}



\begin{references}
\bibitem{stability} This condition needs to be verified at the outset,
being related to the stability
 of the saddle-point solution.

\bbibitem{simplified}
The expression for $M(q)$ could be somewhat simplified since it can be
proved that $\Gam_{rr}^g(q)$ is a symmetric matrix.
\end{references}
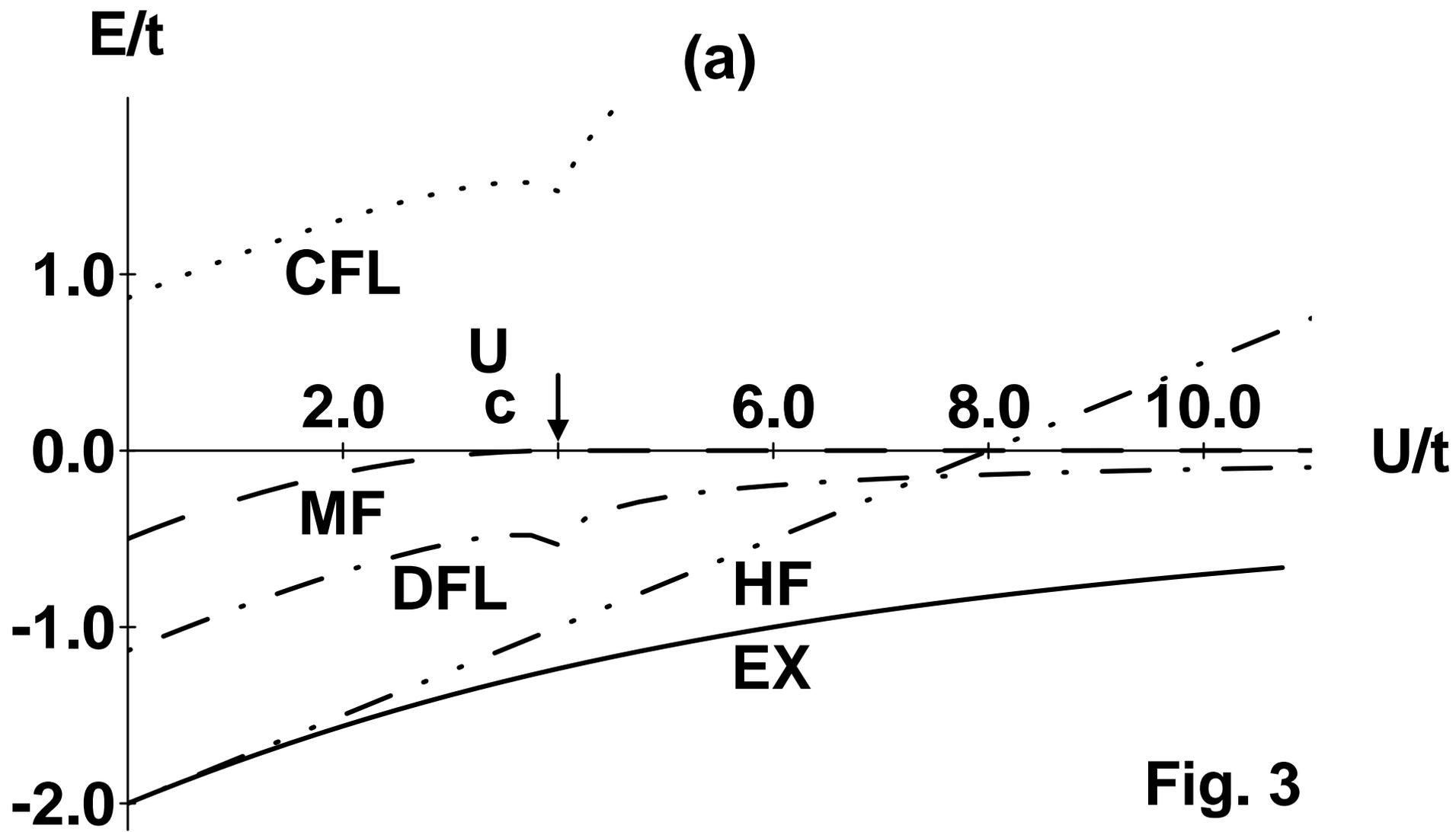
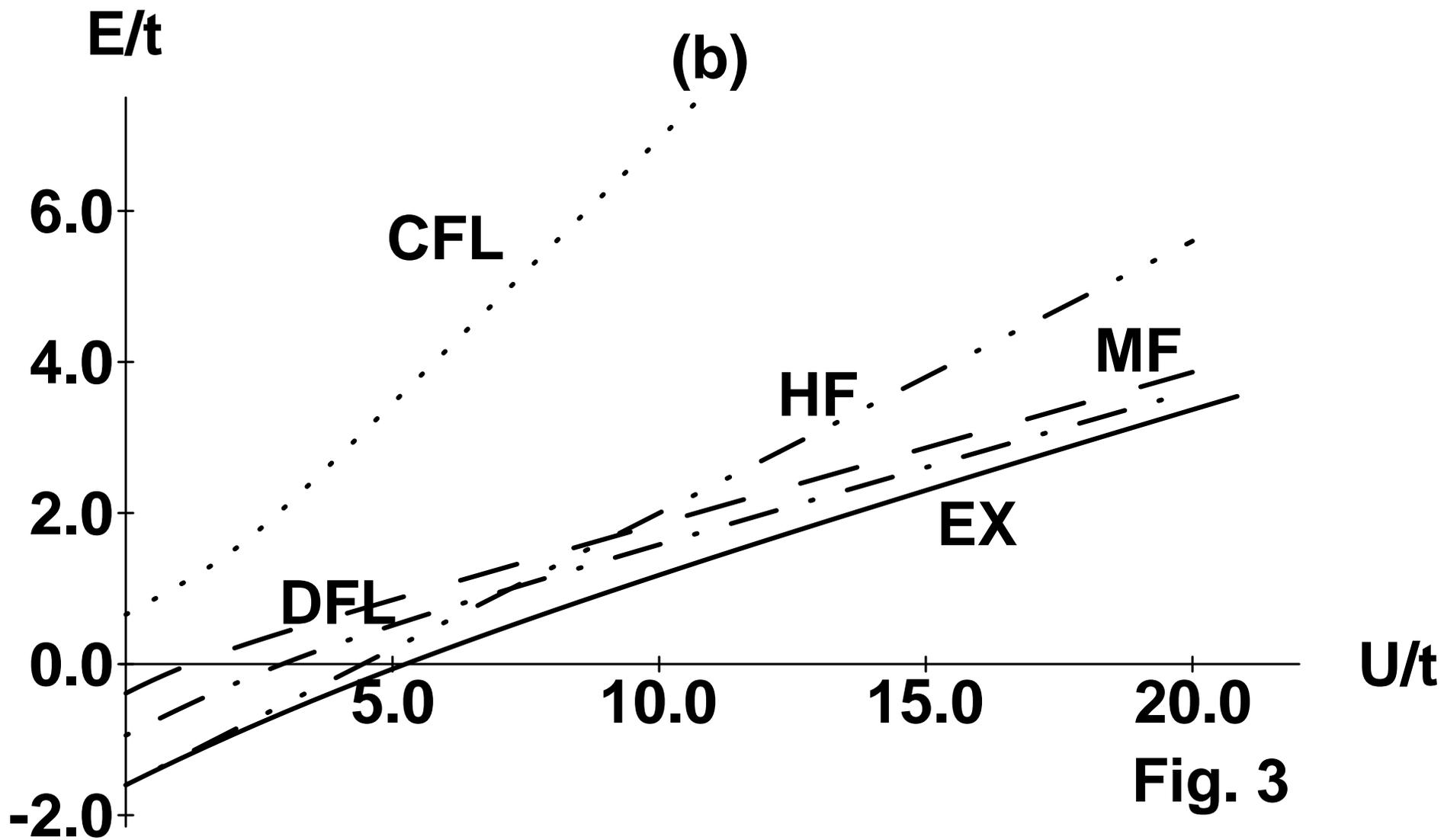
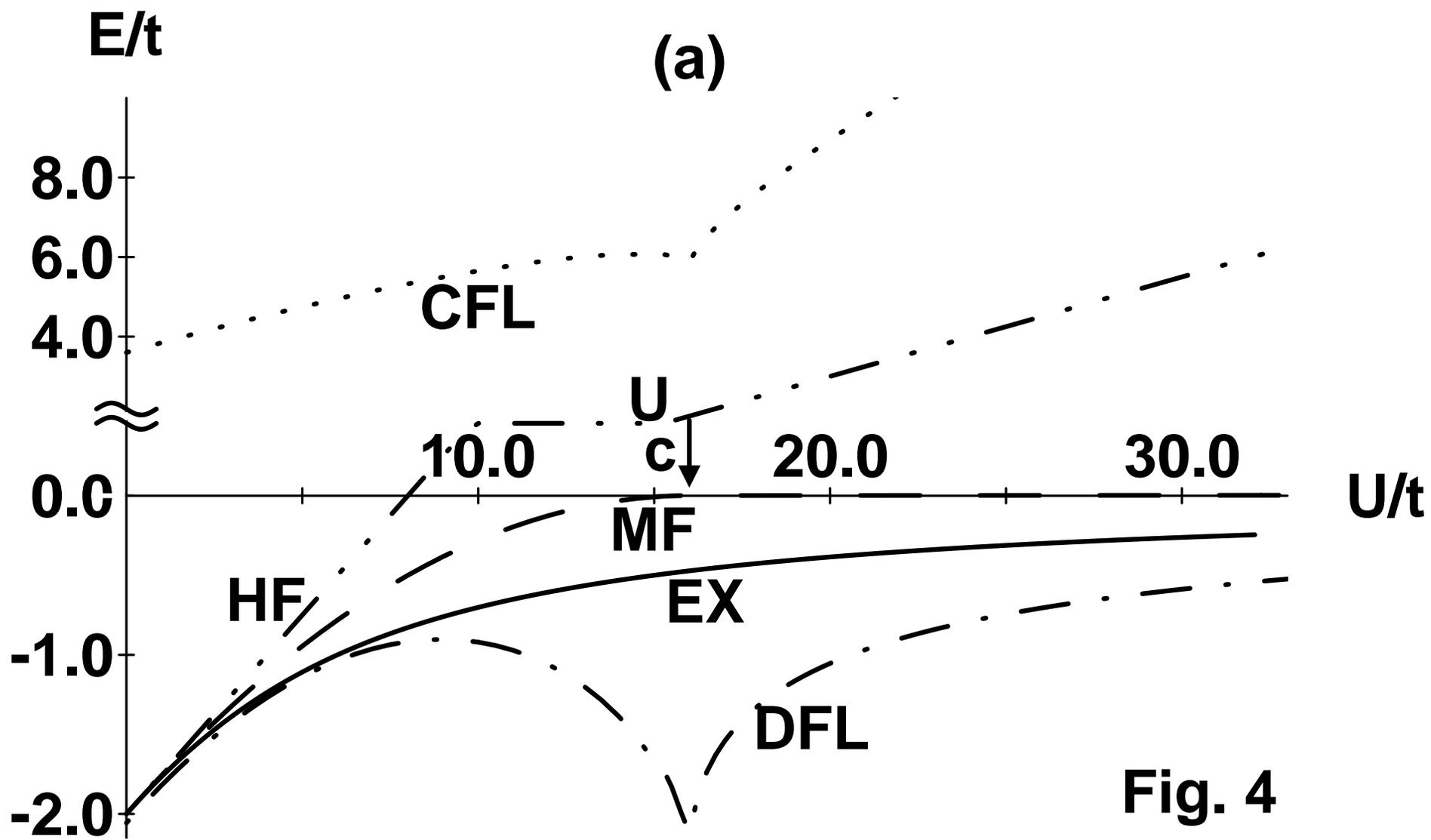
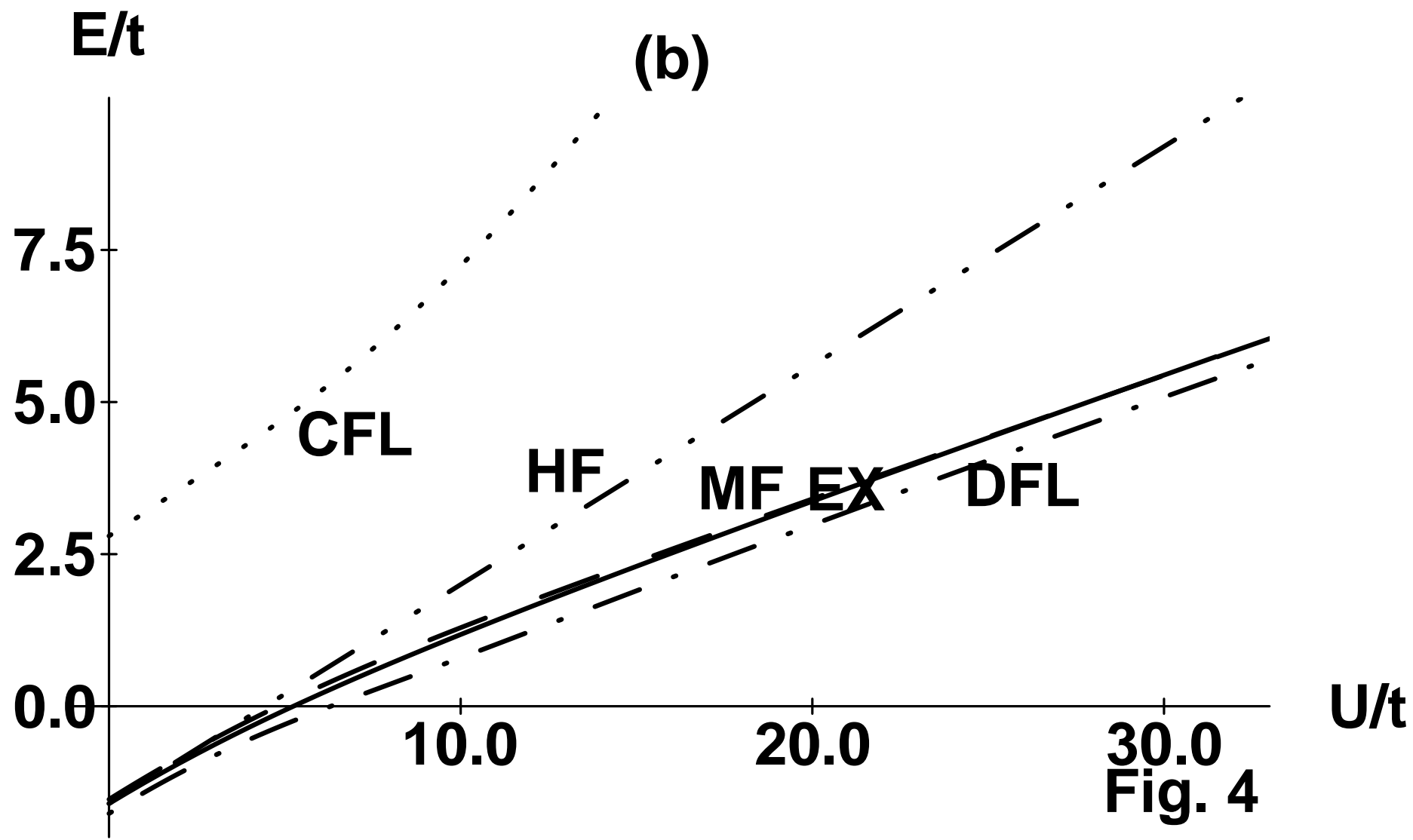
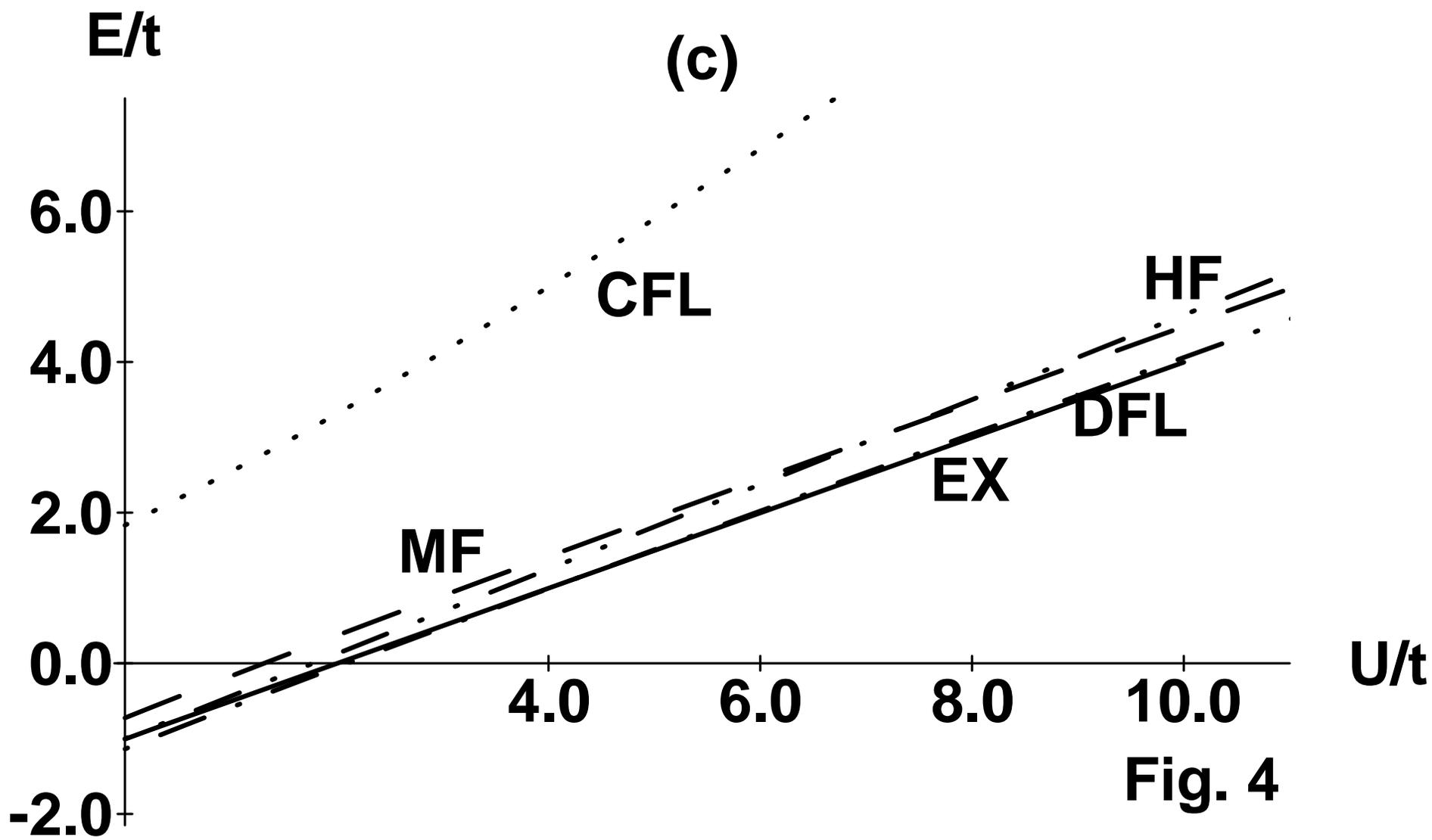
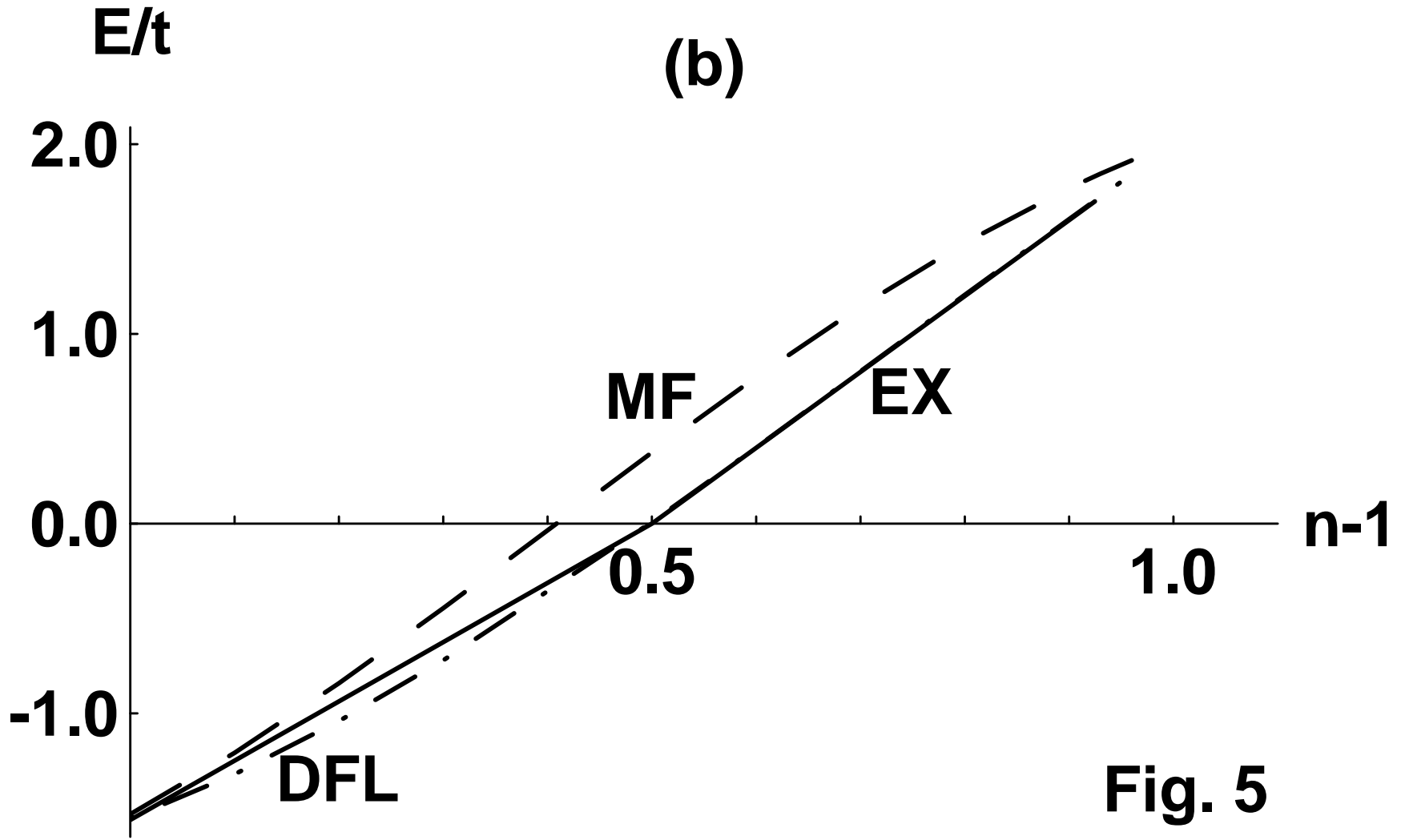
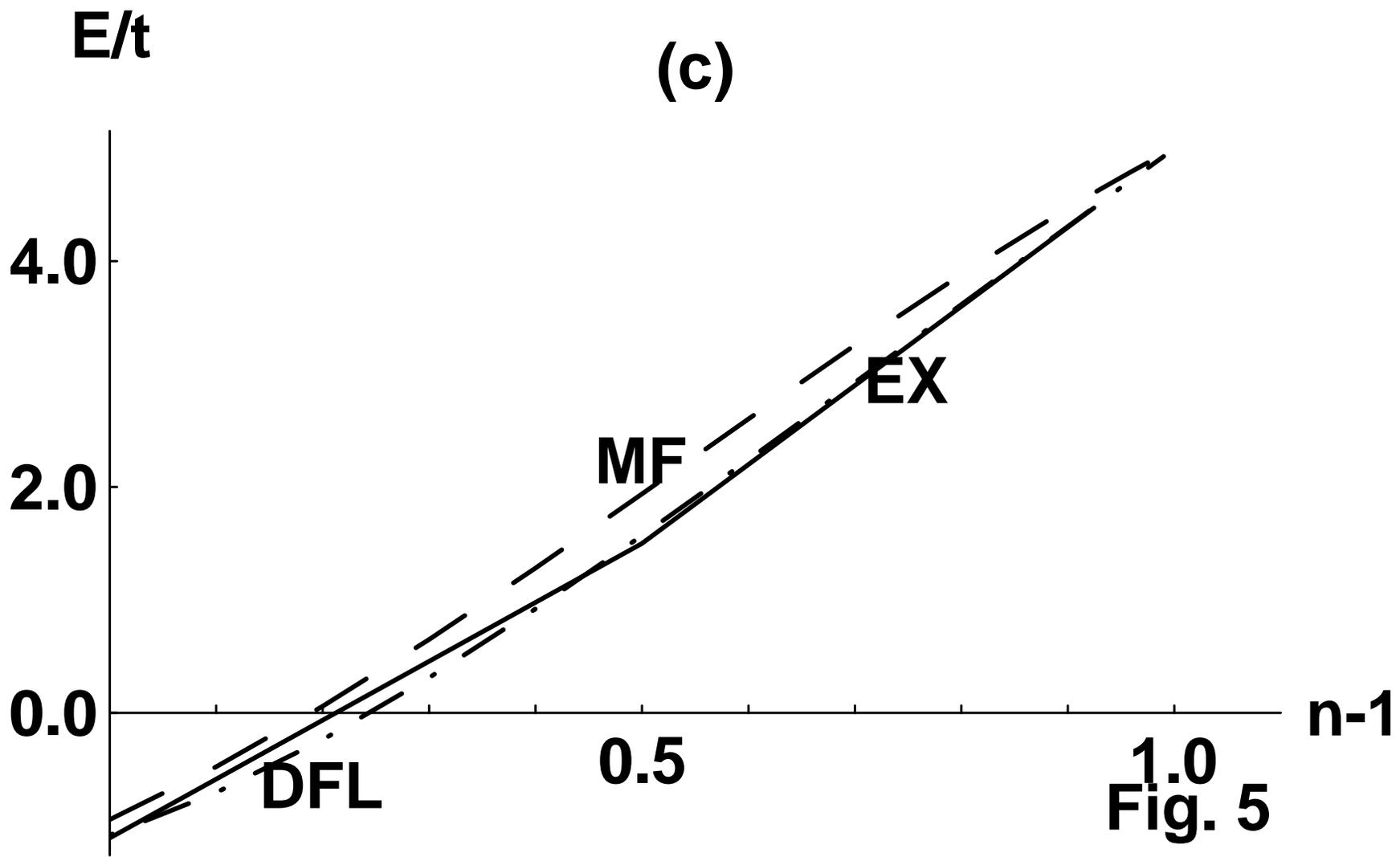
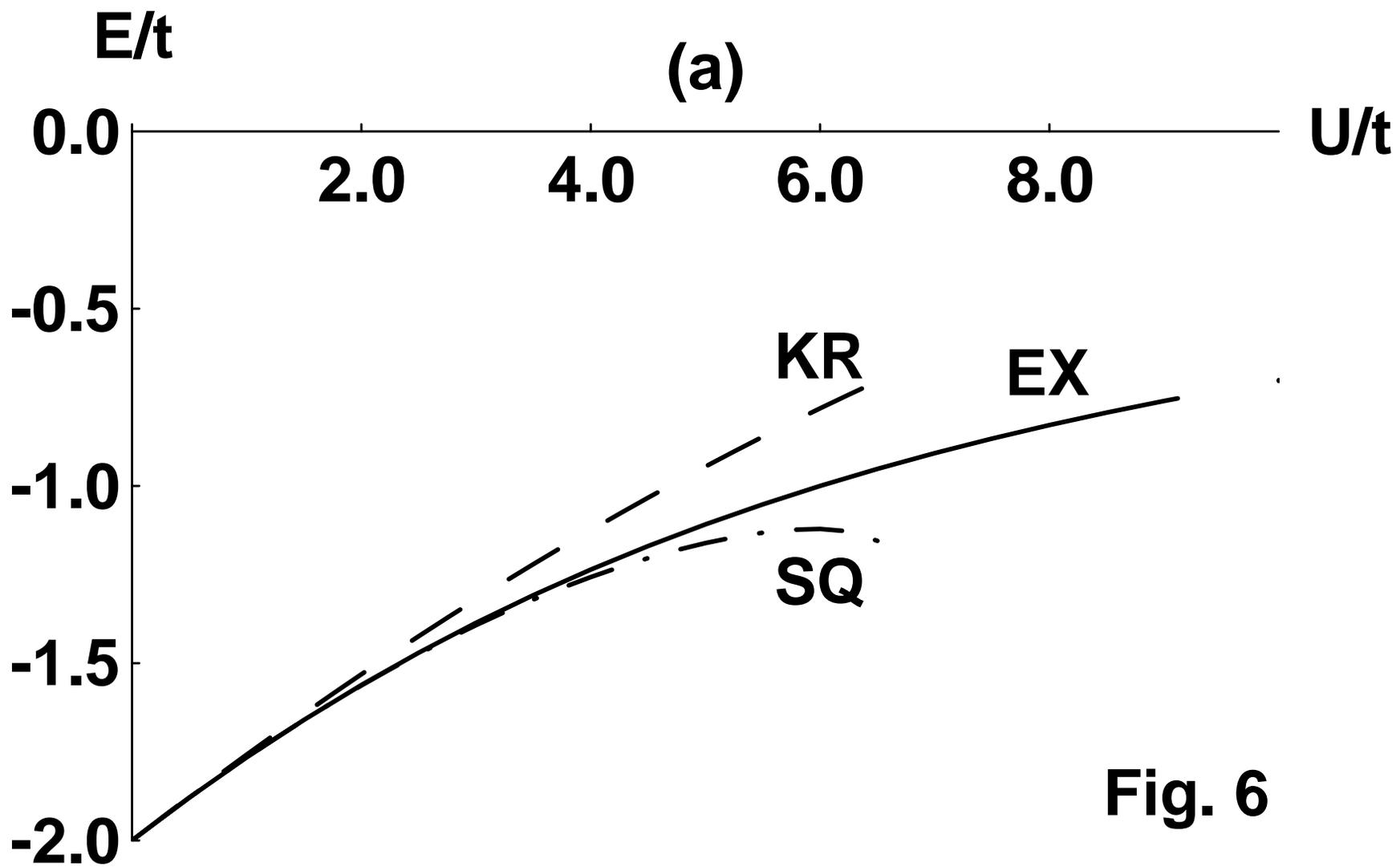
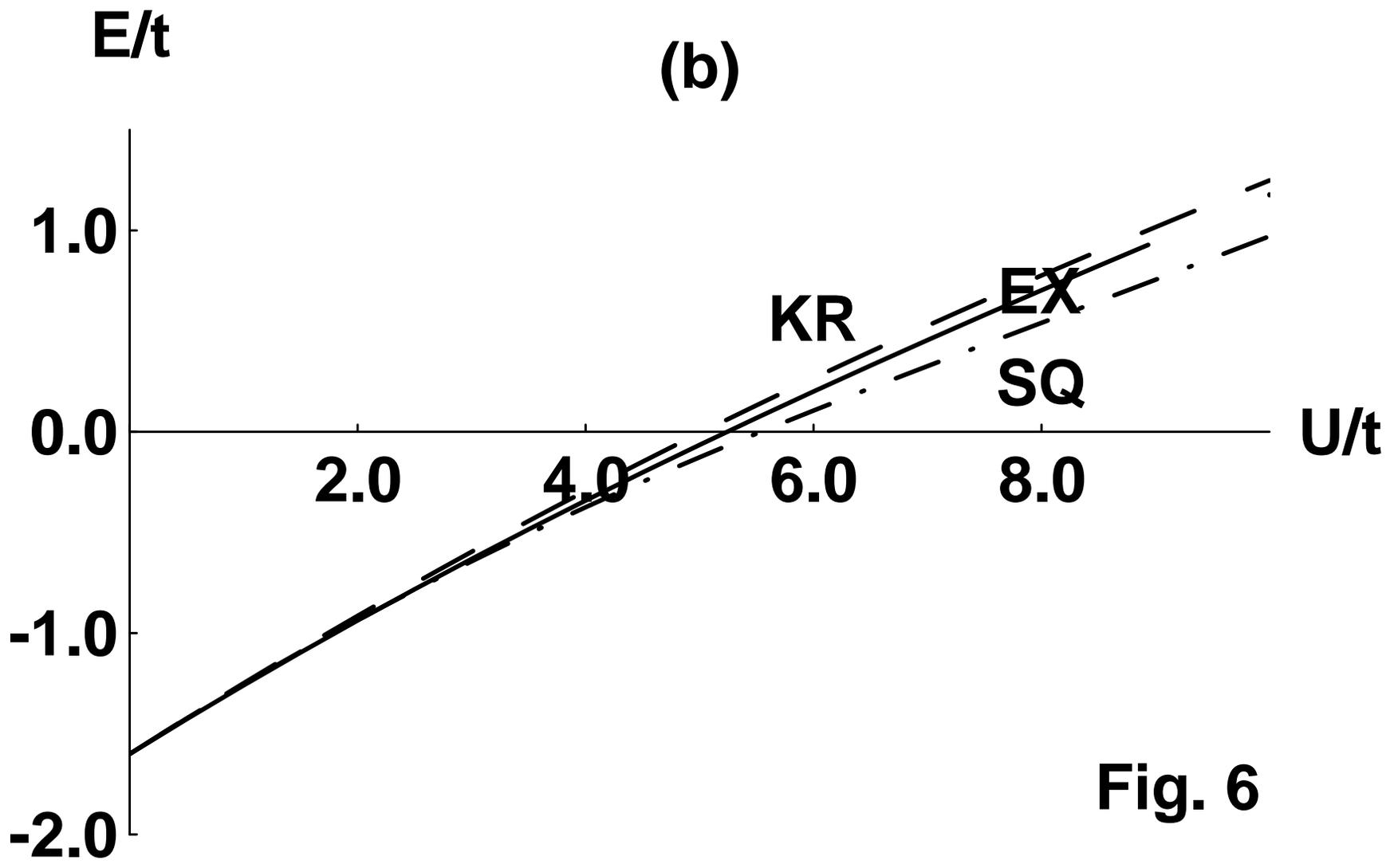
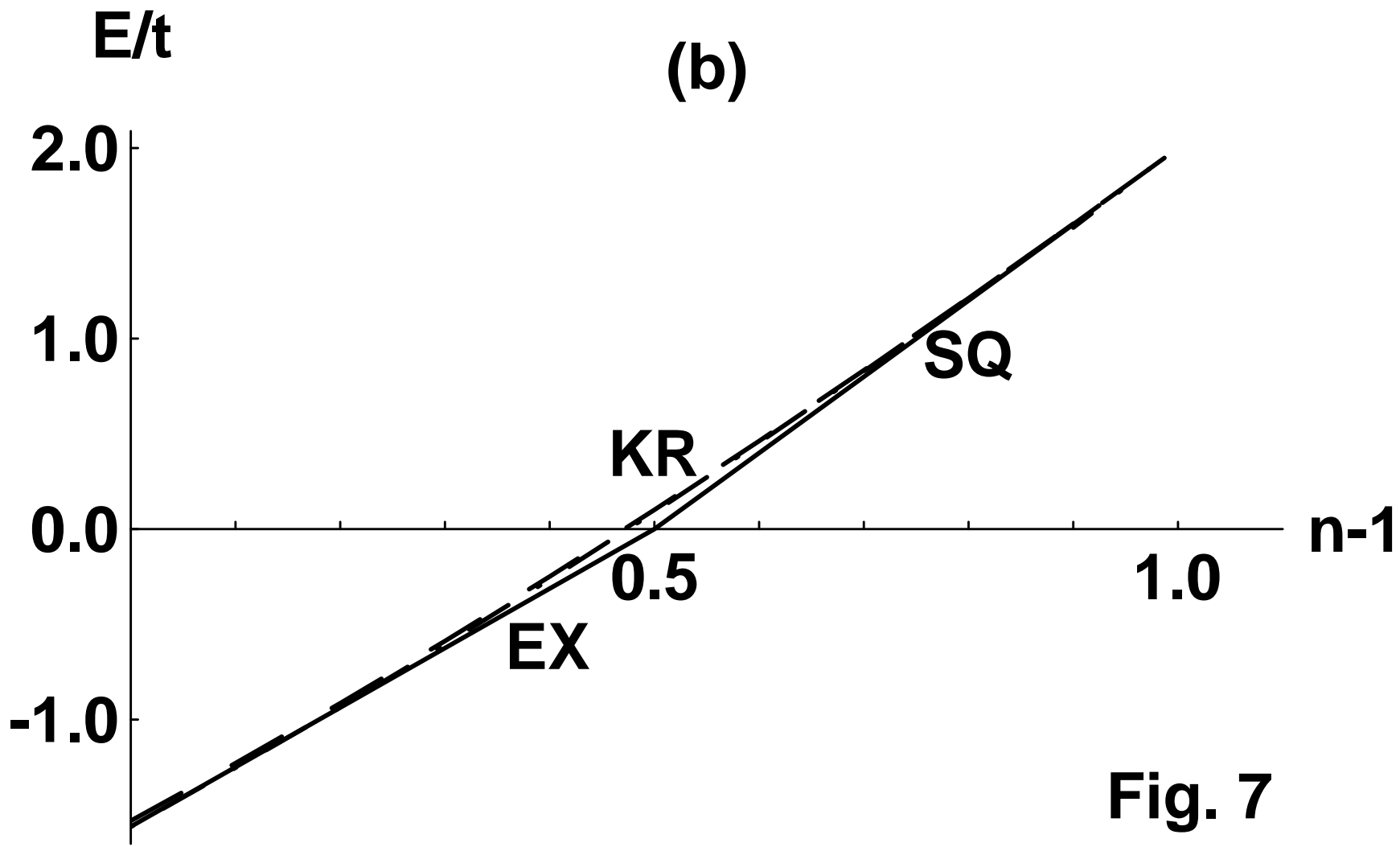
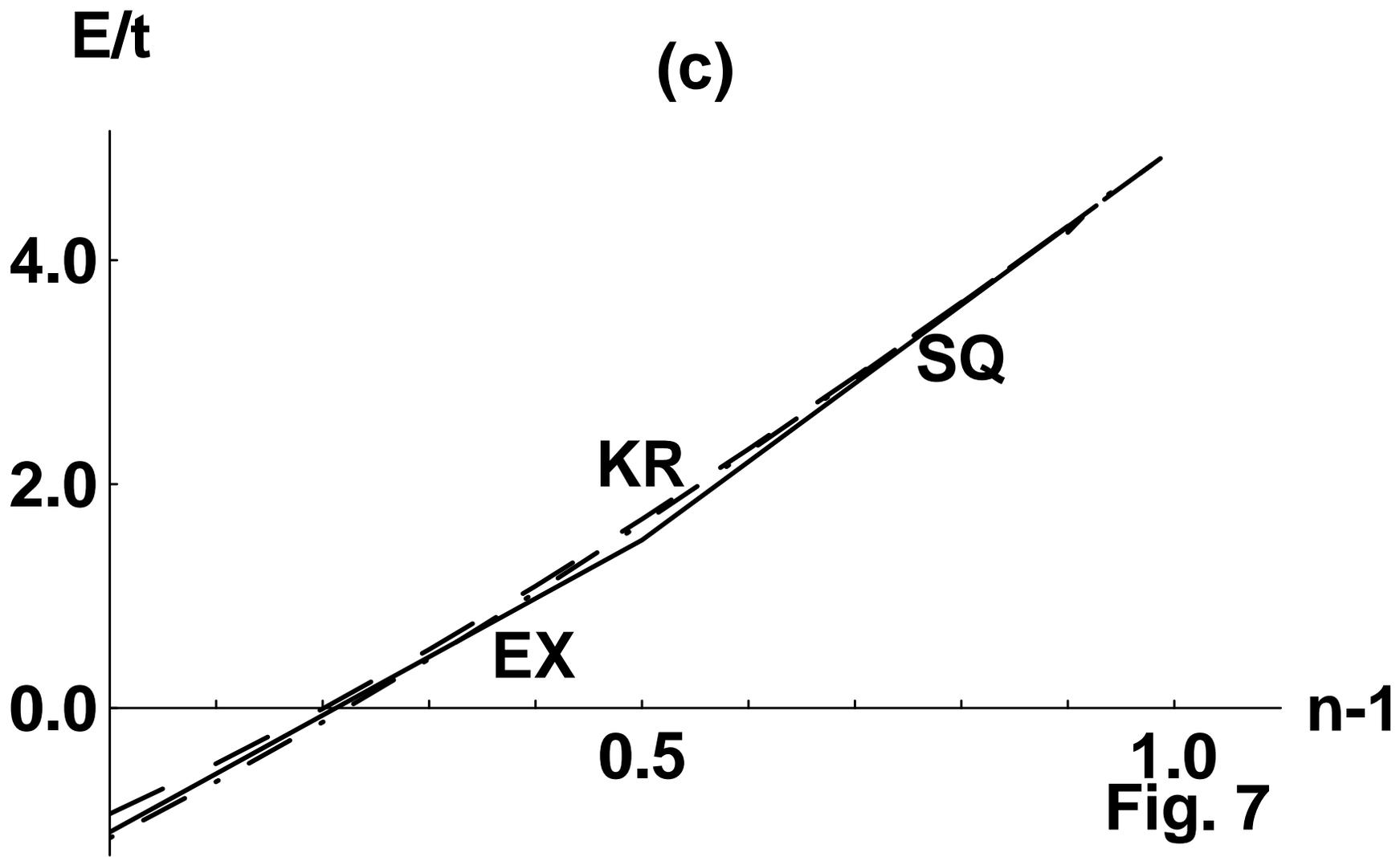
\end{document}